\documentclass[]{article}

\usepackage{listings}
\usepackage{fullpage}
\usepackage{epstopdf}
\usepackage{color}
\usepackage{graphicx}
\usepackage{amsmath}
\usepackage{amssymb}
\usepackage{permute}
\usepackage{tikz}
\usepackage[all]{xy}
\usepackage[utf8x]{inputenc}
\usepackage{import}
\usepackage{setspace}
\usepackage{authblk}
\usepackage{hyperref}
\usepackage{xfrac}

\definecolor{Sion}{rgb}{.45,0.05,.85}

\definecolor{todos}{rgb}{.1,.66,.45}

\begin{document}

\title{On the reorderability of node-filtered order complexes}
\author[1]{Ann Sizemore Blevins}
\author[1,2,3,4,5,6]{Danielle S. Bassett}
\affil[1]{Department of Bioengineering, School of Engineering \& Applied Science, University of Pennsylvania, Philadelphia, PA 19104 USA}
\affil[2]{Department of Physics \& Astronomy, College of Arts \& Sciences, University of Pennsylvania, Philadelphia, PA 19104 USA}
\affil[3]{Department of Electrical \& Systems Engineering, School of Engineering \& Applied Science, University of Pennsylvania, Philadelphia, PA 19104 USA}
\affil[4]{Department of Neurology, Perelman School of Medicine, University of Pennsylvania, Philadelphia, PA 19104 USA}
\affil[5]{Department of Psychiatry, Perelman School of Medicine, University of Pennsylvania, Philadelphia, PA 19104 USA}
\affil[6]{To whom correspondence should be addressed: dsb@seas.upenn.edu}

\maketitle

\begin{abstract}

Growing graphs describe a multitude of developing processes from maturing brains to expanding vocabularies to burgeoning public transit systems. Each of these growing processes likely adheres to proliferation rules that establish an effective order of node and connection emergence. When followed, such proliferation rules allow the system to properly develop along a predetermined trajectory. But rules are rarely followed. Here we ask what topological changes in the growing graph trajectories might occur after the specific but basic perturbation of permuting the node emergence order. Specifically we harness applied topological methods to determine which of six growing graph models exhibit topology that is robust to randomizing node order, termed global reorderability, and robust to temporally-local node swaps, termed local reorderability. We find that the six graph models fall upon a spectrum of both local and global reorderability, and furthermore we provide theoretical connections between robustness to node pair ordering and robustness to arbitrary node orderings. Finally we discuss real-world applications of reorderability analyses and suggest possibilities for designing reorderable networks.

\end{abstract}

\newpage

\newpage
\section{Introduction}

Growing networks can embody myriad developing systems spanning from maturing organisms in ecology and biology, to expanding connections in economics and sociology. For each of these systems, the representative growing network captures the normal growth process by noting when new actors, or nodes, join the mass and where relations between actors, or edges, form. Such a representation could, for example, allow us to understand spreading synchronization of coupled dynamical systems \cite{arenas2006synchronization} and crystallization processes in materials science \cite{rodriguez2016nonclassical,karthika2016review}. Outside of normal growth processes, we can also use growing networks to understand abnormality in maturation, enlargement, or spreading processes such as the propogation of a seizure through a brain network \cite{khambhati2015dynamic} or the dissipation of $\beta$-amyloid in Alzheimer's disease \cite{raj2012network,henderson2019quantitative}. Many of these examples include mechanisms or models that describe the specific growing process and as such inherently suggest a proper order of nodes necessary to achieve the desired final network architecture.

But what if, instead, the growth process was perturbed? Can normal development recover if, for example, those objects that are usually last to emerge were to instead to appear first in the growth order? Or how might the growth process react to a larger perturbation such as an entirely random reordering of the nodal emergence order? Intuition suggests that a networked system's robustness to such perturbations may depend upon some intrinsic property of the system, and that different systems will exhibit differing levels of robustness. In some systems, we may expect robustness to these types of perturbations; for example, a monodisperse physical system undergoing crystallization will achieve its final ordered form regardless of growth process minutiae, due to underlying thermodynamics \cite{chaikin1995principles}. Thus, any small reordering of particles joining the crystal will not change the outcome. In other systems, we expect sensitivity to these types of perturbation; for example, the rate of disease spread across a network can largely depend on the network organization surrounding those nodes that were infected earliest \cite{kitsak2010identification}, and therefore the global system outcome would indeed be susceptible to the order in which nodes were infected. 

Here we aim to understand a piece of the above phenomenon by formalizing and investigating the stability of a growing graph's developing architecture in response to perturbation of node addition order. We will refer to this notion of stability as the \emph{reorderability} of a growing graph. Thinking generally, a reordered growing graph with a developing architecture that is similar to that produced by the original node ordering may share many properties with the original growing graph. Still perhaps the simplest property that the two would share is a similar evolving global topology. Quantitatively tracking and characterizing that evolving global topology is made tractable by emerging tools from applied topology \cite{carlsson2009topology,carlsson2009computing,zomorodian2005computing}, specifically, persistent homology which records the pattern of topological cavity emergence throughout network growth. Such tools allow us to quantitatively compare the evolving topology of two or more growing graphs \cite{cohen2007stability}, facilitating inferences regarding their reorderability. 

In this work we examine the reorderability of six growing graph models chosen to span a range of topologies relevant for social systems, material systems, spatially embedded systems, and dynamical systems. We first ask if each of these growing graph models exhibits stable persistent homology after randomly reordering the node addition order, and we refer to the degree of such stability as the global reorderability of a growing graph model. We next ask which growing graph models show stable persistent homology in the face of temporally localized swaps in the node ordering, and we refer to the degree of stability as the local reorderability of a growing graph. Finally, the third direction of this study investigates how the local reorderability of a growing graph may (or may not) constrain its global reorderability. We observe that a spectrum of both global and local reorderability exists within the graph models tested. Furthermore we find that our notions of local and global reorderability are surprisingly distinct, which provides an opportunity to design growing processes with specific reorderability characteristics.

\section{Methods}
\subsection{Building intuitions}

Suppose we observe a growing graph such as the developing brain \cite{tang2017developmental}, a broadening social network \cite{mercken2012longitudinal,simpkins2013adolescent}, or a spreading vasculature network \cite{ronellenfitsch2016global}. How might this development change if we perturb the growth process slightly? Or if we instead perturb the growth process drastically? Consider for example the birth of neurons in the developing brain, which is a strictly ordered process \cite{yu2010complete,berry1964pattern}. Recent studies revealed that small amounts of stochasticity actually are advantageous for early developing populations of neurons \cite{he2012variable,gomes2011reconstruction,gritti2019random}. These results suggest that if only a few temporally close neurons swap birth times, healthy development will proceed as expected. Still, since overall neuronal development is extremely stereotyped \cite{pearson2004specification,yu2010complete} and since cell function is so linked to birth time \cite{varier2011neural,jacob2008temporal}, we might reasonably speculate that drastic birth order swaps, such as spawning progenitor cells last, would impede proper maturation. Ultimately the response of a growth process to a reordering of node birth times depends on both the network connections (the topology) and the original order of the nodes (our baseline to which we compare). Then our goal in this work is to understand how the connections and node order influence the stability of growth processes.

\begin{figure}
	\centering
	\includegraphics[width=6.5in]{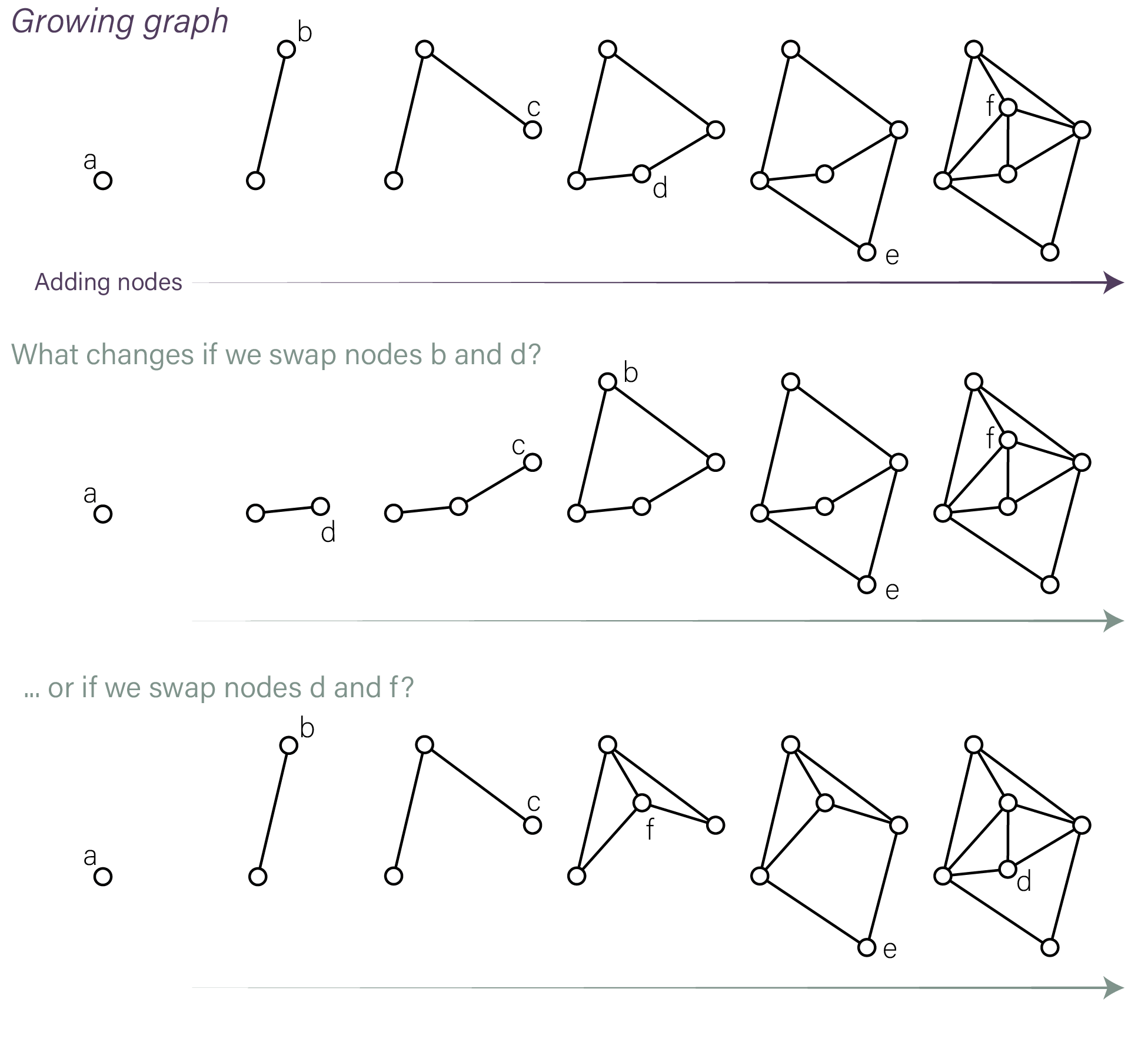}
	\caption{\textbf{A growing graph may or may not exhibit differing topology depending on the order of nodes in the growth sequence.} Given a growing graph (top row), we ask how the evolving topology changes after we swap nodes $b$ and $d$ (middle row) or after we swap nodes $d$ and $f$ (bottom row) in the node birth order. Curiously, the $b \rightleftarrows d$ swap does not affect the growing graph architecture, while the $d\leftrightarrows f$ swap results in a notable effect.}
	\label{fig:0a}
\end{figure}

To make our goal more concrete, we consider the growing graph in the top row of Fig.~\ref{fig:0a}. Each node $a,b,c,d,e$ and $f$ is added in turn, with new edges added only between the new node and nodes already present. After adding the first three nodes $a$, $b$, and $c$, our graph has only exhibited a tree structure. As we move forward in time, node $d$ creates a cycle, node $e$ forms a second loop, and then finally node $f$ tessellates the top cycle. Next we ask how the network grows after a perturbation to node order. Specifically, how would the maturing graph architecture change if we swapped nodes $b$ and $d$ in the ordering so that we build the same final graph, but the nodes are now reordered (row two of Fig.~\ref{fig:0a})? We observe that this perturbation does not seem to have changed the growing graph much -- after the first three nodes we have only seen tree structures, then we see a 4-node loop after the fourth node, a second loop after the fifth node, and a tessellated cycle after the sixth node. At each slice (node addition) the original and perturbed growing graphs are quite similar (isomorphic in this example), and we can say that our growing graph is stable with respect to \emph{this particular} node swap. But perhaps we did not change the evolving graph much because nodes $b$ and $d$ were only two steps apart in the original ordering. Let us instead swap $d$ and $f$ (also two steps apart in the original ordering) and observe the effects. We see in the third row of Fig.~\ref{fig:0a} that this $d\rightleftarrows f$ switch drastically changes the growing graph. We see triangles much earlier than in the original ordering, differing numbers of edges, and by the time node $d$ is added forming the loop with $a$, $b$, and $c$, that loop has already been tessellated. 

\begin{figure}
	\centering
	\includegraphics[width=6in]{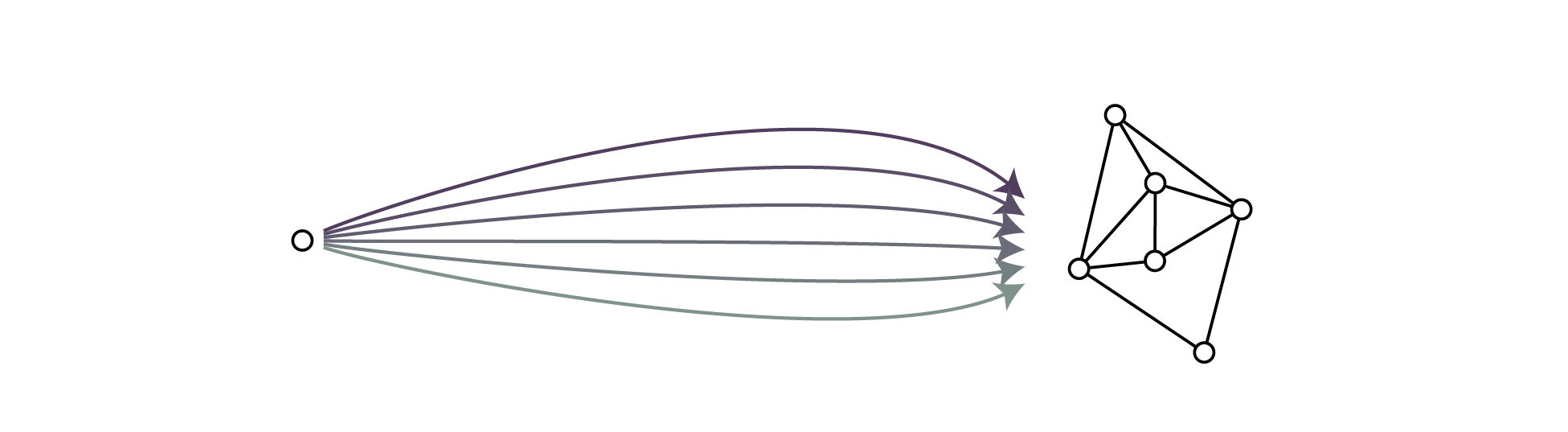}
	\caption{\textbf{Illustrating growing graphs as graph paths.} A possible perspective on growing graphs chooses a path of (unlabeled) graphs of increasing size. Permuting node order allows one to choose different paths from a graph on one node to the final binary graph.}
	\label{fig:0b}
\end{figure}

The example just described suggests that the reorderability problem is at least more complicated than simply understanding how far nodes move between the reordered growing graph and the original. Furthermore, not all growing graphs exhibit change after node reordering. As shown in Fig.~\ref{fig:totally_reorderable} and discussed further in Example 2 in the Supplement, growing trees, cross-polytopes, and cliques all exhibit total reorderability (in at least one dimension, see Example 2 for more details). So indeed some growing graphs can always have their growth reordered without a change to their growing topology, while others (as in Fig.~\ref{fig:0a}) cannot. Generally, if we always begin with one node and end with the same final binary graph, we might also frame this study as an attempt to understand the possible variability in growing a particular graph. One may choose many different node orders for a growing graph, but we can think of each node order choice as a path from one node to the final graph (Fig.~\ref{fig:0b}), and we consider the degree of variation in the growing process as akin to a distance between paths.

Importantly we note that studying how growing graphs may or may not change after reordering node birth times is not a new concept. Indeed from the statistical analysis perspective on networks, great work has been performed in understanding vertex- and edge-exchangeable graphs, which are sequences of graphs whose distributions are invariant with respect to the order of node or edge addition, respectively \cite{orbanz2014bayesian,aldous1985exchangeability,aldous1981representations,hoover1979relations,kallenberg1990exchangeable,crane2016edge}. Results in this area include (but are not limited to) understanding how graphons and point processes relate to exchangeable graphs \cite{borgs2016sparse,lovasz2006limits} and how one can construct exchangeable graphs \cite{cai2016edge,lloyd2012random}. Our work differs from the above literature in that we have a specific topological focus which offers us a unique set of mathematical tools with which to understand growing graphs. Still, we suspect interesting links exist between these ideas of exchangeability and reorderability, and leave the unraveling of such relations for future endeavors.

\subsection{Adding rigor to our intuitions}

In order to explicitly study questions of reorderability, we need to first formalize a few of the above concepts. If we assume that we will add nodes one at a time (for the sake of simplicity), and that we can only add edges -- between the new node and any node already present -- at the time of node addition, then we can define a growing graph with only a binary graph $B$ and node order $s$. Explicitly, we define a growing graph as a pair $(B,s)$ with $B = (V,E)$ a binary graph with vertex set $V$ and edge set $E \subseteq V \times V$, and $s$ being the map $s:V \rightarrow \mathbb{N}$ that defines the order in which we add nodes (Fig.~\ref{fig:Methods_1}a). Note that we could also define $s$ as a map $s:V \rightarrow \mathbb{R}$, and then keep only the ordering of the nodes, as has been done in previous work \cite{sizemore2018knowledge}. In the exposition to follow, we will describe the process by which we will extract the topology of a growing graph via persistent homology. For additional intuition on the relevant methods, we direct the interested reader to several excellent sources for further details \cite{ghrist2018homological,otter2017roadmap,zomorodian2005computing,carlsson2009topology,edelsbrunner2012persistent}. 

\subsubsection*{From graph to simplicial complex}

\begin{figure}
	\centering
	\includegraphics[width=6.5in]{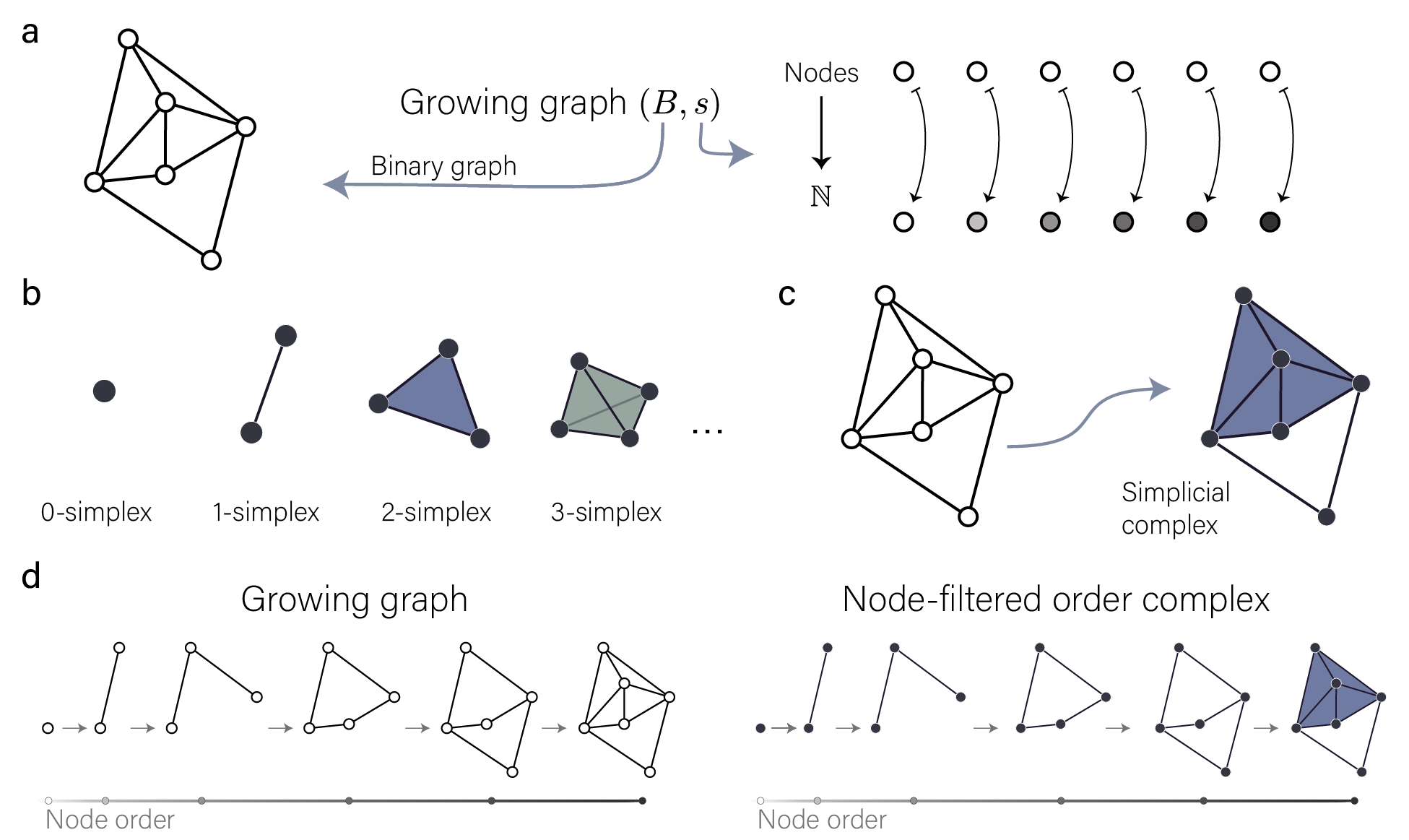}
	\caption{\textbf{Illustration of the steps taken to convert a growing graph to a node-filtered order complex.} \emph{(a)} A growing graph is defined as a pair $(B,s)$ with $B$ a binary graph and $s$ being an ordering of the nodes. \emph{(b)} Examples of simplices defined using 1 to 4 nodes. \emph{(c)} Filling in all cliques of a binary graph (left) with simplices yields a simplicial complex (right). \emph{(d)} An example growing graph (left) and the corresponding filtered simplicial complex (right), here called the node-filtered order complex. }
	\label{fig:Methods_1}
\end{figure}

To examine the topology of a growing graph, we need to chronicle the topological cavities or voids that are absent of connections nested within the evolving architecture, which consists of a sequence of binary graphs (one graph after every node addition). To first capture the cavities of all dimensions within a \emph{single} graph we must translate the graph into a higher-relational structure by (abstractly) filling in all cliques with a (higher-dimensional) volume of matching dimension, called a simplex. Specifically, an $(n+1)$-clique is a set of $(n+1)$ nodes that are all-to-all connected by edges in a graph, and an $n$-simplex is the convex hull of $n+1$ affinely positioned nodes (see Fig.~\ref{fig:Methods_1}b for examples). A collection of simplices nicely glued together is called a simplicial complex, which -- more rigorously -- is a set of vertices $V$ and a set of simplices $K$ such that if $\sigma \in K$ and $\tau \subseteq \sigma$ then $\tau \in K$. We create a simplicial complex from a binary graph by assigning an $n$-simplex to each $(n+1)$-clique in the graph as shown in Fig.~\ref{fig:Methods_1}c. To reiterate, we fill all cliques of a graph with simplices so that we can detect topological cavities of all dimensions.

\subsubsection*{From growing graph to node-filtered order complex}

Note that we can perform the conversion from graph to simplicial complex for \emph{every} binary graph in our growing process. That is, after each node addition to the growing graph, we can construct a simplicial complex by filling in cliques with simplices. Furthermore, since we only add elements to our graph as it grows, the maps from the growing graph (Fig.~\ref{fig:Methods_1}d, left) induce maps between simplicial complexes, now giving us a growing simplicial complex and formally referred to as a filtered simplicial complex (Fig.~\ref{fig:Methods_1}d, right). Previous studies focusing on filtered simplicial complexes arising from adding edges one at a time and translating all cliques to simplices referred to the construction of these edge-filtered simplicial complexes as the \emph{order complex} of a weighted graph \cite{giusti2015clique}. The rationale for this name is that the filtered simplicial complex retains the ordering information given by the edge weights (notably this construction is also called the Weight Rank Clique Filtration \cite{petri2013topological,petri2014homological,otter2017roadmap,stolz2016topological} as it preserves the ranks of edge weights). Since in our growing graph case we have an ordering of the nodes (as opposed to an ordering of the edges), we call our above described filtered simplicial complex the \emph{node-filtered order complex} of the growing graph $(B,s)$ (Fig.~\ref{fig:Methods_1}d), and we note that this complex can be constructed from either a node-ordered or node-weighted network in which node weights induce the node order \cite{sizemore2018knowledge}. For the sake of brevity, here we shorten this name to the n-order complex of $(B,s)$, which we denote as $nord(B,s)$. We suggest that the node-filtered language described here is potentially a natural description of real network growth. 

\subsubsection*{Homology of a simplicial complex}

\begin{figure}
	\centering
	\includegraphics[width=6.5in]{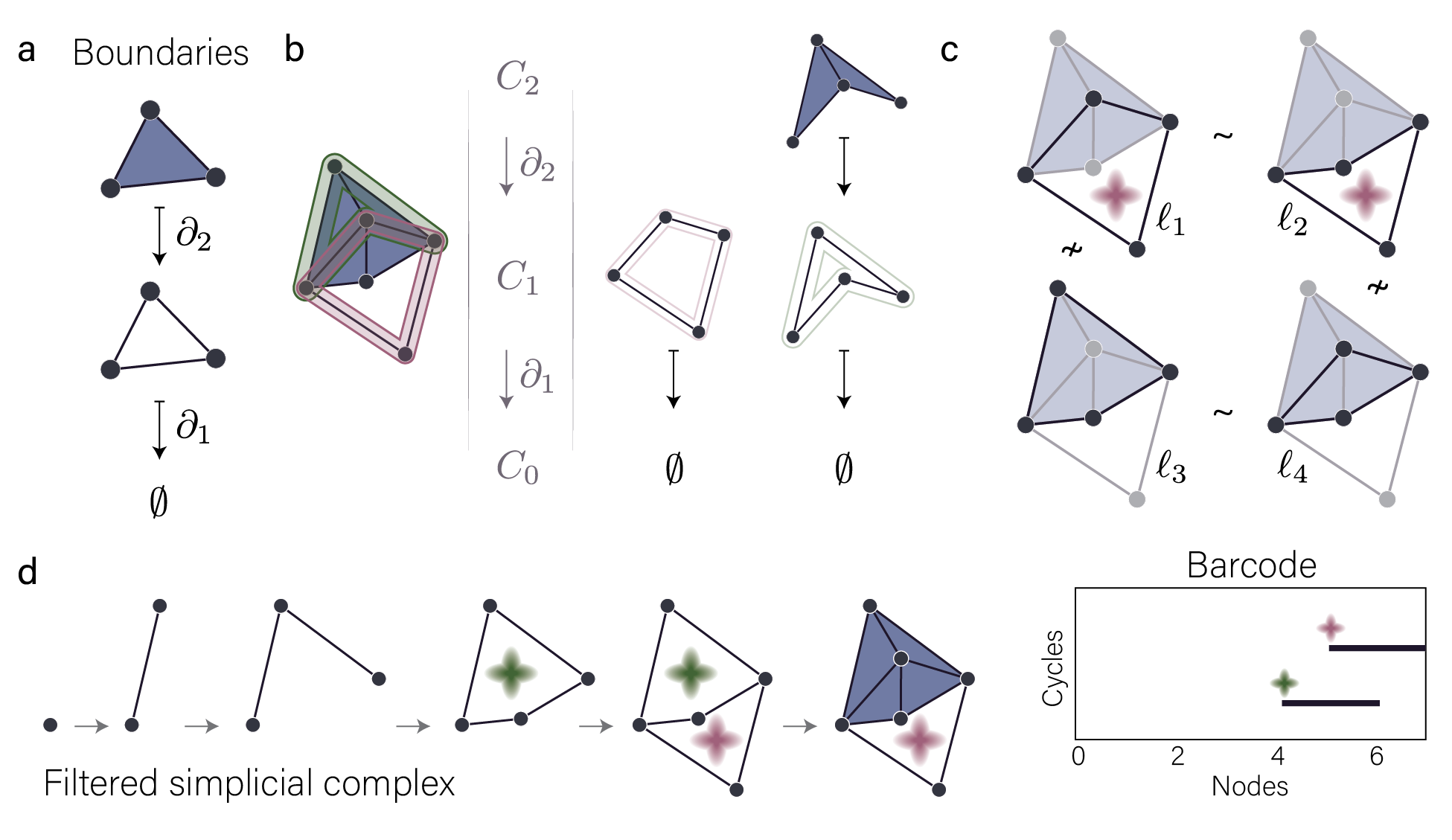}
	\caption{\textbf{Concepts in persistent homology.} \emph{(a)} The boundary of a 2-simplex is three 1-simplices, whose boundary in turn is empty. \emph{(b)} The green and pink 1-cycles both have a zero boundary, but the green cycle is the boundary of a collection of 2-simplices. \emph{(c)} Loops $\ell_1$ and $\ell_2$ are equivalent as they both surround the same cavity (pink). Cycles $\ell_3$ and $\ell_4$ are equivalent as they both surround no cavities. \emph{(d)} The example filtered simplicial complex contains two persistent cavities of dimension 1 (green and pink stars), which are recorded in the barcode shown on the right.}
	\label{fig:Methods_2}
\end{figure}

We aim to understand the topology of the n-order complex as defined by cavities or voids within the structure, as detected by homology. To begin the homology computation, for each dimension $n$, we create the vector space $C_n(K)$ (over $\mathbb{F}_2$, the binary field) in which each basis vector of $C_n(K)$ corresponds directly to a particular $n$-simplex in the simplicial complex $K$. For simplicity we conflate the basis vector and the associated $n$-simplex, and often we will write simply $C_n$ when the simplicial complex under study is obvious. Note that since the physical boundary of an $n$-simplex is a shell of $(n-1)$-simplices, we immediately acquire a relation between simplices of dimension $n$ and $n-1$. This relation allows us to map from $C_n \rightarrow C_{n-1}$ via the boundary operator $\partial_n$, which is explicitly a linear map that sends an $n$-simplex to its boundary $(n-1)$-simplices in $C_{n-1}$ (Fig.~\ref{fig:Methods_2}a). Each $C_n$ is called the chain group of dimension $n$ and the collection of chain groups and boundary maps forms a chain complex 
$$ \dots C_{n+2} \xrightarrow{\partial_{n+2}} C_{n+1} \xrightarrow{\partial_{n+1}} C_n \xrightarrow{\partial_{n}} C_{n-1} \xrightarrow{\partial_{n-1}} ...$$ which contains all of the assembly information for the simplicial complex.

Homology, generally, counts the voids of a particular dimension within a simplicial complex. Naturally, a void within the complex must be both encapsulated by the complex (being surrounded by a cycle or loop of simplices) and have at least some empty space within it (not being filled with higher dimensional simplices). We in fact already have all of this information housed in the chain complex: the space of cycles of dimension $n$ is $\ker\partial_n =: Z_n$ and the space of boundary cycles -- that is, shells of simplices that are boundaries of higher dimensional simplices -- is $\text{im}\partial_{n+1} =: B_n$ (for examples of cycles in different dimensions, see Example 1 and Fig.~\ref{fig:cavities} in the Supplement). Note that by definition $B_n \subseteq Z_n \subseteq C_n$ since the boundary of a boundary is null (Fig.~\ref{fig:Methods_2}b). Two cycles surround the same cavity if they differ by a boundary cycle, which induces an equivalence relation on cycles. That is, for $\ell_1, \ell_2 \in Z_n$, $\ell_1 \sim \ell_2 \iff \ell_1 - \ell_2 \in B_n$ so that all cycles of $n$-simplices that surround the same cavity (or cavities) are equivalent. As an example, in Fig.~\ref{fig:Methods_2}c we see $\ell_1 \sim \ell_2$ and that they surround the same cavity, while $\ell_3\sim\ell_4$ as they both are boundaries of 2-simplices. Intuitively, since any boundary loop (element of $B_n$) surrounds no cavities, we can add a boundary cycle to any other cycle $\ell$ and not change the cavities that the cycle $\ell$ surrounds. We can collect all loops that are equivalent to a particular loop $\ell_0$, which comprise an equivalence class $[\ell_0] = \{\ell \in Z_n | \ell \sim \ell_0\}$. Then to pass from the space of cycles $Z_n$ to the space of equivalence classes in which each basis element corresponds to an independent void in the complex, we take the vector space quotient and finally define $H_n(K) := Z_n / B_n$. 

\subsubsection*{Persistent homology of a filtered simplicial complex}

Moving beyond one simplicial complex, we now consider the computation of persistent homology of a filtered simplicial complex $K^1 \hookrightarrow \dots \hookrightarrow K^m \hookrightarrow \dots \hookrightarrow K^M$ where for our purposes the simplicial complex $K^m$ contains $m$ vertices and $M$ is the total number of nodes. The map $K^m \hookrightarrow K^{m+1}$ defines a map between chain complexes $C_*(K^m) \hookrightarrow C_*(K^{m+1})$ in which basis elements map to basis elements, in a manner consistent with how simplicies map to simplices. Now, the map between chain complexes induces a map between homology groups $H_*(K^m) \rightarrow H_*(K^{m+1})$, since cycles map to cycles and boundaries map to boundaries in $C_*(K^m) \hookrightarrow C_*(K^{m+1})$. Having maps between homology groups allows us to track persistent cavities (equivalently called persistent cycles) along the filtration (Fig.~\ref{fig:Methods_2}d green and pink stars). A persistent cycle will be born at some point in the filtration, called the \emph{birth}, will live for some time (i.e. mapped via $H_*(K^m) \rightarrow H_*(K^{m+1})$ to a non-trivial equivalence class), and then may finally die (when it maps to the trivial equivalence class, as does the green cavity in Fig.~\ref{fig:Methods_2}d) via tessellation by higher dimensional simplices. Those persistent cycles that do not die are formally given a death time of $\infty$. Then the persistent homology can be summarized as a (multi)-set of pairs $(b,d)$, $b\leq d$, marking the birth and death time of each persistent cycle. We can visually display these pairs as intervals along the filtration axis, drawing a bar from $b$ to $d$ for each persistent cavity. The resulting plot is called a barcode and is illustrated pictorially in Fig.~\ref{fig:Methods_2}d. Finally, to more easily summarize the persistent homology, we might only record the number of persistent cavities of dimension $n$ alive at each filtration value $m$ with a function called the Betti curve $\beta_n(m)$. The area under $\beta_n(m)$ is given by $\overline{\beta_n} = \sum d_i - b_i$ with $i$ indexing over persistent cavities of dimension $n$, and intuitively describes the total amount of persistent homology seen throughout the growth process. To summarize, we now can take a growing graph model, compute the evolving topology via persistent homology, and record the output as a barcode or Betti curve.

We now have the necessary background to extract the evolving topology from a growing graph $(B,s)$ using persistent homology. With this background we can reexamine the example in Fig.~\ref{fig:0a} and see that the persistent homology of the growing graphs in the top and middle rows will be the same, while the persistent homology of the growing graph in the last row will be quite different. Explicitly, the top two rows show two persistent cavities: one born after node 4 and killed with the addition of node 6, and a second born with the addition of node 5 that persists to $\infty$. Conversely, the bottom row shows only one persistent cavity born with the addition of node 5 that persists to $\infty$. Note also that this observation suggests that nodes $b$ and $d$ contribute similarly to the persistent homology -- both $b$ and $d$ participate in the top persistent cycle, while nodes $d$ and $f$ serve distinct functions as the former usually births a persistent cavity, while the latter usually kills the cavity. The quantification of differences between the persistent homology of two growing graphs will then allow us to make our notions of reorderability more rigorous, and therefore we next focus on the practical consideration of measuring formal distances between topological summaries.

\subsection{Distances between barcodes and between Betti curves}

Since we wish to compare the persistent homology of growing graphs, we need to define a distance between Betti curves and a distance between barcodes. First considering the former, we note that Betti curves record the homology of the growing graph slice by slice, and we are therefore motivated to compare Betti curves in a slice-wise fashion. Specifically, we define the distance between two Betti curves of dimension $n$, $\beta_n$, $\beta_n'$ to be 

\begin{equation} \label{eq:0}
d_{\beta}^n(\beta_n,\beta_n') = max_i(|\beta_n(m)-\beta_n'(m)|)
\end{equation}

\noindent with $m$ indexing the number of nodes added. Intuitively, this distance records the maximal vertical distance between the Betti curves at any point in the filtration. Second, considering persistence diagrams, we use the bottleneck distance \cite{cohen2007stability} for its simplicity and interpretability, although we recognize that alternatives exist \cite{bubenik2015statistical,chevyrev2018persistence,edelsbrunner2012persistent}. Generally, the bottleneck distance measures the maximal left or right difference between matched bars in two barcodes after optimally matching the bars. More specifically, for two barcodes $P_1$, $P_2$ of dimension $n$, 

\begin{equation} \label{eq:1}
d^n_{BN}(P_1,P_2) = min_{\gamma}(max_{p\in P_1}(|p_{birth}-\gamma(p)_{birth}|,|p_{death}-\gamma(p)_{death}|)) \end{equation}

 \noindent where $p = (p_{birth},p_{death})$ is a bar in $P_1$ and $\gamma$ is a matching of bars so that $\gamma(p)$ is a bar of $P_2$. Note that bars are allowed to match to any point $(r,r)$ for $r\in \mathbb{R}^+$ as well. Importantly, as shown in \cite{cohen2007stability}, the persistent homology with respect to the bottleneck distance is as stable as we could hope under perturbation. That is, if $w_1,w_2$ are weight functions on nodes determining node order, then 
 
 \begin{equation} \label{eq:2}
 d^n_{BN}(P_1,P_2) \leq ||w_1 - w_2||_{\infty} \hspace{10px} ,
 \end{equation}
 
\noindent with $\|\cdot\|_{\infty}$ the $L_{\infty}$-norm. Said another way, if we take the $i^{th}$ and $j^{th}$ node in the ordering and swap them, the bottleneck distance between the original and swapped persistent homology is bounded above by $|j-i|$. In the Supplement we include a section that provides more detail and intuition for this upper bound (see Fig.~\ref{fig:Supp_stability}).

Specifically in the context of our study, we wish to compare the persistent homology of a growing graph under its original ordering $(B,s_0)$ and under a new ordering $(B,s_{i,j})$ in which we have swapped only the $i^{th}$ and $j^{th}$ nodes, $v_i$ and $v_j$. Then the bottleneck distance between the persistent homology of these growing graphs, $P_0$ and $P_{i,j}$, respectively, describes the effective perturbation of the persistent homology caused by the $v_i,v_j$ swap. Furthermore, $d^n_{BN}(P_0,P_{i,j})$ then gives a measure of how similarly $v_i$ and $v_j$ contribute to the persistent homology in dimension $n$. If $d^n_{BN}(P_0,P_{i,j}) = 0$, then the $v_i,v_j$ swap does not change the persistent homology. If instead $d^n_{BN}(P_0,P_{i,j}) = |j-i|$, then the $v_i$, $v_j$ node swap causes the maximal effective perturbation to the persistent homology. Combining the bottleneck distance with the theoretical upperbound, we can create a measure of similarity between two nodes in terms of their roles in the persistent homology. We define the \emph{topological similarity} in dimension $n$ as 

\begin{equation} \label{eq:3}
T_n(v_i,v_j) = 1- \frac{d^n_{BN}(P_0,P_{i,j})}{|j-i|}
\end{equation}

\noindent so that $v_i$ and $v_j$ have a topological similarity of 1 if their swap in the node ordering does not change the persistent homology, and a topological similarity value of 0 if the persistent homology changes maximally as a result of the node swap. We can more generally discuss the (weighted) topological similarity of $v_i$ and $v_j$ by averaging over dimensions, that is 
\begin{equation} \label{eq:4}
T(v_i,v_j) = \frac{1}{D}\sum_{n=1}^{D}T_n(v_i,v_j).
\end{equation}

\subsection{Graph summaries and statistics}

In this work we will ask how topological similarity might form motifs described by commonly used graph summaries. First recall that the \emph{degree} of node $v_i$ is the number of edges incident to node $v_i$, and that it is commonly denoted $k_i$. If we also have edge weight information, we can calculate the \emph{strength} of node $v_i$ as 

\begin{equation} \label{eq:5}
k^w_i = \sum_{i=0, i\neq j}^N w_{i,j}
\end{equation}

\noindent where $w_{i,j}$ is the weight of edge $(v_i,v_j)$. Additionally we will ask how the topological similarity between nodes relates to the topological overlap of the nodes in the original graph. We acknowledge that different definitions of topological overlap exist \cite{yip2007gene,ravasz2002hierarchical}, but for our purposes we will adapt the definition from \cite{ravasz2002hierarchical} slightly and define topological overlap between nodes $v_i$ and $v_j$ as

\begin{equation} \label{eq:6}
O(v_i,v_j) = \frac{l_{i,j} + B_{i,j}}{\max(k_i,k_j)}
\end{equation}

\noindent with $B_{i,j} = 1$ if $v_i,v_j$ are connected by an edge in binary graph $B$ and 0 otherwise, and where $l_{i,j}$ is the number of nodes $v_k$ with $k\neq j, k\neq i$ that are neighbors of both nodes $v_i$ and $v_j$.  Intuitively, the numerator counts the number of shared neighbors between nodes $v_i$ and $v_j$, and the denominator normalizes by the larger of the two node degrees. Then two nodes who share exactly the same set of neighbors (and no more) will have a topological overlap of 1 and nodes that are not connected and share no neighbors will have a topological overlap of 0. Finally we ask about the community structure of the topological similarity graph, by applying the Louvain method for modularity maximization \cite{blondel2008fast} implemented in iGraph \cite{csardi2006igraph}.

\subsection{Growing graph models}

Thus far we have developed our intuitions on a single growing graph. Yet, in order to understand reorderability at a larger scale we must work with models that can generate ensembles of growing graphs. Here we will test the reorderability of six growing graph models that span a range of topologies including both embedded and non-embedded models. For each model we will generate a growing graph (and later an n-order complex) by adding nodes $v_1, v_2, \dots, v_N$ one at a time for a final count of $N=70$ nodes. For clarity we refer to the order $v_1, v_2, \dots, v_N$ as $s_0$, where the subscript reflects the fact that this order is the original ordering for the growing graph. Note that the current study differs in kind from those that focus on edge weighted networks, in which both the number of nodes and number of edges are easily held constant. Here in the growing graph case, we seek to compare growing processes across the number of nodes added, and thus we do not restrict our study to graph topologies with a fixed edge density. 

The six growing graph models that we study are described in greater detail below.

\begin{itemize}
	\item \textit{Constant probability.} We begin with a random model from \cite{gilbert1959random,erdos1959random} that aims to capture random topology with few constraints. At the addition of node $v_i$, edges are added between node $v_i$ and all previously added nodes $v_*$ with a constant probability $p(v_i,v_*) = c \in [0,1]$. We report results for $c=0.4$ in the main text, and note that results for $c = 0.3$ can be found in Ref. \cite{sizemore2018knowledge}.
	
	\item \textit{Proportional probability.} Next we include a random model in which connection probabilities increase throughout growth, instead of being held constant. As in \cite{sizemore2018knowledge}, when adding node $v_i$ we add each edge between $v_i$ and previously added nodes $v_*$ with a probability proportional to the node number added, $p(v_i,v_*) = i/N$. 
	
	\item \textit{Oscillating probability.} Here we examine the effects of a non-monotonic function determining connection probabilities on the growing topology. In the oscillating probability model, we determine the edge probability using an oscillatory function; that is, we add edges between the new node $v_i$ and previous nodes with probability $p(v_i,v_*) = |sin(\alpha\pi\frac{i}{N})|$. Here we chose $\alpha = 2$ in order to ensure that oscillations in topology that match the periodicity are clearly observable by eye. The $\alpha$ parameter controls the frequency of connection probability oscillations and as such, larger values of $\alpha$ would be appropriate for larger growing graphs (i.e. those comprised of more nodes) so that the effects of each peak can be more easily determined.
	
	\item \textit{Preferential attachment.} The preferential attachment model famously constructs a graph with a heavy-tailed degree distribution via its ``rich-get-richer" growth algorithm. Following \cite{barabasi1999emergence,price1976general} we begin with a connected random graph on $m_0$ nodes. We then add nodes one at a time and at the addition of node $v_i$, we connect $v_i$ with $m$ previously added nodes with a probability proportional to the degree of each node already present. These rules produce a growing graph in which high degree nodes are likely to continue receiving more connections as nodes are added, in comparison with low degree nodes. We show results for $m=m_0 =4$ in the main text. 
	
	\item \textit{Random geometric.} Next we move to embedded graphs and begin with a common random graph designed to capture topology arising only from constraints imposed by the embedding. Directly inspired by \cite{kahle2011random,yogeshwaran2015topology,owada2017limit}, we sample $N$ points uniformly at random from $[0,1]^d$. We choose a threshold $\epsilon$ and create a binary random geometric graph with edges existing if $d(v_i,v_j) < \epsilon$. To mimic an observer moving through the embedded graph, we order nodes by the value of the first coordinate. For this study, we chose $d=3$ and $\epsilon=0.15$, in order to ensure that at least some persistent homology would be present by the end of the filtration \cite{sizemore2016classification,kahle2011random}.

	\item \textit{Spatial growth.} Alternatively, an embedded graph may randomly grow out from a particular point, spawning new nodes that attach most often to nearby nodes. Following \cite{kaiser2004spatial}, we begin with one node at $(0.5,0.5)$ in $[0,1]^2$. We add nodes $v_2,\dots,v_N$ one at a time, first choosing a location in the unit square uniformly at random, and then connecting the new node $v_i$ to previously added nodes $v_*$ with probability $p(v_i,v_*) = \beta e^{-\alpha d(v_i,v_*)}$. Here $d(v_i,v_*)$ is the Euclidean distance between node positions, $\beta$ is the density parameter, and $\alpha$ controls the spatial range. We show results for $\beta = 1$ and $\alpha = 4$ in the main text, which was shown to produce graphs in the small-world regime \cite{kaiser2004spatial}.

\end{itemize}

For each growing graph model, we first generate 1000 growing graphs $(B_{\alpha},s_0)$, $\alpha = 1,\dots, 1000$ and then we construct the n-order complex as described above by filling in cliques with simplices. Since we can uniquely associate the pair $(B,s)$ with $nord(B,s)$, we will use the growing graph language for simplicity in the Results section when it does not cause confusion. We compute the persistent homology using Eirene \cite{henselmanghrist16} and we use the TDA package in R (https://CRAN.R-project.org/package=TDA) for bottleneck distances. For the sake of the calculations, any persistent cavity that persists through the end of the filtration (or formally has a death time of $\infty$) is assigned a death time of $N+1$. Code for all analyses and growing graph models can be found at https://github.com/BassettLab/Reorderability\_scripts.

\section{Results}

\subsection{Meet the team}

\begin{figure}
\centering
\includegraphics[width=6.5in]{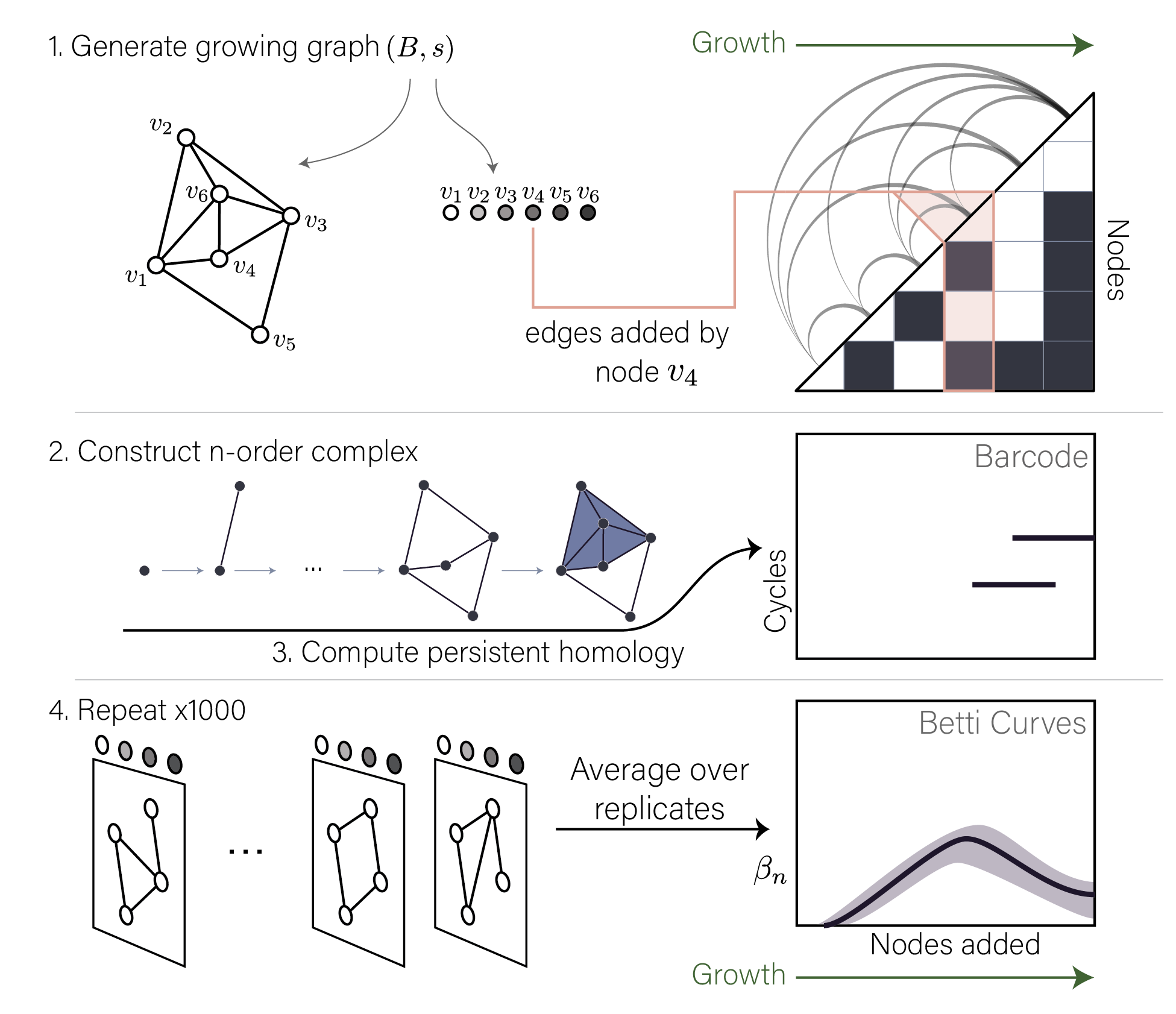}
\caption{\textbf{Outline of the computation of persistent homology for growing graph models.} (1) Following the model growth rules we first construct a growing graph, illustrated by the adjacency matrix and fingerprint graph (right). In the fingerprint graph, we order node positions via ordering $s_0$ and display edges between nodes in $B$ as gray arcs, so that at the addition of $v_4$, for example, we see two arcs reaching leftwards to connect to nodes $v_1$ and $v_3$. We then (2) translate the growing graph to a n-order complex and (3) compute the persistent homology to recover the barcode (right). (4) We repeat this process over 1000 growing graphs constructed from the same generative model, and we present the average Betti curves (solid line) and one standard deviation (shaded).}
\label{fig:team_intro}

\end{figure}

Before tackling how the topology of a growing network model might (or might not) be susceptible to node reordering, we first ask how the evolving topology differs between six growing network models chosen to span growth mechanisms relevant for social systems, material systems, spatially embedded systems, and dynamical systems. Because we are interested in the growing global topology of these networks, we use persistent homology, which describes how topological cavities emerge, evolve, and collapse throughout the growth of a network (see the Methods section and \cite{carlsson2009topology,zomorodian2005computing,ghrist2018homological,otter2017roadmap} for more details). Persistent homology returns the birth and death times for each persistent cavity of a particular dimension in the form of a barcode, with each bar corresponding to a persistent cavity. Additionally we can summarize the barcode by only recording the number of persistent cavities of a given dimension present after each node addition, and we refer to the data recorded in this way as the Betti curve of dimension $n$ or $\beta_n$. Specifically for our growing graph models, we follow the rules of a given generative model to construct a growing graph $(B,s_0)$ with binary graph $B$ and node ordering $s_0 = v_1, v_2, \dots, v_N$ (Fig.~\ref{fig:team_intro}, 1). From this growing graph we construct the node-filtered order complex (n-order complex) and compute its persistent homology, displayed as a barcode (Fig.~\ref{fig:team_intro}, 2 and 3). We repeat this process for 1000 instantiations of the generative model, and then we calculate the average Betti curves across instantiations (Fig.~\ref{fig:team_intro}, 4). We show the results of this analysis for all growing graph models in Fig.~\ref{fig:team1},\ref{fig:team2} and in the Supplement we also include expanded figures showing how the average edge density, average degree distributions, and average number of persistent cavities born or killed evolve as the graph matures (Fig.~\ref{fig:meet_the_team_extd1},\ref{fig:meet_the_team_extd2}).

\subsubsection*{Constant probability.} 
\begin{figure}
	\centering
	\includegraphics[width=6in]{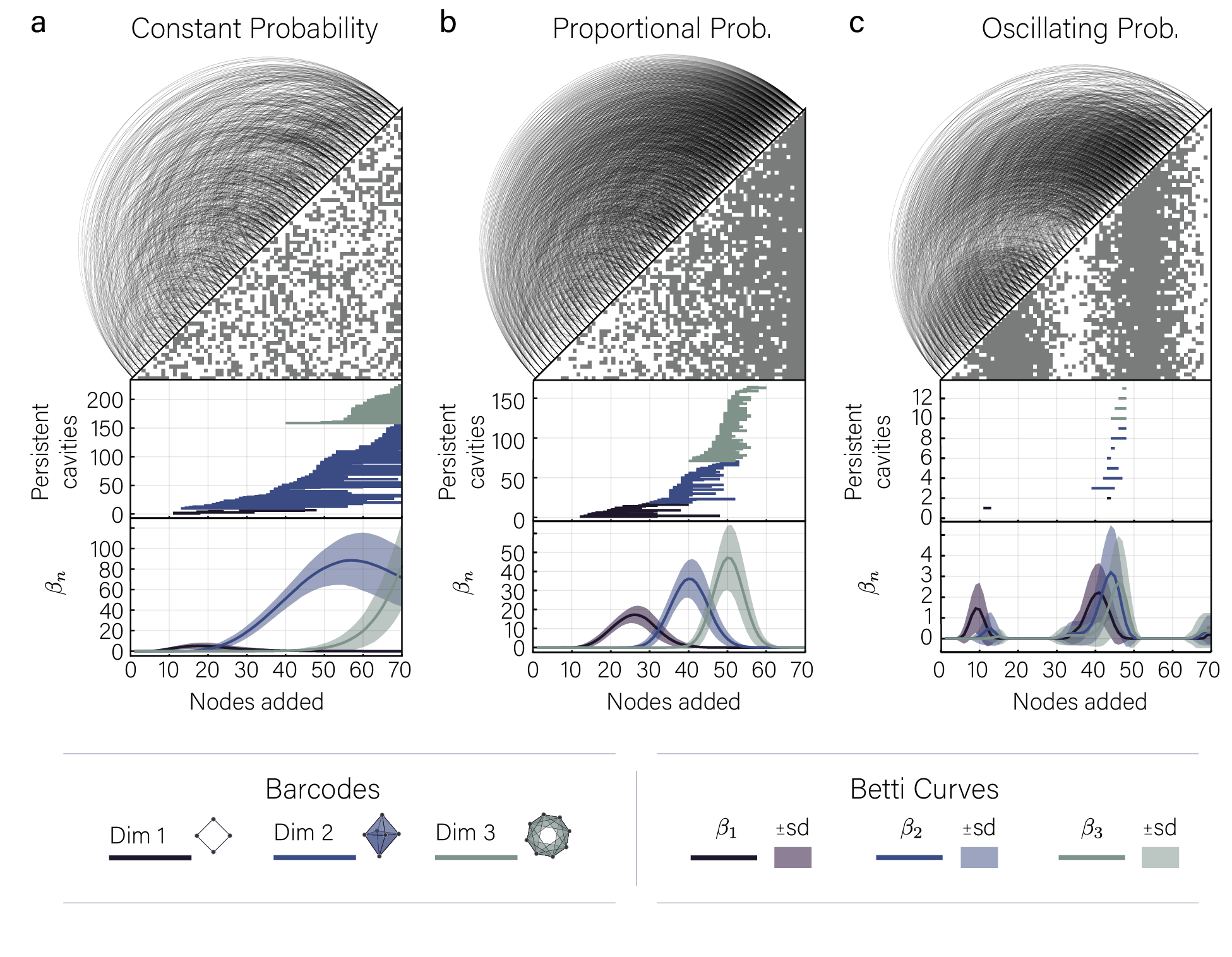}
	\caption{\textbf{Growing graph models display a wide range of evolving topologies.} Persistent homology of the \emph{(a)} constant probability, \emph{(b)} proportional probability, and  \emph{(c)} oscillating probability models. Each panel shows one example adjacency matrix and fingerprint graph (top), the corresponding barcode (middle), and averaged Betti curves across replicates (bottom) with standard deviation (shaded).}
	\label{fig:team1}
\end{figure}

We first begin with the constant probability model as it is one of the most basic random growing graph models. For this model at each node addition we add edges with constant probability $c\in [0,1]$ (here $c = 0.4$). Then if we slice the filtration at node $v_i$, we will have a random graph on $i$ nodes in which each edge was added with probability $c$ \cite{gilbert1959random}. In the average Betti curve (Fig.~\ref{fig:team1}a) we see first a wide peak of persistent cavities of dimension 1 followed by a sudden growth of persistent cavities of dimension 2. Later, as the dimension 1 persistent homology dies, dimension 3 begins to sharply increase. Interestingly the barcode shows that once persistent cavities in dimensions 2 and 3 are born, they are extremely unlikely to be killed by a later node addition. 

\subsubsection*{Proportional probability.} We next consider the proportional probability model, which adds edges between the new node $v_i$ and previous nodes with probability $i/N$. Note that this rule requires that the last node connect to all other nodes in the graph. Consequently we see no homology can live through the end of the filtration. Still, as shown by the example barcode, there tends to be a wide distribution of lifetimes, particularly in dimension 1. The Betti curves show an increasing pattern of peaks, each taller and narrower than the last. Additionally, the pattern observed in these lower dimensions suggests that the support of each $\beta_n$ overlaps approximately half of $\beta_{n-1}$ and $\beta_{n+1}$, so that as the $\beta_{n-1}$ tends to 0, $\beta_{n+1}$ begins to increase.

\subsubsection*{Oscillating probability.} While the previous two functions describing the probability of adding edges emanating from the new node at each node addition have been constant or strictly increasing as the graph grows, we next ask what type of topology we can achieve using a non-monotonic function. Specifically here we choose an oscillating function, with edges added between a new node $v_i$ and a previously added node with probability $|sin(2\pi\frac{i}{N})|$. This choice leads to the evolution of two groups of nodes with high degrees: generally nodes added between node 20 and node 30 will have a large degree, and those added between node 50 and node 60 will have an even larger degree, reflecting the periodicity of the edge probability function (shown in Supplementary Fig.~\ref{fig:meet_the_team_extd1}c). Although we generally do not see many persistent cycles, on average we see the Betti curves reflect the oscillations in edge density (Fig.~\ref{fig:team1}c). That is, we generally observe a burst of persistent cycle activity (birth and death) while the edge density curve has a positive second derivative.

\subsubsection*{Preferential attachment model.} 
\begin{figure}
	\centering
	\includegraphics[width=6in]{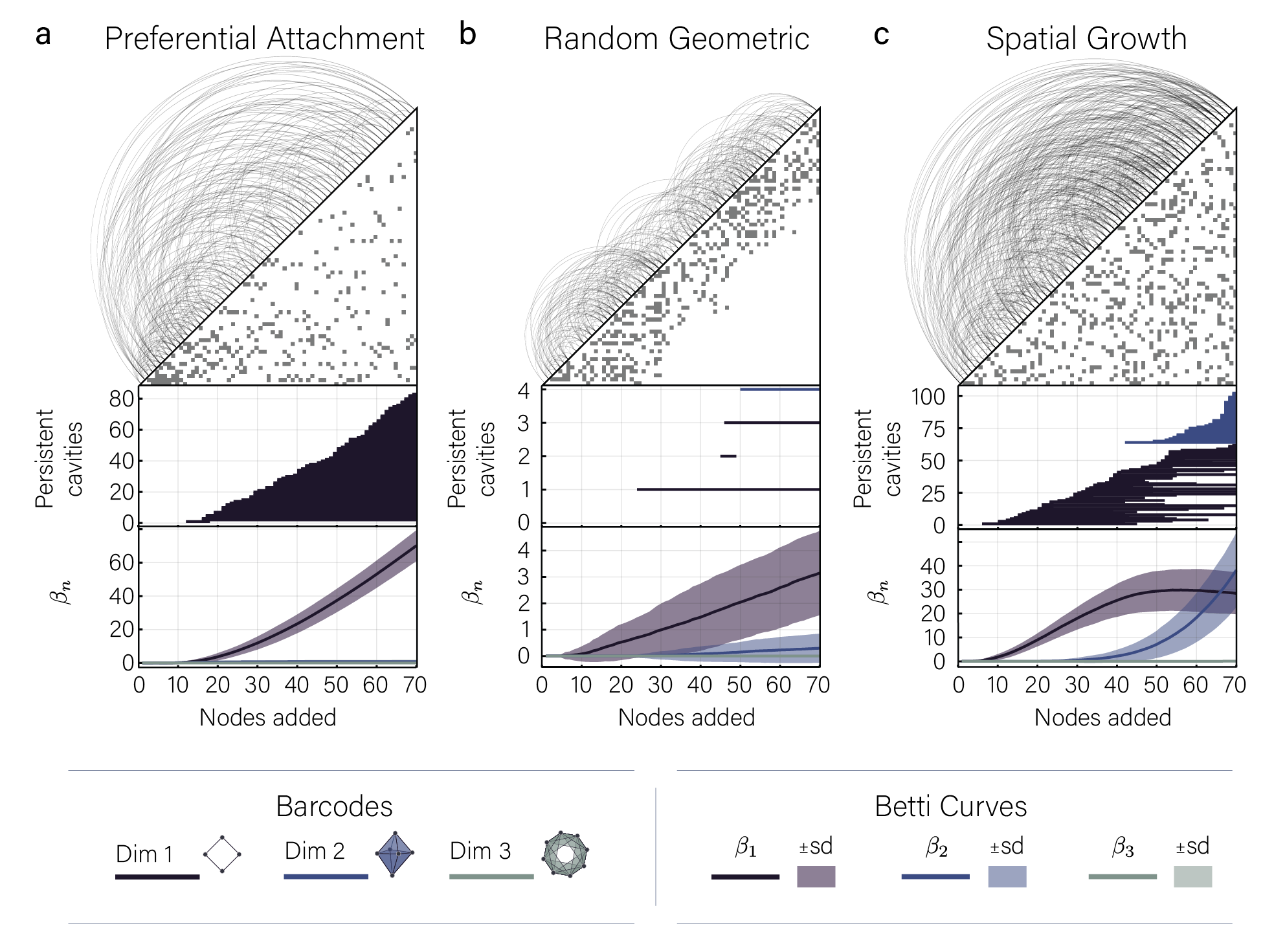}
	\caption{\textbf{Growing graph models display a wide range of evolving topologies.} Persistent homology of the \emph{(a)} preferential attachment, \emph{(b)} random geometric, and  \emph{(c)} spatial growth models. Each panel shows one example adjacency matrix and fingerprint graph (top), the corresponding barcode (middle), and averaged Betti curves across replicates (bottom) with standard deviation (shaded).}
	\label{fig:team2}
\end{figure}

One of the most studied graph models in network science -- the preferential attachment model \cite{price1976general,barabasi1999emergence} -- adds a fixed number of edges with each node addition, but connects these edges preferentially to high-degree nodes in the network. For our parameters chosen ($m=m_0 = 4$) we see that the first few nodes end with the highest degrees and that most other nodes have low degrees, as expected (Fig.~\ref{fig:meet_the_team_extd2}a). We also observe a large abundance of persistent cavities in dimension 1 with very little cavity death (Fig.~\ref{fig:team2}a). The low frequency of persistent cavity death is unsurprising, since to kill a persistent cavity an added node must connect to all nodes of a generating cycle, and a cycle likely involves both high degree and low degree nodes.

\subsubsection*{Random geometric model.} All models that we have considered thus far are not embedded into any particular Euclidean (or non-Euclidean) space. Yet, many real world systems including the brain, transportation networks, and granular materials exist within and are often constrained by some embedded space such as $\mathbb{R}^3$. Our next model -- the random geometric model -- incorporates information about how the growing graph lives in the embedded space. To construct this growing graph, we randomly choose locations in $[0,1]^3$, connect all nodes with distance less than 0.15, and then filter by moving along the $x$ direction through the embedded graph. We imagine this process as walking through a point cloud in which nearby nodes are connected. We find only a small number of persistent cavities emerge (Fig.~\ref{fig:team1}), and often we only observe persistent homology in dimension 1 (as expected for these parameters \cite{kahle2011random}). Curiously, this growing graph model produces the only (near) linear average Betti curves for these parameters out of all the models tested. Importantly, we do not include periodic boundary conditions, which is reflected in the heatmap of node degree evolution (Fig.~\ref{fig:meet_the_team_extd2}b).

\subsubsection*{Spatial growth model.} Finally we investigate a spatial growth model that simulates a growing process in $[0,1]^2$. In this model we spawn nodes randomly in the unit square, and we add edges with a probability based on the Euclidean distance between nodes, so that nodes placed near each other will connect with higher probability. In contrast to the random geometric model, we see large amounts of persistent homology in dimensions 1 and 2, and more S-shaped Betti curves that are qualitatively more similar to the non-embedded constant probability model than the random geometric model. Additionally, the increase of persistent homology in dimension 2 aligns with the beginning of the slow decrease of persistent homology in dimension 1, which suggests a predictable transition from a lower dimensional to a higher dimensional persistent homology regime similar to that seen in the proportional probability or constant probability model.

To summarize, the six growing graph models that we consider here display a wide range of persistent homology signatures arising from the differences in growth processes.

\subsection{Random permutations and global reorderability}

While the above section considered the evolving topology of the originally ordered growing graphs, here we ask how resilient the topology is to any random ordering of node addition. In real systems, such random ordering would be observed if a group of neurons randomly fired instead of properly propagating a signal \cite{ju2018network}, or if diseases were to randomly spawn instead of diffuse predictably along a network \cite{pastor2001epidemic,newman2002spread,granovetter1983threshold,barrett2008episimdemics}. If indeed a growing process was to produce the same final binary graph but with a random node addition order, as discussed in the Methods, then there are two bounding possibilities: either the persistent homology of the growing graph could be independent of the order in which nodes are added, or it could be highly dependent on that order. We call the resiliency of the growing graph's persistent homology to random node addition order the \emph{global reorderability} of the growing graph, and in the following experiments we examine the extent to which each of the above growing graph models is globally reorderable.  

\begin{figure}
	\centering
	\includegraphics[width=6in]{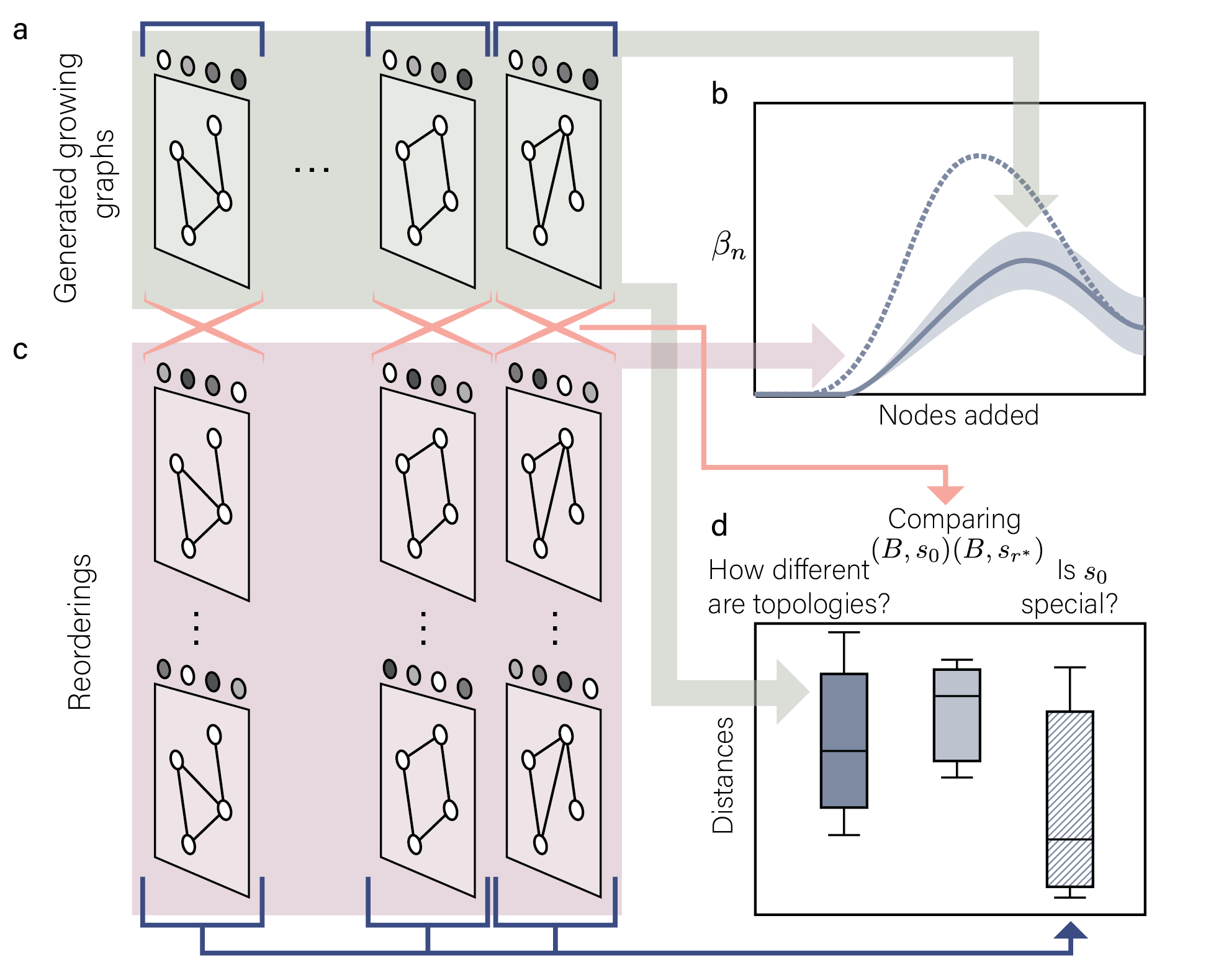}
	\caption{\textbf{Global reordering process and comparisons.} \emph{(a)} Following a growing graph model we generate 100 growing graphs and \emph{(b)} calculate their persistent homology and average Betti curves (solid line). \emph{(c)} For each growing graph in \emph{(a)} we randomly permute the node order, then compute the persistent homology and average Betti curve (dashed line in \emph{(b)}) to compare with that of the original growing graphs. \emph{(d)} To determine reorderability, we plot distributions of the distances between persistent homology outputs of randomly sampled generated graphs (green arrow, dark box), distances between persistent homology outputs of the generated graph and that of their reordered growing graphs (orange lines, light box), and distances between any pair of node orderings on the same binary graph (blue lines, patterned box). For this example, since the lighter box generally sits higher than the dark, we would say this growing graph model is not globally reorderable. Additionally since distances between the persistent homology of randomly sampled orderings on the same topology (striped) are smaller than that between $(B,s_0)$, $(B,s_{r^*})$ distances, we conclude that the $s_0$ ordering has a special significance.}
	\label{fig:global_intro}
\end{figure}

In order to examine the global reorderability of each growing graph model, we first generate 100 growing graphs $(B_{\alpha},s_0)$ for each $\alpha = 1,\dots, 100$ following the model rules (Fig.~\ref{fig:global_intro} green shaded). Next for each generated growing graph we create reordered growing graphs in which we add nodes uniformly at random using the same underlying binary network (Fig.~\ref{fig:global_intro}, pink shaded). More specifically, we construct $(B_{\alpha},s_0)$ for $\alpha = 1,\dots, 100$, and then for each $\alpha$ we randomly permute $s_0$ to give a new node order $s_{r^m}$ for $m = 1,\dots, 100$. All random permutations of $s_0$ are generated anew for each $\alpha$. We then compute the persistent homology of each generated and reordered growing graph, and we show the average Betti curves of the originally generated and reordered growing graphs (Fig.~\ref{fig:global_intro}c solid and dashed lines).

Now before directly comparing the persistent homology of the reordered growing graphs to that of the originally generated growing graphs, we must first understand the intrinsic variability of the persistent homology with respect to the range of growing topologies created by the generative model. For example, if we later find that the reordered graphs show a very different persistent homology than that of the originally ordered graphs, could this be simply due to the fact that the underlying graph model produces such a wide range of topologies? In order to answer this question we sample 10,000 pairs of generated growing graphs with node order $s_0$ and compute distances between their persistent homology summaries (Fig.~\ref{fig:global_intro}d, dark box). Then when we calculate the distance between the persistent homology summaries (barcodes or Betti curves) of the originally ordered graph and each of its reorderings for all generated growing graphs (Fig.~\ref{fig:global_intro}d, light box, orange lines), we can determine if these distances are smaller (or larger) than expected given the variability of the growing graph model topology. If the distances between the persistent homology of growing graphs and that of their reorderings is smaller than the distances between the persistent homology of randomly sampled generated graphs, then we say that the growing graph model is globally reorderable, since the distance after reordering is smaller than the distance between generated graph topologies. 

Next for illustrative purposes let us imagine that the persistent homology output of the original and random ordered growing graphs are points in some high dimensional space. It may be that the original ordering $s_0$ results in an evolving topology that sits within the masses of persistent homology generated by random orders, or on the other extreme the persistent homology from $s_0$ may be extremely different from that generated from any random ordering. If the former, then we might conclude that the $s_0$ ordering offers no distinct difference in terms of the evolving topology than any random ordering. But if the latter, then we can conclude that the particular $s_0$ ordering does hold a special significance in terms of the persistent homology for the growing graph model, as the originally ordered persistent homology is far different than that from careless orderings. To assess this behavior, for each set of growing graphs with the same binary graph (of which there are 101), we randomly sample 100 pairs of these growing graphs and compute distances between their topological summaries (Fig.~\ref{fig:global_intro}d, blue lines, striped box). If the distribution of random orderings within a topology (striped box) is larger than or equal to the distribution of distances between the original and reordered growing graphs (light box), $s_0$ would not appear to generate a persistent homology signature distinct from any random node order permutation. Conversely, if we see that distances between persistent homology outputs within a topology (patterned box) are generally smaller than distances between persistent homology from $s_0$ and that from a random reordering (light box), then the evolving topology from $s_0$ must be quite distinct compared to the evolving topology generated by any random ordering.

\subsubsection*{Differences in Betti curves after reordering}

\begin{figure}
	\centering
	\includegraphics[width=6.5in]{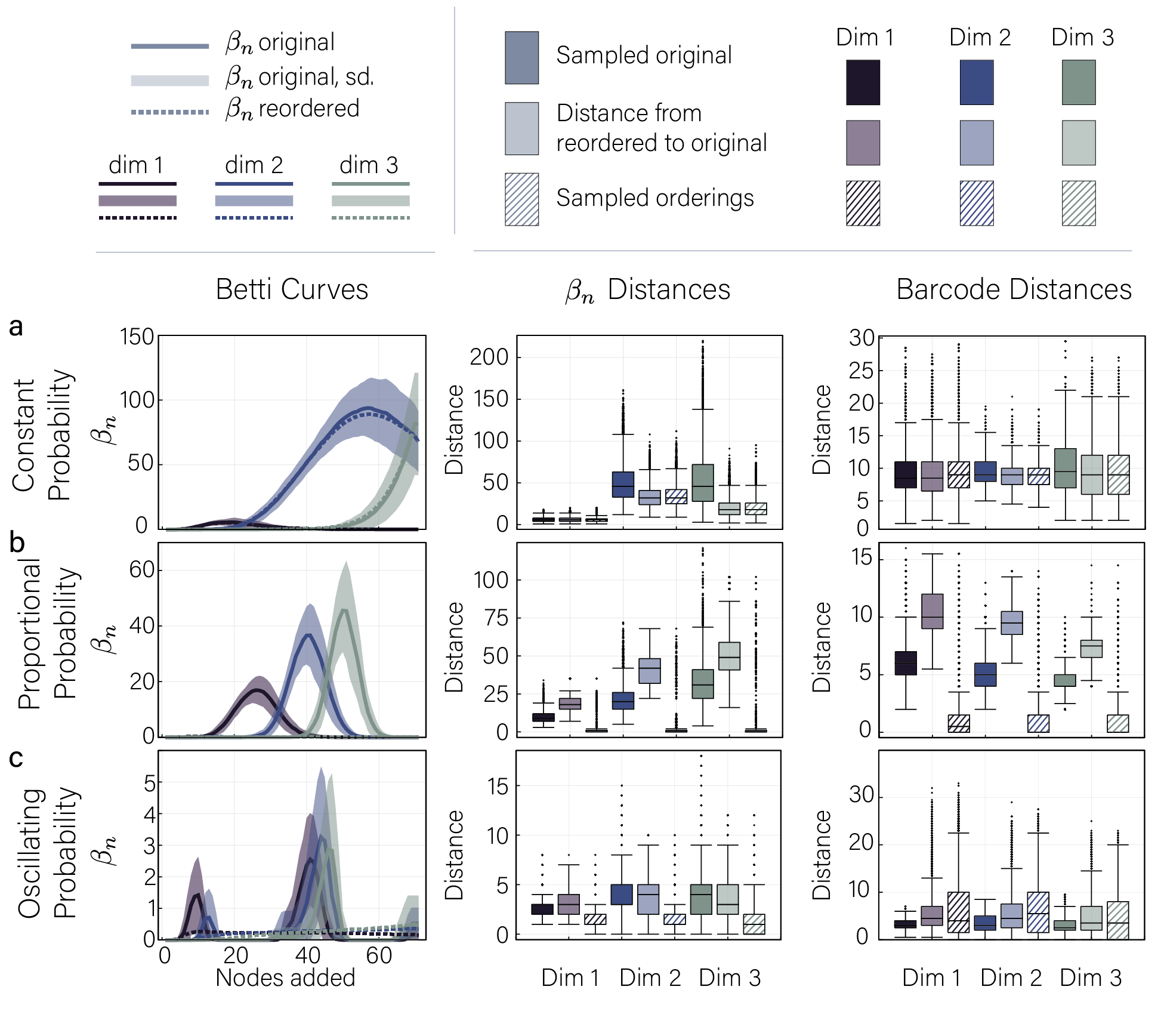}
	\caption{\textbf{Global reorderability varies across growing graph models.} Results of the random reordering analyses for the \emph{(a)} constant probability, \emph{(b)} proportional probability, and \emph{(c)} oscillating probability models. Within each panel we show the average Betti curves (left) produced by the originally ordered (solid line, standard deviation shaded) and randomly reordered (dashed line) growing graphs, as well as box plots for Betti curves (middle) and barcodes (right) showing the within-model distribution of distances (solid), distances between the topological summary of the original and randomly reordered growing graphs (lighter shade), and the distribution of differences between randomly sampled pairs of growing graphs ending in the same binary network (striped).}  
	\label{fig:global_results1}
\end{figure}

\begin{figure}
	\centering
	\includegraphics[width=6.5in]{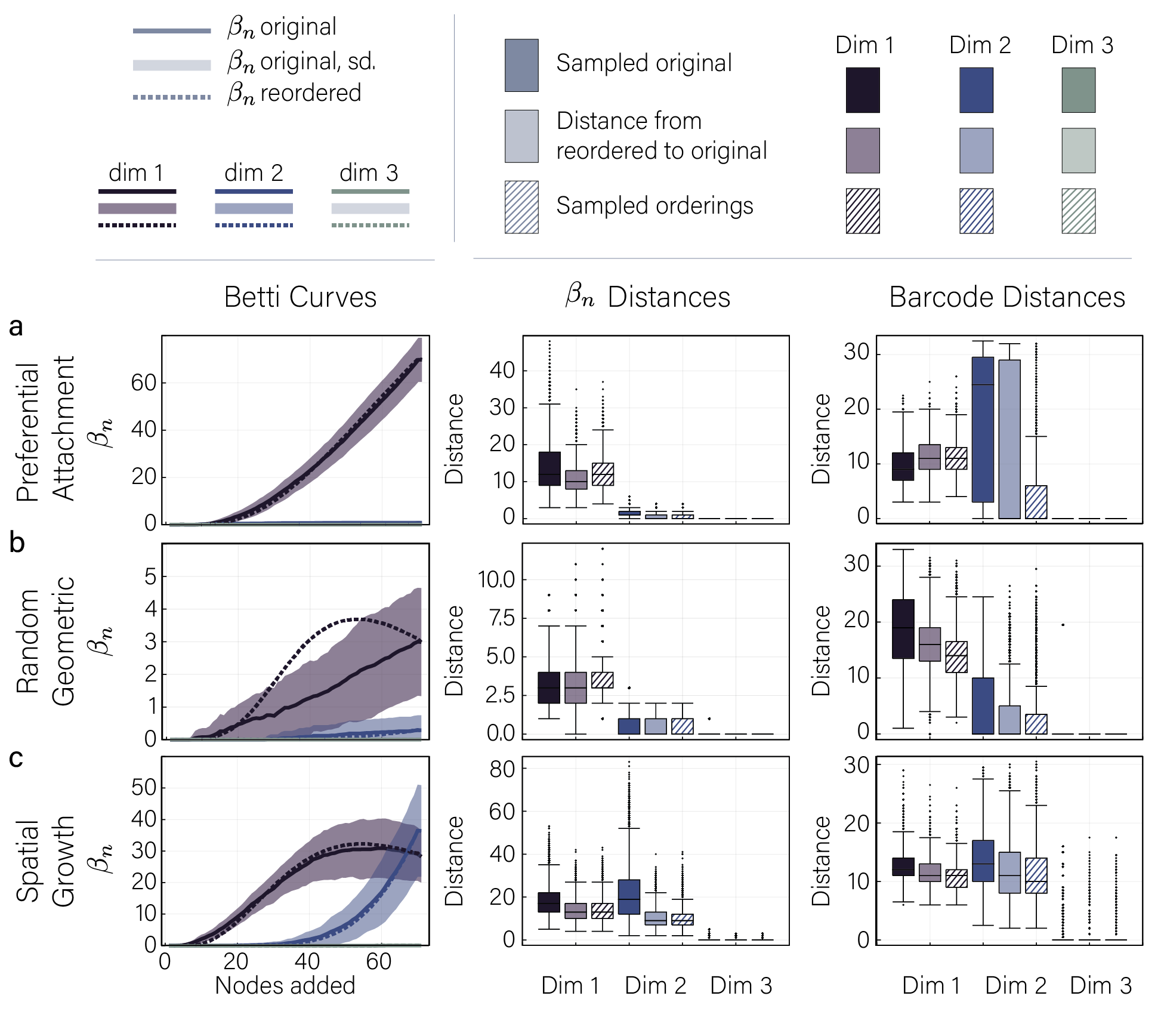}
	\caption{\textbf{Global reorderability varies across growing graph models.} Results of the random reordering analyses for the \emph{(a)} preferential attachment, \emph{(b)} random geometric, and \emph{(c)} spatial growth models. Within each panel we show the average Betti curves (left) produced by the originally ordered (solid line, standard deviation shaded) and randomly reordered (dashed line) growing graphs, as well as box plots for Betti curves (middle) and barcodes (right) showing the within-model distribution of distances (solid), distances between the topological summary of the original and randomly reordered growing graphs (lighter shade), and the distribution of differences between randomly sampled pairs of growing graphs ending in the same binary network (striped).}  
	\label{fig:global_results2}
\end{figure}

We aim to determine if the Betti curve changes as a result of reordering the nodes in our growing graphs. As a first look we show the average Betti curves for the original growing graph models $(B_*,s_0)$ and for all reorderings $(B_*,s_{r^*})$ in the left column of Fig~\ref{fig:global_results1},\ref{fig:global_results2} (solid and dashed lines, respectively). We find little difference in the average Betti curves before and after node-reordering for the constant probability, preferential attachment, and spatial growth models. Conversely the Betti curves of the random geometric and oscillating probability models show marked differences in shape between the original and globally reordered growing graphs. More drastic still, on average we observe that all of the persistent homology seen in the original proportional probability model is killed when we randomly reorder the node addition sequence.

In order to quantify these results we can compute a distance between Betti curves in each dimension from any pair of growing graphs. For simplicity we define the distance between two Betti curves as the maximum vertical difference between the curves at any point (see Eq.~\ref{eq:0}). With this Betti curve distance definition, we now show the distribution of Betti curve distances between randomly sampled pairs of generated graphs (Fig.~\ref{fig:global_results1},\ref{fig:global_results2}, middle column, solid boxes), between different orderings while holding the binary graph constant (Fig.~\ref{fig:global_results1},\ref{fig:global_results2}, middle column, striped boxes), and between the original growing graph and each of its reorderings (Fig.~\ref{fig:global_results1},\ref{fig:global_results2}, middle column, light boxes) as discussed above (see Fig.~\ref{fig:global_intro}). If the distances between the original and reordered graphs are significantly smaller than that of any sampled pair of generated graphs (comparing solid and light boxes), we call the model globally reorderable with respect to the Betti curves. We see that the constant probability, preferential attachment, and spatial growth models are clearly globally reorderable with respect to the Betti curves in all three dimensions tested. Additionally most models show a smaller or near-equal distance between the original and reordered models in comparison to any sampled order of a particular graph $B$ (comparing light to patterned boxes), suggesting that a perturbation from the original growth order will result in a smaller or similar change to the evolving topology than a perturbation to any random order. The proportional probability (and to a lesser extent the oscillating probability) model, however, exhibits notably different behavior, as the Betti curves from any pair of random orderings on a binary graph are likely very similar but the distance between the Betti curves of the original ordering and any reordered growing graph can be an order of magnitude larger.

\subsubsection*{Effect of global reordering on barcodes}

The Betti curves still do not capture all of the details of a growing graph's persistent homology. Indeed, multiple barcodes could produce the same Betti curve, and additionally there is no guarantee that the distance (as defined here) between Betti curves will be small as a result of a small perturbation. So we next ask how the barcodes change as a result of global reordering (Fig.~\ref{fig:global_results1}, \ref{fig:global_results2} third column). Recall that the bottleneck distance between two barcodes $P_1,P_2$ is colloquially given by the maximum distance that either end of a bar in $P_1$ has to move in order to align with its match in $P_2$ (see Eq.~\ref{eq:1}). As with the Betti curves, we first compute the distributions of barcode distances between the persistent homology of randomly sampled pairs of growing graphs generated by the same model (Fig.~\ref{fig:global_results1},\ref{fig:global_results2}, right column, solid boxes). Interestingly, although the random geometric model shows only small differences in within-model Betti curve distances, the barcode distances are some of the highest of all six models, suggesting a large change to only a few highly persistent cavities. In contrast, the oscillating probability model shows low variability in both the Betti curves and barcodes by our measures.

Using the same intuition as for the Betti curves, we can extract the global reorderability of the growing graph models with respect to barcode distances by comparing between barcode distances from random samples of original graphs and barcode distances from reorderings of the original binary graph (Fig.~\ref{fig:global_intro}). We see that the preferential attachment model does not exhibit reorderability in dimension 1 and that the random geometric model appears more reorderable in both dimensions 1 and 2 than when we considered Betti curves as the persistent homology summary. Still, we find that the proportional probability model shows extreme sensitivity to reordering, and to a lesser extent so does the oscillating probability model. We observe that the constant probability model, the random geometric, and the spatial growth models display global reorderability with respect to the barcode distance. Finally we note that for both distances, the proportional probability model shows a marked preference for the original $s_0$ ordering. Then for this model if we had let $s_0$ be any random ordering and performed the same analysis, based on the boxplots in Fig.~\ref{fig:global_results1}b we might expect the growing graph to be reorderable. 

But why is it that we see such shifts in Betti curves and barcodes after reordering? For example we may see smaller Betti curves if we increase persistent cycle birth times on average. Or perhaps we might see the same shift in Betti curves but decreased average persistent cycle death times. In Supplementary Fig.~\ref{fig:Supp3} we show distributions of differences between average persistent cycle birth times, average death times, and $\overline{\beta_n}$ values between the original and randomly reordered growing graphs. In particular for the proportional probability model, we see that the large barcode distances between global reorderings and the original ordering is likely due to both a decrease in average birth time and a larger decrease in average death time (Fig.~\ref{fig:Supp3}) so that any persistent cavity that does form does so early in the filtration and lives for only a short time. Additionally after reordering the proportional probability model we see that $\overline{\beta_n}$ is generally lower, indicating that we simply see less homology in total, and that when we do see persistent homology it is shorter lived, likely due to the larger probability of early edge addition. Based on the differences in $\overline{\beta_n}$, we also expect that shifts in the reordered random geometric persistent homology are likely caused by changes in the overall amount of homology and that on average the persistent cavities in the reordered growing graphs die earlier.

\begin{figure}
	\centering
	\includegraphics[width=6.5in]{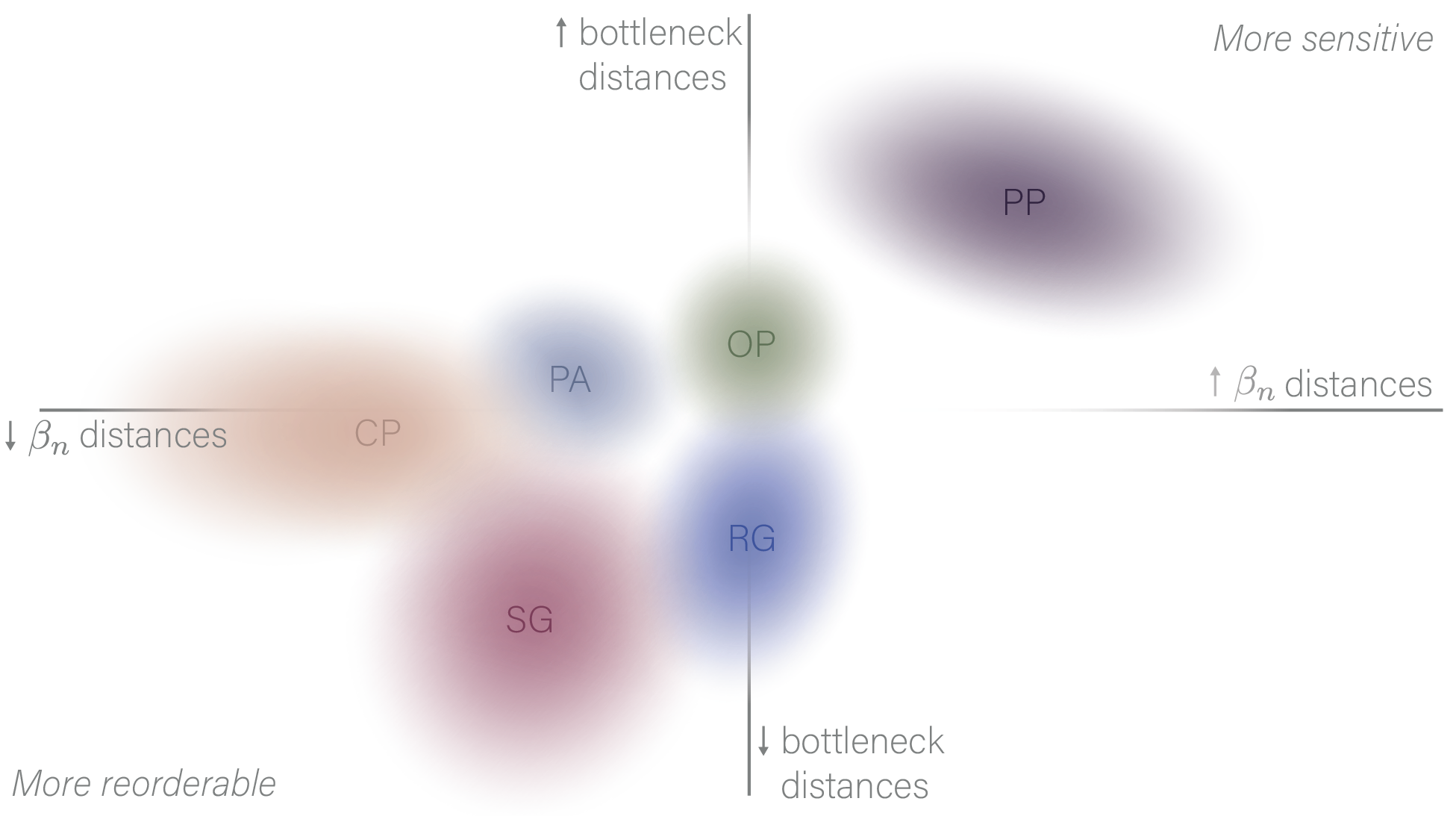}
	\caption{\textbf{Schematic of the space of global reorderability.} We imagine this two-dimensional space as being parameterized by the distances between Betti curves $\beta_n$ and barcodes between the reordered and original growing graphs. That is, a model in which the bottleneck distances between the reordered and original growing graphs were higher than expected by random sampling of original graphs would be on the top half of the plane, suggesting that the model is not globally reorderable with respect to this measure. Similarly, a growing graph model in which the Betti curve distances between reorderings and original growing graphs were less than expected by random sampling of original growing graphs would lie on the left half of the plane, suggesting that the model is reorderable with respect to Betti curves. Based on our definitions, we describe global reorderability with respect to both tested distances as the lower left quadrant. Abbreviations: CP = constant probability, PP = proportional probability, OP = oscillating probability, PA = preferential attachment, RG = random geometric, SG = spatial growth.}
	\label{fig:5}
	
\end{figure}

When we compare the barcode distances between generated growing graph pairs and their reordered networks, we gain a different perspective on which growing graph models are globally reorderable. Those models that exhibit reorderability with respect to Betti curves but not barcodes are, in a way, similar only at a per-slice level. That is, if we consider any of the models at a given node number $i$, then the topology in terms of the numbers of cavities in each dimension will be similar between the original and reordered growing graphs. However, if we look across time and take the fruits of the growing process as a collective, the evolving topology has changed drastically. These models illustrate the difference between slicing the growing graph at each node addition and comparing topology at time points, versus comparing the collective, evolving topology that emerges over the course of graph growth. Conversely, a model that is reorderable by the barcode description but not by Betti curves (though less likely) will show a similar evolving topology throughout the entire growth process but may not show the same homology when comparing slices. In Fig.~\ref{fig:5} we show a schematic illustrating where our six studied models with our chosen set of parameters fall on the reorderability axes based on our results. We note that all four quadrants of this graph are accessible, and that creating models that optimize one reorderability scheme versus another is an interesting direction for future research.

\subsection{Pairwise swaps and local reorderability}

While above we considered randomly swapping all nodes in the ordering, at the other end of the spectrum we can also investigate the effect on the evolving topology of swapping only one pair of nodes in the growing graph. Instead of a growth process progressing randomly, this process would instead be akin to a pair of neurons swapping firing order in a neuronal population \cite{lee2004combinatorial} or a student switching the order in which they learn a pair of vocabulary words \cite{sizemore2018knowledge}. Here performing node swaps in our models will offer a measure of the resiliency of a growing graph to a small perturbation of only two nodes as opposed to a complete reordering. As we will see, we gain not only an understanding of each growing graph model's local reorderability, but also a deeper appreciation for node participation in the persistent homology.

\begin{figure}
	\centering
	\includegraphics[width=6.5in]{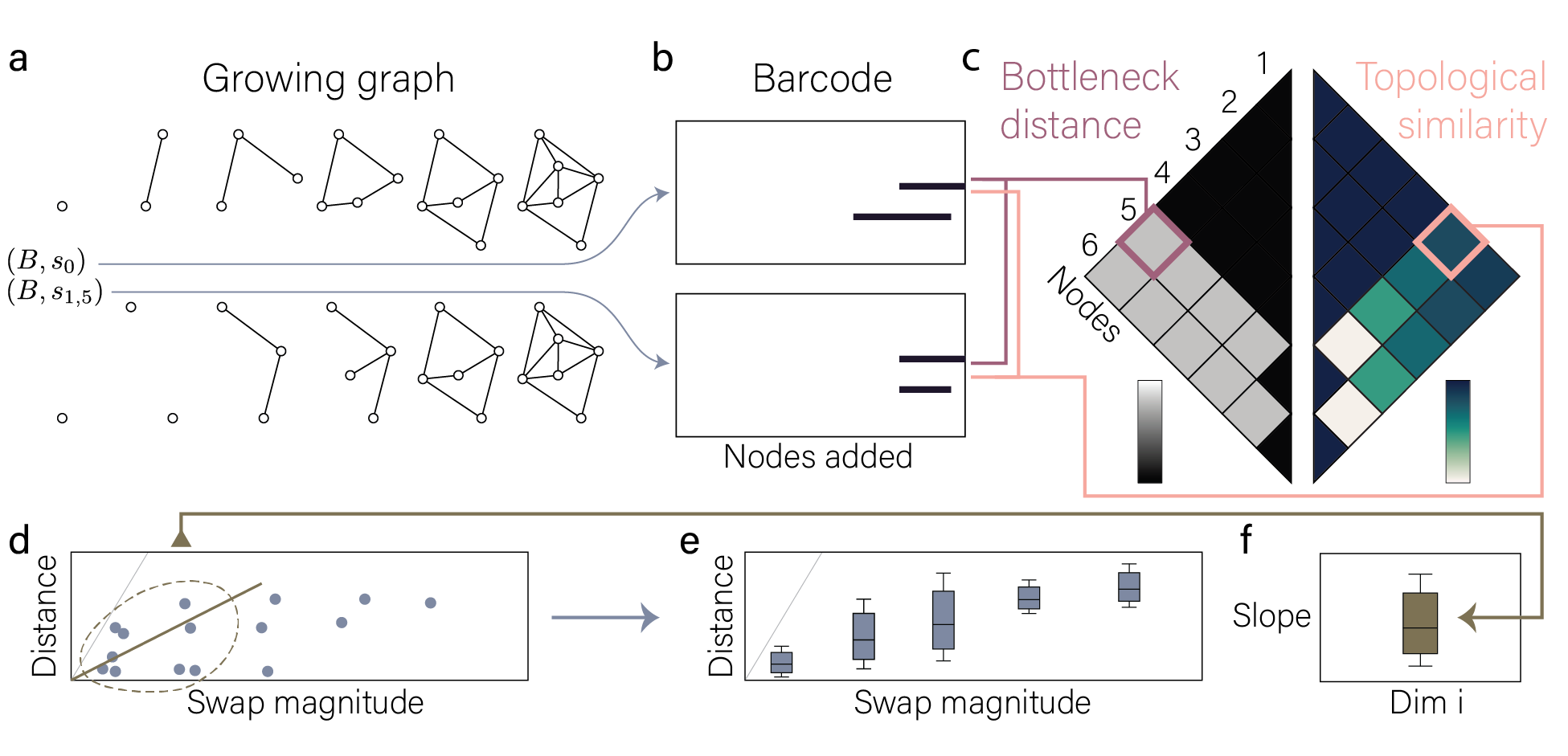}
	\caption{Analysis flow for node swap studies. \emph{(a)} Filtration of a growing graph $(B,s_0)$ and $(B,s_{1,5})$, the latter of which being constructed by swapping nodes $v_1$ and $v_5$ in the ordering. \emph{(b)} The barcode calculated from $(B,s_0)$ and $(B,s_{1,5})$. \emph{(c)} Calculating the bottleneck distance between the persistent homology of $(B,s_0)$ and $(B,s_{v_i,v_j})$ for all $i,j$ gives a bottleneck distance matrix (left) and, after dividing by the swap magnitude $|j-i|$ and subtracting from 1, the topological similarity matrix (right). \emph{(d)} We can investigate the relation between swap magnitude and bottleneck distance by \emph{(e)} creating boxplots of average bottleneck distance against swap magnitude and \emph{(f)} determining the slope of a line of best fit to points with small swap magnitude and the resulting distribution of such slopes over replicates.} 
	\label{fig:local_intro}
\end{figure}

In order to test how swapping a pair of nodes in the node addition order will alter the persistent homology, we first generate a growing graph $(B,s_0)$ and a node-swapped version $(B,s_{i,j})$ in which the binary graph $B$ is kept unchanged but we swap the $i^{th}$ and $j^{th}$ node in the ordering (Fig.~\ref{fig:local_intro}a). Recall since our original ordering $s_0 = v_1,v_2, \dots, v_N$ that the $i^{th}$ node in this sequence is $v_i$. As shown in Fig.~\ref{fig:local_intro}b we then compute the persistent homology and recover the barcodes from each growing graph $(B,s_0)$ and $(B,s_{i,j})$, and next calculate the bottleneck distance between these barcodes. Repeating this process for every node pair, we can construct a matrix of distances in which the $(i,j)$ entry is the bottleneck distance in dimension $n$ between barcodes arising from $(B,s_0)$ and $(B,s_{i,j})$ (Fig.~\ref{fig:local_intro}c, left for dimension 1). Now importantly recall that the bottleneck distance between barcodes recovered from $(B,s_0)$ and $(B,s_{i,j})$ is bounded above by the largest distance any node has moved between the orderings. In our case, this upper bound then is the magnitude of the node swap, or $|j-i|$. Then finally to calculate topological similarity between nodes, we normalize the bottleneck distance by this swap magnitude and subtract from 1. Explicitly, we define the topological similarity in dimension $n$ of a node pair $v_i$, $v_j$ as $T_n(v_i,v_j) = 1-\frac{d^n_{BN}(P_0,P_{i,j})}{|j-i|}$ and the average topological similarity over dimensions as $T(v_i,v_j) = \frac{1}{D}\sum_{n=1}^{D}T_n(v_i,v_j)$ where here our maximum dimension $D=3$. If a pair of nodes $(v_i,v_j)$ has topological similarity close to 1, then the bottleneck distance between barcodes from $(B,s_0)$ and $(B,s_{i,j})$ must be small compared to the swap magnitude. Said another way, the swap could have caused a large change in persistent homology but did not. Again we can record the topological similarity of each node pair in a matrix as shown in the right of Fig.~\ref{fig:local_intro}c. Note that for both sides of the split matrix, a lighter color indicates node pairs that when swapped will alter the persistent homology more drastically while darker colors indicate node pairs that when swapped have little effect on the persistent homology. Though computed and described here, we will make the impact of the topological similarity matrix apparent even further in the last section of the Results. In Fig.~\ref{fig:local_intro} we show our process for one growing graph, but for growing graph models we generate 20 replicates and then average bottleneck distances and topological similarity values across replicates and dimensions unless otherwise specified.

Once we have calculated bottleneck distances and topological similarity for each node pair, we next ask how the bottleneck distance varies with the swap magnitude. We can plot the bottleneck distance calculated from each node pair, averaged over dimensions, in one growing graph against the swap magnitude as in Fig.~\ref{fig:local_intro}d. Then, we average across dimensions and collate across 20 replicates from a growing graph model and show the results as a sequence of boxplots as in Fig.~\ref{fig:local_intro}e. Note that neither the boxplots in Fig.~\ref{fig:local_intro}d nor the boxplots in Fig.~\ref{fig:local_intro}e can surpass the $y=x$ line due to the upper bound on bottleneck distance; thus, for reference we plot the $y=x$ line in gray. Finally recall that the goal of this section is to determine the robustness of each growing graph model to small perturbations. In order to quantify this robustness we use the bottleneck distances from all swaps with a magnitude in the smallest 20\% of possible swap magnitudes (magnitudes $\leq 14$) for each of the 20 replicates to calculate a linear best fit and record the distribution of slopes in a boxplot for each dimension (illustrated in Fig.~\ref{fig:local_intro}f). Smaller slope values indicate that temporally local perturbations to the node addition order have only a small effect on the persistent homology, while slope values close to 1 suggest that nodes temporally close in spawning order contribute very differently to the persistent homology of the growing graph.

\subsubsection*{Node swap analyses on growing graph models}

\begin{figure}
	\centering
	\includegraphics[width=6in]{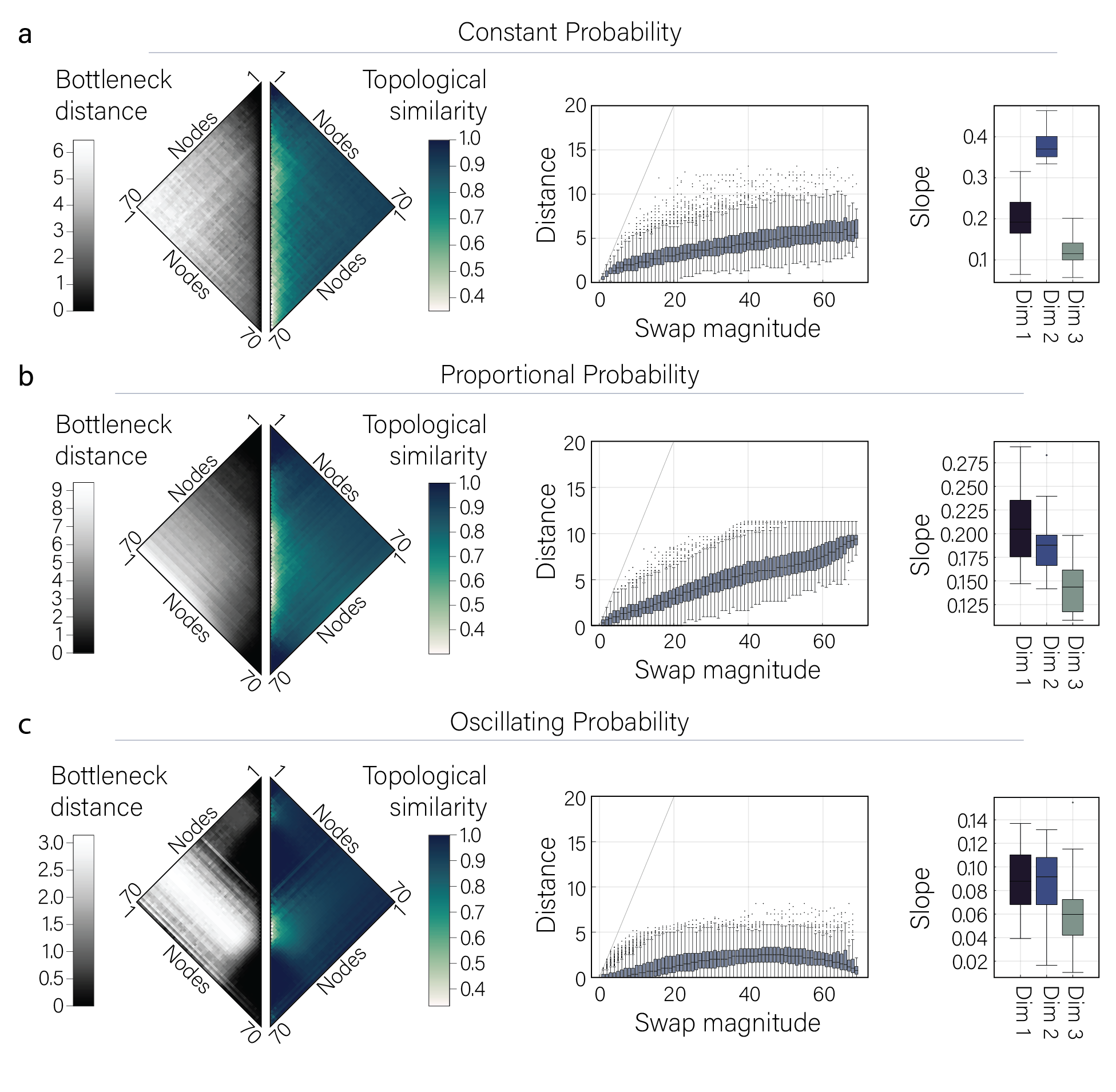}
	\caption{\textbf{Node swaps generate a range of effects on persistent homology: Part I.} Results of node swap analyses for the \emph{(a)} constant probability, \emph{(b)} proportional probability, and \emph{(c)} oscillating probability models. \emph{(Left)} The average bottleneck distances across dimensions 1-3 and across replicates (gray colormap), and the average topological similarity across dimensions 1-3 and across replicates (green colormap). \emph{(Middle)} Boxplots showing distributions of averaged bottleneck distances across dimensions in relation to swap magnitude for all replicates, and distributions of slope values calculated from bottleneck distance and swap magnitude ($< 14$) pairs from each replicate and for each dimension.} 
	\label{fig:local_results1}
\end{figure}

When we use the above analyses to study our six growing graph models, we find again a diverse range of profiles from the models, but we also observe that node properties can vary widely. For the constant probability model we see in Fig.~\ref{fig:local_results1}a that the bottleneck distance increases sharply with swap distance for small swaps (magnitude $\approx <7$) but levels out at the largest swap magnitudes $(\approx>45)$. Interestingly when broken down by dimension, the distributions of slopes calculated from small swap magnitude ($\leq 14$) versus bottleneck distance vary between an average of 0.379 (dimension 2) and an average of 0.122 (dimension 3) across dimensions. In contrast, the proportional probability model shown in Fig.~\ref{fig:local_results1}b shows a strong linear increase of bottleneck distance with swap magnitudes as expected from the definition of the model. The latest points added will connect to nearly every node; thus, if shifted earlier, this behavior will prevent non-trivial homology from forming. Slope distributions for each dimension have relatively similar averages, reflecting the fact that the increase in bottleneck distance associated with the interval of swaps is similar for each dimension. In Fig.~\ref{fig:local_results1}c we observe that the oscillating probability model has two large regions of zero bottleneck distance between swaps and that these zeroed regions align with periods of no homology in the originally ordered network (Fig.~\ref{fig:team1}c). This phenomenon contrasts with the globally reordered oscillating probability model in which we observed homology through nearly the entire growth process (Fig.~\ref{fig:global_results1}c). We only observe bottleneck distances $<10$ in the locally reordered oscillating probability model, and as a result we see slopes of the lines fit to these data are small.

\begin{figure}
	\centering
	\includegraphics[width=6.5in]{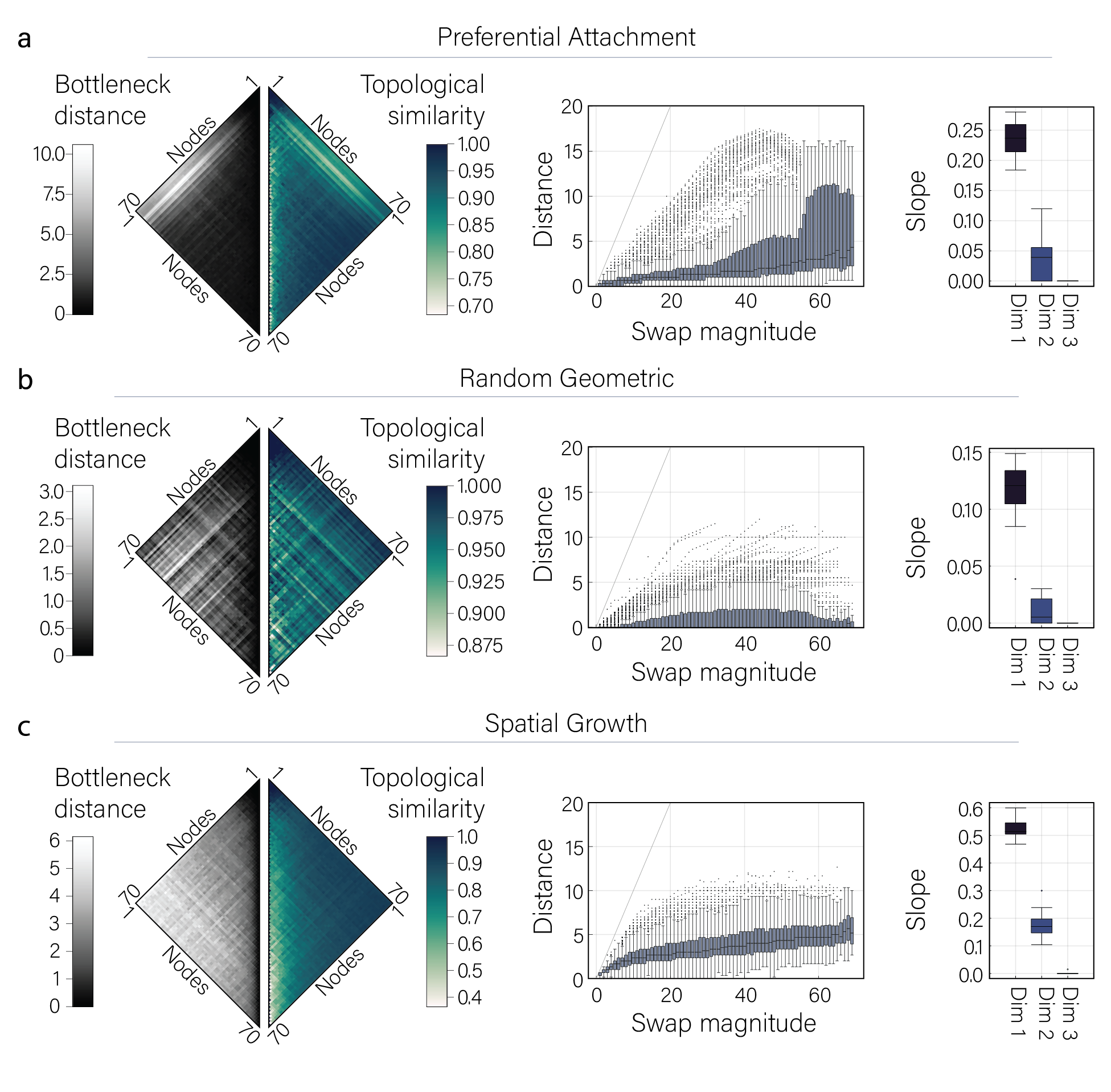}
	\caption{\textbf{Node swaps generate a range of effects on persistent homology: Part II.} Results of node swap analyses for the \emph{(a)} preferential attachment, \emph{(a)} random geometric, and \emph{(c)} spatial growth models. See Fig.~\ref{fig:local_results1} for complete caption.}
	\label{fig:local_results2}
\end{figure}

The other three models show generally less uniform results across nodes. Figure \ref{fig:local_results2}a shows that the preferential attachment model exhibits a large shift in bottleneck distances between small and large node swaps. As evident in the heatmaps in the left of Fig.~\ref{fig:local_results2}a, we find that the earliest nodes cause a large shift in bottleneck distances when swapped. Since these early nodes become the few high-degree nodes in the system, we observe by eye at least for this model that swapping high-degree nodes will often cause a large change in the persistent homology in comparison to the remaining, low-degree nodes. In Fig.~\ref{fig:local_results2}b we see that the random geometric model shows a striped pattern in the averaged topological similarity and averaged bottleneck distance matrices, suggesting marked inhomogeneity in the graph architecture. We mainly see homology in dimension 1 for this model, resulting in generally low average bottleneck distances. Still when we restrict ourselves to dimension 1 we see that the average slope of the line fitting the relationship between small swaps and their associated bottleneck distances is 0.116, which is larger than that of the oscillating probability model. Finally the spatial growth model displays a surprisingly homogeneous topological similarity matrix (Fig.~\ref{fig:local_results2}c). However the boxplots of average bottleneck distance display a sharp rise in bottleneck distance for small swaps, a slow rise for mid-range swaps, and finally a quick upturn again for the largest swaps. Indeed, in dimension 1 the slopes of the line fit to these data have an average $>0.5$, which is the highest average slope seen across all graph models.

We can also use these results from the local reordering analyses to uncover finer details of growing graph architecture. First, if a growing graph model experiences a period of no homology and if when we swap the nodes that are added within this period we still see no homology, then the earlier node could not cone a cavity that forms before the later node. If the earlier node did cone such a cavity, then moving this node to a later position would allow the cavity to be born thus making the homology of the resulting growing graph after swapping non-zero. We observe this phenomenon most strikingly in the oscillating probability model and to a lesser extent in the beginning and end of the proportional probability model. Second, we note that the proportional probability and preferential attachment models are loosely inverses of each other; that is, in the proportional probability model we add the highest degree nodes last but in the preferential attachment model we add the highest degree nodes first. Indeed we observe that the bottleneck distance and topological similarity matrices match when flipped vertically by eye. Third and last, we only examine the behavior of lines fit to results from the smallest swaps. The extent to which the bottleneck distance of large swaps plateaus suggests a degree of homogeneity in neighborhood structure between nodes that are temporally distant. For example the constant probability model plateaus near an average bottleneck distance of 5, which is unsurprising given the random nature of this model; in contrast, the proportional probability model shows steady increases in bottleneck distance across all swap magnitudes. 

Additionally we note that in the global reordering experiments we discover that the spatial growth and constant probability models exhibited similar behavior. When considering the averaged bottleneck distance and averaged topological similarity matrices, this again looks to be the case -- at least initially -- as the matrices are nearly indistinguishable. However, we find a drastic difference in the values of calculated slopes after fitting the swap distance versus bottleneck distance points. The spatial growth model has the steepest slope in dimension 1 but then is shallower in dimension 2, while the constant probability model shows the opposite effect. Still, these two models are the only growing graph models examined here that display average slopes above 0.3, suggesting that nodes added closely in time contribute very differently to the persistent homology. This property contrasts most with the small slopes seen from the preferential attachment, random geometric, and oscillating probability models, suggesting that nodes added within small time-spans form and kill similar cavities.

\begin{figure}
	\includegraphics[width=6in]{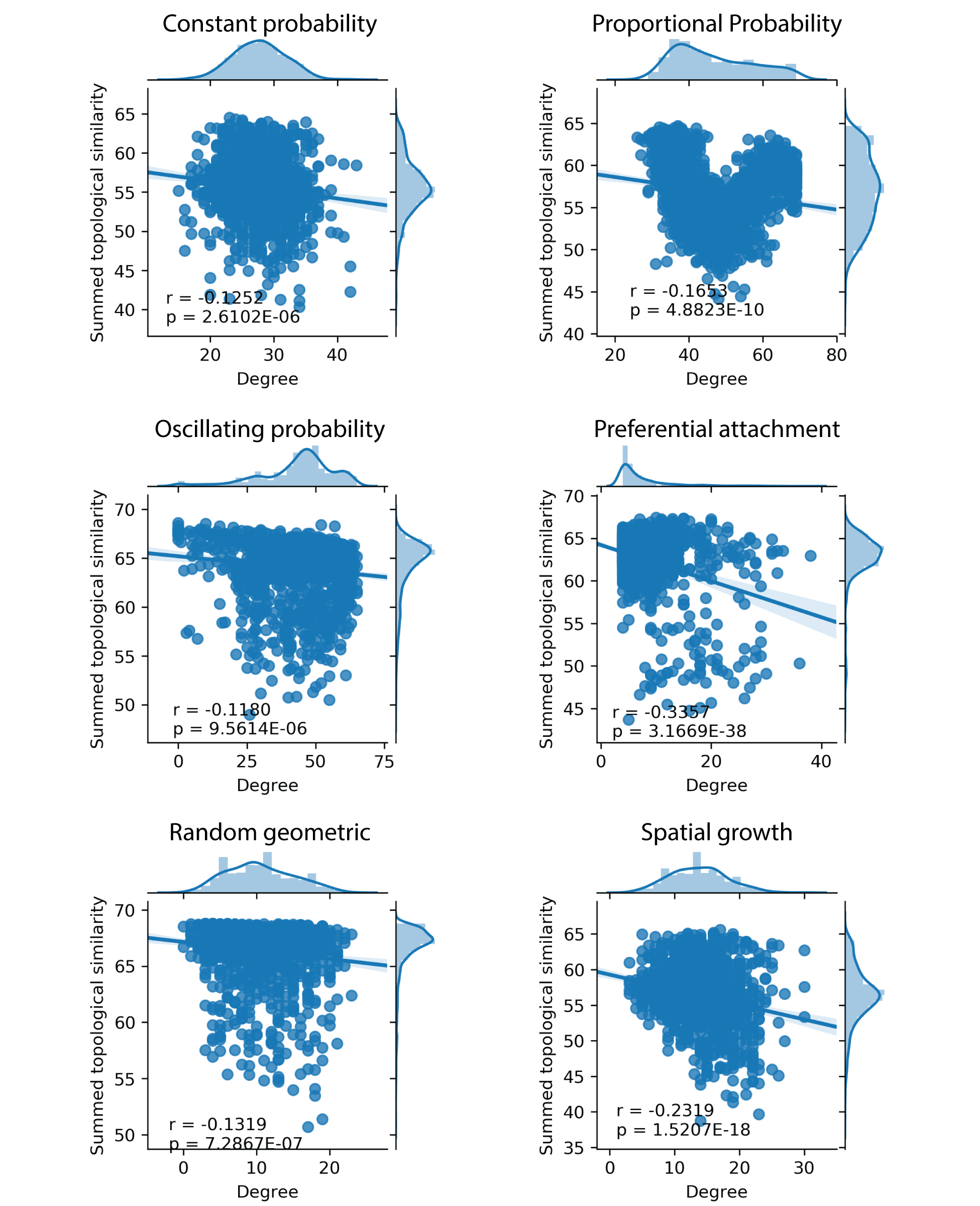}
	\caption{\textbf{Scatterplots and distributions of degree and summed node averaged topological similarity.} Plots of node degree versus summed topological similarity with line of best fit overlayed and marginal distributions for each of the six growing graph models. Additionally the Pearson correlation coefficient $r$ and associated $p$-value are included.}
	\label{fig:Supp2}
\end{figure}

We have discussed that node swaps with high topological similarity suggest node-neighborhood similarities, but it is interesting to ask whether the topological similarity between nodes could be driven mostly by simply the neighborhood overlap. Certainly if two nodes have exactly the same neighborhoods in $B$ then swapping these nodes in the ordering could not change the persistent homology. But more generally could topological similarity be explained by topological overlap? Similarly, we intuitively expect that nodes of high degree should produce a larger effect on the growing topology when swapped than nodes with small degrees (as we generally saw by eye in the preferential attachment model), but is this indeed quantitatively true? To answer the first question, we examine the relationship between the average topological similarity between nodes and the average topological overlap between the same nodes in the original binary graphs $B_{\alpha}$, $\alpha = 1, \dots 20$ (Fig.~\ref{fig:Supp1}). We see that for all but the proportional probability model, most growing graphs show a weak relation between these two variables, suggesting that the topological overlap alone does not predict the topological similarity between nodes. To answer the latter question, we examine the relation between summed node topological similarity and average degree in the original binary graphs (Fig.~\ref{fig:Supp2}). Again we find that for most models (preferential attachment excluded) the average degree poorly predicts the topological similarity of a node to all others. These results imply that topological similarity can capture information about nodal roles in the larger architecture that is distinct from the graph metric properties tested here.

Together these node swap experiments suggest a spectrum of local reorderability for growing graph models. Of the models that we examine, we observe that the random geometric and oscillating probability are most resilient to small reorderings in all dimensions tested, followed by the proportional probability and preferential attachment models, and finally the spatial growth and constant probability models, which exhibit the largest topological changes after temporally local reorderings. Furthermore we find that in most cases the topological similarity cannot be well described by either topological overlap or node degree.

\subsection{Influence of local reorderability and pairwise swaps on global reorderability}

After investigating topological resilience to large perturbations (global random reordering) and small perturbations (swapping one node pair) we naturally wonder: Are these two properties related? On even just one growing graph, if \emph{every} node pair was topologically similar (swapping does not change the persistent homology), then could we deduce that any possible ordering of nodes to result in the same growing topology? To answer this question and similar questions we need to dive a little deeper into the node swap analysis results. Let us first restrict ourselves to a binary form of topological similarity and say that two nodes are topologically similar if their swap has no effect on the persistent homology, and two nodes are topologically dissimilar otherwise. Then as shown in the Supplement (Fig.~\ref{fig:transitive}), topological similarity between nodes is a dependency relation: symmetric and reflexive but not transitive (see Counterexample 1 in the Supplement). Thus, we can define the (binary) topological similarity graph of dimension $n$, denoted $T_n$, as the graph with the same nodes as our original graph $B$ but with edges between $v_i,v_j$ if $v_i,v_j$ are topologically similar in dimension $n$. The topological similarity graph allows us to condense the information generated from the $\dbinom{N}{2}$ node swaps and further unravel the growing graph structure at the mesoscale level. We next can ask questions about the derived topological similarity graph that may offer novel insights into the growing process. For example, do communities suggesting groups of similar nodes exist in the topological similarity graph? Do we find that all nodes possess a similar capability of swapping without changing persistent homology, or will variation emerge? We will show that from small motifs in $T_n$ we can glean a deeper understanding of the global reorderability of a growing graph, but that in total these two concepts remain individually manipulable.

\subsubsection*{Intuition on one growing graph}

To begin let us consider one growing graph $(B,s_0)$ on $N$ nodes with topological similarity graph $T_n$ of dimension $n$ and reduce our above questions accordingly. Are the edges in $T_n$ the totality of the information contained in $T_n$, or does the structure imply more about the reorderability of $(B,s_0)$? First, we know that if our graph is totally reorderable (see Fig.~\ref{fig:totally_reorderable}) then $T_n$ will be a complete graph by definition. So if we can identify that the binary graph has this reorderable property then we know information about $T_n$.

Can we also go the other direction and infer information about large node order permutations from the $T_n$ structure? In possibly the simplest case, does finding an $m$-clique in $T_n$ indicate that all swaps on those nodes admit the same persistent homology? Let us examine what happens when nodes $v_i,v_j,$ and $v_k$ with $i <j <k$ all connect in $T_n$. By definition the barcodes (and consequently the Betti curves) produced from the $s_0$, $s_{i,j}$, $s_{i,k}$, and $s_{j,k}$ orderings on $B$ are all equal in dimension $n$. If we write out the graph filtration arising from each of these orderings with the same binary graph $B$ we have

\begin{displaymath}
\xymatrixcolsep{0.7pc}\xymatrixrowsep{1pc}\xymatrix{
	G_{v_1} \ar[r]  & \dots \ar[r] & G_{v_{i-1}} \ar[r] & G_{v_i} \ar[r]  & G_{v_{i+1}} \ar[r] & \dots \ar[r] & G_{v_{j-1}}   \ar[r] & G_{v_j} \ar[r]  & G_{v_{j+1}} \ar[r] & \dots & G_{v_{k-1}} \ar[r] & G_{v_k} \ar[r] & G_{v_{k+1}}\ar[r] & \dots \ar[r] & G_N\\
	G_{v_1} \ar[r]  & \dots \ar[r] & G_{v_{i-1}} \ar[r] & G_{v_j} \ar[r]  & G_{v_{i+1}} \ar[r] & \dots \ar[r] & G_{v_{j-1}}   \ar[r] & G_{v_i} \ar[r]  & G_{v_{j+1}} \ar[r] & \dots & G_{v_{k-1}} \ar[r] & G_{v_k} \ar[r] & G_{v_{k+1}}\ar[r] & \dots \ar[r] & G_N \\
	G_{v_1} \ar[r]  & \dots \ar[r] & G_{v_{i-1}} \ar[r] & G_{v_k} \ar[r] & G_{v_{i+1}} \ar[r] & \dots \ar[r] & G_{v_{j-1}}   \ar[r] & G_{v_j} \ar[r] & G_{v_{j+1}} \ar[r]  & \dots \ar[r] & G_{v_{k-1}} \ar[r] & G_{v_i} \ar[r]  & G_{v_{k+1}}\ar[r] & \dots \ar[r] & G_N \\
	G_{v_1} \ar[r] & \dots \ar[r] & G_{v_i-1} \ar[r] & G_{v_i} \ar[r] & G_{v_{i+1}} \ar[r] &\dots \ar[r] & G_{v_{j-1}} \ar[r] & G_{v_k} \ar[r] & G_{v_{j+1}} \ar[r] & \dots \ar[r] & G_{v_{k-1}} \ar[r] & G_{v_j} \ar[r] & G_{v_{k+1}}\ar[r] & \dots \ar[r] & G_N,
}
\end{displaymath}

\noindent in which each $G_{v_*}$ is the binary graph after the addition of node $v_*$. More specifically, $G_{v_*}$ is the induced subgraph of $B$ on all nodes added up to and including $v_*$. This observation is useful, but if we dive deeper we find that we in fact gain more information. Let us examine the graph filtration constructed by swapping nodes $v_i,v_j,v_k$ all together to form the ordering $s_{i,k,j}$ such that now our graph filtration reads as follows:

\begin{displaymath}
\xymatrixcolsep{0.7pc}\xymatrixrowsep{1pc}\xymatrix{
	G_{v_1} \ar[r] & \dots \ar[r] & G_{v_i-1} \ar[r] & G_{v_j} \ar[r] & G_{v_{i+1}} \ar[r] &\dots \ar[r] & G_{v_{j-1}} \ar[r] & G_{v_k} \ar[r] & G_{v_{j+1}} \ar[r] & \dots \ar[r] & G_{v_{k-1}} \ar[r] & G_{v_i} \ar[r] & G_{v_{k+1}}\ar[r] & \dots \ar[r] & G_N
}.
\end{displaymath}

\noindent Notice that here we add the first $j-1$ nodes in exactly the same order as row 2 above (the $s_{i,j}$ row). After we have added node $v_k$ (the $j^{th}$ node in this $s_{i,k,j}$ sequence), we have added the same subset of nodes as in row 3 above (the $s_{i,k}$ row) at the $j^{th}$ node addition. Then, not only does $G_{v_k}$ from the $s_{i,k,j}$ row equal $G_{v_j}$ in the $s_{i,k}$ row, but we see that the entire rest of the filtrations are equal to each other. Thus our new tri-swap graph filtration is constructed from parts of two pair-swap graph filtrations cut and pasted together\footnote{Excluding the map from the first to second part, which will cause trouble for us later.}. Since the graphs are equal at each point (node addition), the homology is also equal at each point, and thus $\beta_n$ of the tri-swap must be equal to $\beta_n$ of the original ordering.

For the above argument we only used two pairwise swaps to show that the tri-swap had to produce the same Betti curve. Indeed, there are only two tri-swaps on three letters and the other swap (adding $v_k$, $v_i$, then $v_j$) can similarly be shown to have the same Betti curve as the original ordering using only the topological similarity of node pairs $v_k$, $v_i$ and $v_k$, $v_j$. Then having a 3-clique of nodes $v_i$, $v_j$, and $v_k$ in $T_n$ guarantees that any tri-swap of $v_i$, $v_j$, and $v_k$ will also share the same $\beta_n$. Does the intuition gained regarding 3-cliques in $T_n$ generalize to higher cliques? In Counterexample 2 in the Supplement, we construct a growing graph in which four nodes are all topologically similar but we observe different Betti curves when we swap all four nodes together. Thus, an arbitrary clique of size 4 or more in $T_n$ does not imply that \emph{all} swaps of the participating nodes will yield the same Betti curves. Still, we can use the above cut-and-paste method with the pairwise-swapped filtrations to construct rules for which permutations of nodes involved in a clique in $T_n$ will produce the same Betti curves. We briefly discuss and explore these rules in Example 3 of the Supplement. Finally, we ask whether the intuitions gained here for Betti curves translate to the bottleneck distance between barcodes. In Counterexample 3 in the Supplement, we construct a growing graph in which $v_i$, $v_j$, and $v_k$ are each pairwise topologically similar, but neither tri-swap produces the same barcode. This finding is perhaps unsurprising, as the barcodes record much more detailed information about the evolving topology of the growing graph than do the Betti curves.

To summarize, we find that the local and global reorderability, though related, are distinct properties of a growing system. A growing graph may be both globally and locally reorderable, locally but not globally reorderable, or not reorderable in either sense. Additionally our results describing information gained from cliques in $T_n$ reemphasize the difference in perceived reorderability based on slices (Betti curves) versus longevity (barcodes).

\subsubsection*{For growing graph models}

\begin{figure}
	\centering
	\includegraphics[width=6in]{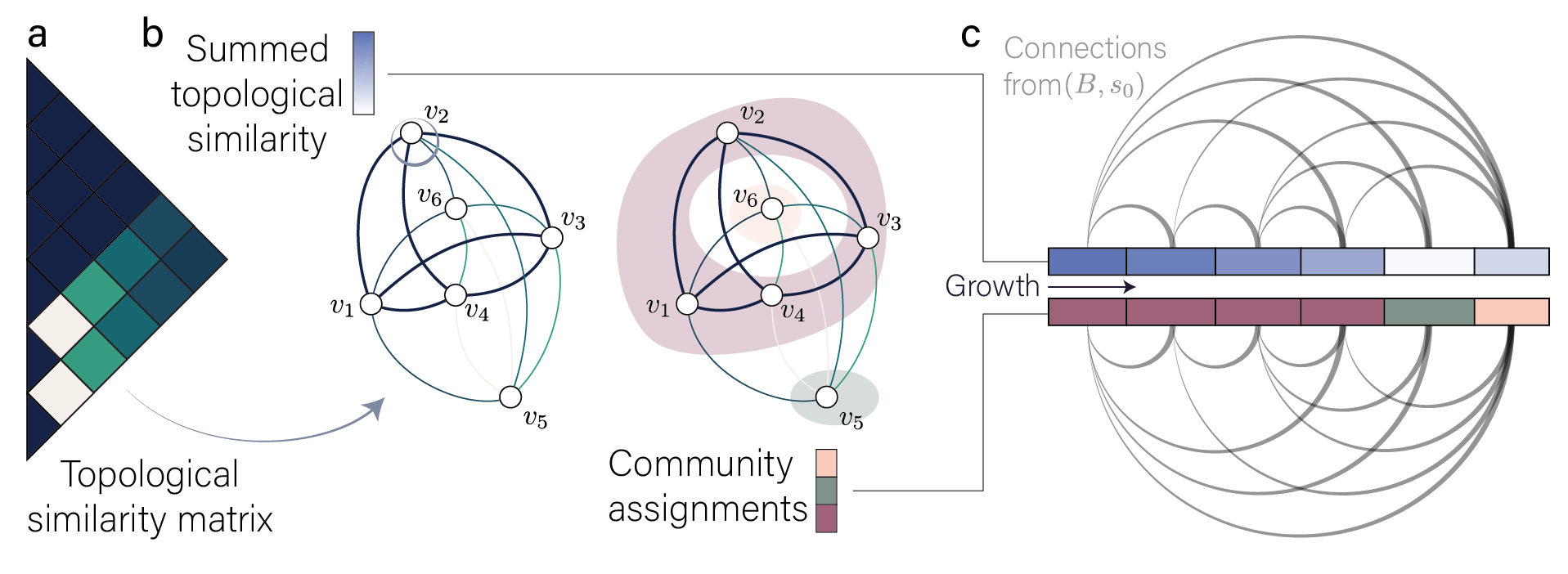}
	\caption{\textbf{Analyzing average topological similarity weighted graphs formed from pairwise node order swaps.} \emph{(a)} The topological similarity matrix forms a weighted graph. \emph{(b)} On the weighted network, node strength corresponds to the summed topological similarity, and community assignments describe sets of nodes that can be pairwise swapped with small consequences on the persistent homology. \emph{(c)} The summed topological similarity (top) and the community assignments (bottom) shown as vectors paired with the fingerprint graphs from $(B,s_0)$.} 
	\label{fig:LtoG_intro}
\end{figure}

\begin{figure}
	\centering
	\includegraphics[width=6in]{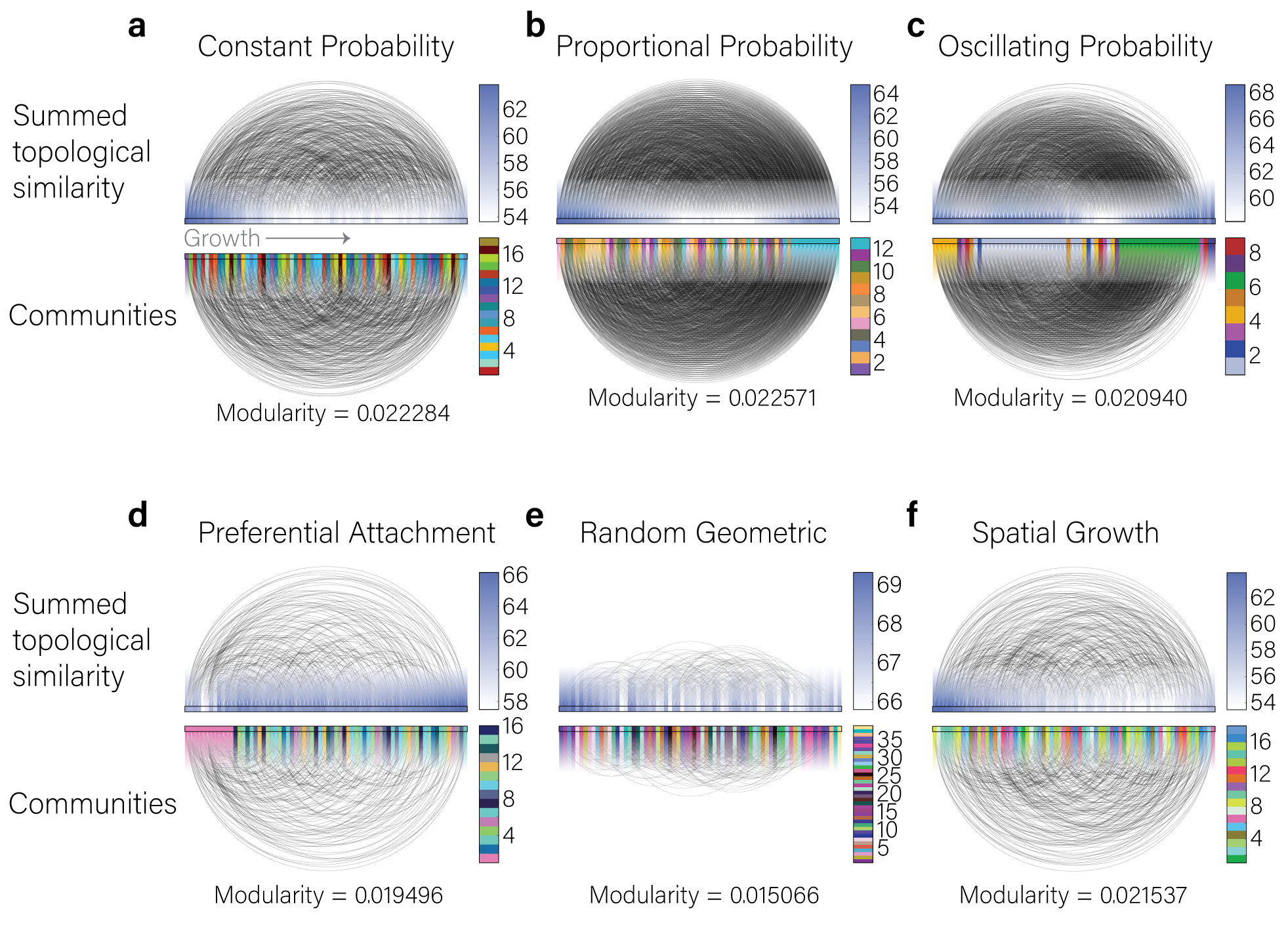}
	\caption{\textbf{Community detection and summed topological similarity overlaid on original binary networks.} We show the value of the summed topological similarity for each node with the blue colormap (top), and calculated community assignment (bottom) overlaid on a representative fingerprint graph for the \emph{(a)} constant probability, \emph{(b)} proportional probability, \emph{(c)} oscillating probability, \emph{(d)} preferential attachment, \emph{(e)} random geometric, and \emph{(f)} spatial growth models.}
	\label{fig:LtoG_results1}
\end{figure}

Now we return to our six growing graph models and ask our questions surrounding what the structure of the averaged topological similarity graph tells us about the original growing system. We begin by averaging the topological similarity of node pairs across replicates and across dimensions (Fig.~\ref{fig:LtoG_intro}a, see also Fig.~\ref{fig:local_results1},\ref{fig:local_results2}). Intuitively, if a node pair has a resulting high average topological similarity, then generally when constructing a growing graph from the particular model, those two nodes will contribute similarly to the persistent homology. We then ask if there indeed exist strongly connected subgroups of the average topological similarity graph using community detection by modularity maximization (Fig.~\ref{fig:LtoG_intro}b, see Methods for details). If a group of nodes forms a community in this averaged topological similarity graph, then often (but not necessarily, as shown above) we expect their pairwise swaps to result in small changes in persistent homology compared to swap magnitude. We represent each of the communities in the detected partition using a different color on the fingerprint graph of an exemplary binary graph $B$ (see Fig.~\ref{fig:LtoG_intro}b,c) for each of the six studied growing graph models in Fig.~\ref{fig:LtoG_results1}. We observe that the proportional probability, oscillating probability, and preferential attachment networks show at least one relatively large, temporally clustered community in their average topological similarity networks.  A temporally clustered community found in the averaged topological similarity graph suggests that as the graph develops, the evolving topology will likely change little in response to permutation in node order within the interval in which that community spawns.

Naturally, we next ask when those most or least swappable nodes emerge in the growing graph model. That is, do nodes with the highest summed topological similarity (averaged across dimensions) occur at the beginning of the growth process or at the end? For example, based on the connection patterns of the growing \textit{C. elegans} cellular nervous system we might expect that neurons born prior to hatching will often have lower summed topological similarity values than neurons born after hatching \cite{nicosia2013phase}. We calculate the strength (weighted degree of $T$, or summed topological similarity) of each node within the average topological similarity graphs (Fig.~\ref{fig:LtoG_intro}b,c) and we depict the summed topological similarity using node color within the fingerprint graphs shown in Fig.~\ref{fig:LtoG_results1}. We see that for most growing graph models, the earliest added nodes show a relatively high topological similarity, with the exception of the preferential attachment model. Interestingly, the proportional probability, oscillating probability, and to a small extent the spatial growth model show a U-shaped curve of summed topological similarity values. That is, the earliest and latest nodes can be swapped with most other nodes without changing the persistent homology much, but those nodes added in the middle of the growth process are less topologically similar to all other nodes. Then for these growing graph models, the nodes added in the middle are most sensitive to perturbation, suggesting that they exhibit distinct roles in the persistent homology of the growing network.

To summarize, our results suggest that the notion of local reorderability as well as our general analysis approach to studying growing graph processes via node swaps can provide insight into sections of the graph maturation course that may be more topologically robust to node perturbations than other sections. Additionally, these analyses can also expose nodes that are relatively essential for the proper topological evolution of the graph by revealing which nodes on average often create relatively large changes in the evolving topology when swapped.

\section{Discussion}

In this paper we have explicitly defined a convenient framework for studying the evolving topology of growing graphs and the robustness of that topology in six growing graph models. We perturbed each growing process by exchanging the node addition order and tested the ability of each growing graph to retain the same evolving topology after perturbation. We found that the examined graph models display a wide range of resiliency in their topology, here termed \emph{reorderability}, as a result of local and global perturbations. Finally, we found that reorderability at the pairwise level exerts limited restrictions on the global reorderability of a growing network, and suggest analyses to understand this influence in the graph model case. In the remainder of this discussion section, we consider the implications of our work first in terms of theoretical matters, and then in terms of applications to real world systems, and finally leave the reader with a series of yet unanswered questions.

\subsection{Persistent homology of growing graph models and their connections to edge filtered complexes}

Most commonly in network science, persistent homology has been applied to study edge-weighted networks \cite{petri2013topological,sizemore2016classification}. The edge-weighted network models that have been most well-studied from the persistent homology perspective are the i.i.d. random edge weight model \cite{kahle2009topology} and the (edge-filtered) random geometric model \cite{kahle2011random}. Here we intentionally use similar ideas to construct the constant probability model and the (node-filtered) random geometric model. Indeed, at the end of these growing processes, we have constructed a binary graph $B$ that could be instead grown from adding edges to a set of $N$ vertices. Specifically, the homology at the end of the filtrations generated by the constant probability model or random geometric model, must match, on average, the homology of their edge-weighted counterparts at that same edge density and number of nodes. Importantly the connection between these two models is complicated by the variability of edge densities after each node addition in our growing graph models. Still, we strongly suspect that a formal connection between node-filtered and edge-filtered models does exist. If true, a particularly interesting direction for future work is to address the question of whether one could solve the intricacies of an edge-filtered model and apply that solution directly to the node-filtered case, or \emph{vice versa}.

Additionally, as noted in \cite{sizemore2018knowledge}, while our definition of the constant probability model matches the nature of the random i.i.d. edge-filtered complex, the Betti curves of the proportional probability model instead qualitatively match those of the random i.i.d. edge-filtered case \cite{kahle2014topology}. We leave the exploration of the connection between these models for future work, but we speculate that it may be related to the similar progression of randomly attaching higher dimensional simplices. In the i.i.d. edge-filtered model, each new edge added randomly completes larger (higher-dimensional) simplices on average, while in the dual of the proportional probability model we randomly add higher dimensional simplices at each step on average. Additionally, both models end with the coning off of all persistent homology, tying at least the homology of the filtration end-points to one another.

\subsection{Extending to models of growing simplicial complexes}

In this work we restricted our attention to node-filtered simplicial complexes constructed from clique complexes of graphs. Importantly many other growing simplicial complex models exist, including those that add new 2-simplices at each filtration step \cite{wu2015emergent}, growing simplicial complexes that display specific geometry \cite{bianconi2016emergent,bianconi2016network}, weighted growing simplicial complexes \cite{courtney2017weighted}, point processes \cite{yogeshwaran2015topology,owada2017limit}, and more \cite{fowler2019homology,courtney2016generalized}. Some growing simplicial complexes may fit directly into our node-filtered perspective, but others may add exclusively larger simplices at each step. Regardless of simplex size (or distribution of sizes) added at each step, one can ask similar reorderability questions to those asked in this work. For example, how might the persistent homology change if we reordered the growth process to add the smallest simplices first, or instead the largest simplices first? Which generative simplicial complex models exhibit local or global reorderability? Can we determine more reorderable motifs in order to allow us to simplify or reduce the computation of local reorderability? Or perhaps can we induce constraints on a developing network that select for reorderability, as with flow networks \cite{rocks2019topological}? These questions on growing simplicial complexes and others raised in the present work additionally build upon previous investigations of the effect of removing vertices or shifting simplex weights on the persistent homology \cite{memoli2018quantitative}.

\subsection{Applications of node-filtered systems}

While the reorderability of node-filtered order complexes raises many interesting theoretical questions, we also emphasize the prevalence of node-ordered systems in applications. Certainly many biological systems grow and spread including vasculature networks \cite{gavrilchenko2019resilience}, fungal networks \cite{papadopoulos2018comparing,heaton2012analysis}, and the developing connectome \cite{quadrato2017cell,mansour2018vivo}, but we also see growing processes in non-biological systems as well such as evolving mass transportation networks in cities \cite{sperry2016rentian} and the evolution of ideas in research journals \cite{dworkin2018landscape,dworkin2018emergent}. Moving further still we can encode signal transduction with the n-order complex formalism, such as occurs in the propagation of a focal seizure along the connectome \cite{kramer2010coalescence}, phosphorylation cascades in response to an extracellular signal \cite{bray1993computer}, and contagion \cite{iacopini2019simplicial,petri2018simplicial}. Finally in graph learning one traverses a network of topics or ideas and relations between them, making studies involving the learnability of graphs \cite{lynn2018structure} and semantic memory \cite{hills2012optimal} interesting applications for reorderability investigations as well. Finally and perhaps most interestingly, in certain applications one may be able to experimentally test and manipulate reorderability of complex systems. For example, one might activate genes in different sequences or introduce topics in diverse orders to test the impact on learnability of the material. Or instead if the growing topology of a seizure spread suggests severity, one could use the presented analyses in order to determine which nodes when stimulated would most drastically disrupt the unfavorable topology. In general we speculate that the reorderability of a complex system will reveal inherently new information about how the system responds to change.

\subsection{Reorderablity derived from Betti curves versus barcodes in applications} 

As discussed in the previous sections, reorderability from the Betti curve perspective captures a different level of information about the system than reorderability from the barcode perspective. If a graph displays reorderability based on the Betti curves, then slice-by-slice the global reorderings must be similar in their topology. In contrast, if a graph displays reorderability based on the barcodes, then the entire topological evolution of each reordered filtration must be similar. Differences in perceived reorderability based on Betti curves versus barcodes mimic differences seen in cross-sectional versus longitudinal studies \cite{pantelis2003neuroanatomical,giedd1999brain}. While we might tend towards desiring the latter, ofttimes in experiments we might find ourselves constrained to the former \cite{schmidt2005issues}. For example, if studying brain development we may perform a cross-sectional study instead of a longitudinal study due to the availability of participants or technology. With cross-sectional slices, one would generally be confined to studying reorderability based on Betti curves. However, if we instead performed a longitudinal study, the presence of natural maps from one timepoint into the next would readily support a comparison of topology based on the barcode. Overall our work underscores the importance of longitudinal investigations in unearthing information about a growing system.

\subsection{Why might we design a reorderable system?}

The results from the local-to-global studies showed that the local reorderability and global reorderability are surprisingly two rather distinct properties of a growing system. This result raises the question of in what contexts we might wish to design a system that exhibits, for example, local but not global reorderability. One such context might be cell proliferation. Here, marked stochasticity exists at the gene transcription level, but only a large fluctuation in particular gene sets will alter phenotype \cite{shaffer2017rare,shaffer2018memory}. It is intuitively plausible that cell regulatory networks exhibit local reorderability to account for the small variations in gene expression, but do not demonstrate global reorderability if the cell is differentiated, making cells difficult to reprogram \cite{ieda2010direct}. Indeed testing the reorderability of such systems is a potentially fruitful avenue for future research. On the engineering side, if we seek to design a growing system that performs a particular function from scratch and that also needs to display robustness to perturbation, the reorderability framework presented here may offer a plausible test of proper network function.  Additionally further investigations into reorderability and specifically motifs that support or prevent reorderability may lead to better predictions of system responses to perturbation.

\subsection{Open questions not addressed in this paper}

Often research projects hatch more questions than they answer. We take this last opportunity to highlight a few additional open questions not addressed in the current work. First, the constant probability model, as discussed, grows into a random i.i.d. network that has a particular Betti curve signature \cite{kahle2009topology}. In Fig.~\ref{fig:team1} we see the constant probability model Betti curves may follow a similar yet stretched pattern that is truncated by the number of nodes. One might ask whether one would continue to see increasing peaks of increasing dimensions as the number of nodes grows (see \cite{kahle2014topology}), and generally if the persistent homology signatures seen here scale with graph size. Second, our examples and intuition suggest that many biological networks may be locally but not globally reorderable. We leave open the question of which, if any, biological or developmental networks indeed exhibit such properties. Third, in the current work we only consider the topological similarity graph, although we in fact recover a topological similarity simplicial complex in which $k$ nodes form a simplex if any permutation of those $k$ nodes does not alter the persistent homology. What more can one learn about the growing graph architecture from more thoroughly studying the simplicial complex formed by the topological similarity relation? Is it even true that permutations of nodes within a clique of $T_n$ alter the persistent homology to a smaller degree than permutations including nodes within and outside of the clique? Finally, we note that while we studied reorderability of growing graphs with respect to one node ordering, removing the base ordering and investigating the variability of the persistent homology generated from one binary graph and any random node order would reveal the topological regularity of a binary graph. That is, given a binary graph the persistent homology may vary greatly or insignificantly across random growth orders, suggesting a sameness in the topology across the binary graph as observed previously in a semantic feature network \cite{sizemore2018knowledge}. Previous studies have linked network homogeneity with synchronizability \cite{shi2013searching,shi2019totally}, but it remains to be shown how topological regularity based in the stability of a graph's persistent homology may also contribute to optimal synchronizability.

\section{Conclusion}

In this work we explore the evolving topology of growing graph models and test the robustness of this evolving topology with respect to perturbations of node order. We find that both spatially embedded and non-embedded models can exhibit reorderability at a global level, and that globally reorderable growing graphs may not be the most locally reorderable. We determine that reorderability at the pairwise level does not necessarily imply global reorderability, but still can be used to determine more information about the growing graph architecture. Finally we proffer suggestions for deepening the theory behind node-filtered networks and speculate that real-world systems will be found to display a range of reorderability properties based on system function.

\section{Acknowledgments}

We are grateful to Jakob Hansen, Darrick Lee, Chad Giusti, Zoe Cooperband, Jason Kim, Lia Papadopoulos, Chris Lynn, Erin Teich, and Zhixin Lu for helpful discussions and brainstorming sessions. We are also grateful to Lia Papadopoulos, Jason Kim, Erin Teich, Chris Lynn, Darrick Lee, Nico Christianson, and Jennifer Stiso for helpful comments on early versions of this manuscript. This work was funded by the Army Research Office through contract number W911NF-16-1-0474. DSB and ASB also acknowledge additional support from the John D. and Catherine T. MacArthur Foundation, the Alfred P. Sloan Foundation, the ISI Foundation, the Paul Allen Foundation, the Army Research Laboratory (W911NF-10-2-0022), the Army Research Office (Bassett-W911NF-14-1-0679, DCIST- W911NF-17-2-0181), the Office of Naval Research, the National Institute of Mental Health (2-R01-DC-009209-11, R01-MH112847, R01-MH107235, R21-M MH-106799), the National Institute of Child Health and Human Development (1R01HD086888-01), National Institute of Neurological Disorders and Stroke (R01 NS099348), and the National Science Foundation (BCS-1441502, BCS-1430087, NSF PHY-1554488 and BCS-1631550). The content is solely the responsibility of the authors and does not necessarily represent the official views of any of the funding agencies.

\bibliography{bibfile}
\bibliographystyle{plain}

\newpage

\section{Supplementary Material}

We organize this Supplement into two main sections. First we provide additional useful examples and supporting arguments for statements made in the main text. Second, we report the results from any additional experiments run on the growing graph models.

\subsection{Intuition for the stability of barcodes in the node-filtered case}

The original stability theorem for persistence diagrams (equivalently barcodes) \cite{cohen2007stability} was expertly crafted in more general terms of weight functions on filtrations. In this work we restrict ourselves to a very specific type of filtered simplicial complex, which is one generated by filling in cliques as simplices within a growing graph. This special case that we consider here lends itself to a simple interpretation of the stability theorem for node-filtered order complexes, which we include below. We wish to emphasize that the following section contains no novel results, only interprets the stability theorem from \cite{cohen2007stability} from the perspective of node-filtered order complexes. 

\begin{figure}
	\centering
	\includegraphics[width=\textwidth]{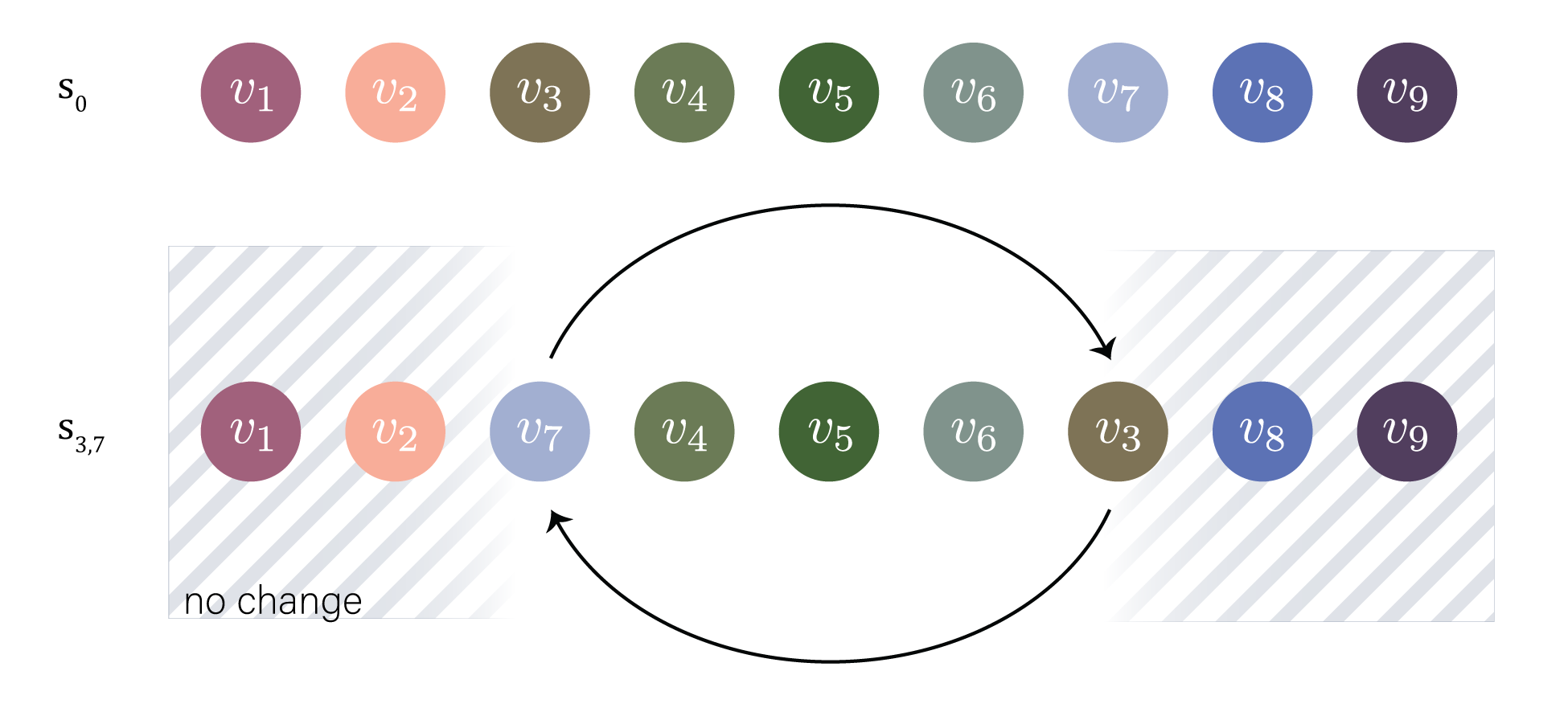}
	\caption{\textbf{A node swap defines the maximum extent to which the resulting persistent homology can change.} When swapping the original order (top) of nodes $v_3$ and $v_7$ (bottom), the filtration from nodes $v_1$ to $v_2$ and the filtration from the addition of the seventh node to the end do not change. Thus, outside the swap bubble the persistent homology cannot be affected. It is only inside the swap bubble that the persistent homology could change in response to the node swap shown.}
	\label{fig:Supp_stability}
\end{figure}

In the main text, we state that the stability theorem gives us an upperbound for the bottleneck distance between $P_0$ and $P_{i,j}$ of $|j-i|$. How do we see this upperbound in our growing graph models? If we grow the same binary network $B$, then recall that we only need to know the ordering $s$ to specify the growing graph. In Fig.~\ref{fig:Supp_stability} we assume the same binary graph $B$ and show only the original ordering of nodes, $s_0$ (top). Then when we swap nodes $v_i$ and $v_j$, $i < j$, we obtain the $s_{i,j}$ ordering shown in the bottom row of Fig.~\ref{fig:Supp_stability}. Notice that from the addition of the first node through node $v_{i-1}$, we do not change the nodes added or the order in which nodes are added, so the filtration from node $v_1$ through $v_{i-1}$ must be equal between $(B,s_0)$ and $(B,s_{i,j})$. Additionally, note that if we have added a subset of nodes $\nu \subseteq V$, the order in which nodes of $\nu$ were added cannot change the subgraph of $B$ induced by $\nu$. Therefore, the binary graph after the addition of $v_i$ in $s_{i,j}$ must be equal to the binary graph after the addition of $v_j$ in $s_0$ since at this point we have added all nodes $v_1, \dots, v_j$, and furthermore the filtrations must match from the addition of the $j^{th}$ node on. We illustrate these regions of the filtration in which no changes to the filtration -- and consequently to the persistent homology -- could have occurred as gray shaded regions of the orderings in Fig.~\ref{fig:Supp_stability}. 

Now, we can imagine a situation in which the addition of node $v_i$ begins a cavity and that all nodes $v_{i+1}, \dots, v_j$ do not participate in the cavity formation. So when we swap nodes $v_i$ and $v_j$, then we move a persistent cavity birth time from $i$ to $j$, and if no other persistent cavities emerge, then $j-i$ will be the bottleneck distance between barcodes generated by $s_0$ and $s_{i,j}$. Thus we could swap nodes $v_i$ and $v_j$ and produce a barcode distance of $j-i$, but no more. To see these points directly from the definition given in \cite{cohen2007stability}, here our weight function on simplices is defined by the node order so that the weight of a node is the number of that node in the sequence, and the norm of the weight function on the simplicial complex is the same as the norm of the function on the vertices. From this definition the interpretation above follows. 

\subsection{Examples and counterexamples}

\subsubsection*{Example 1: Cycles } \label{Cavities}

\begin{figure}
	\centering
	\includegraphics[width=6.5in]{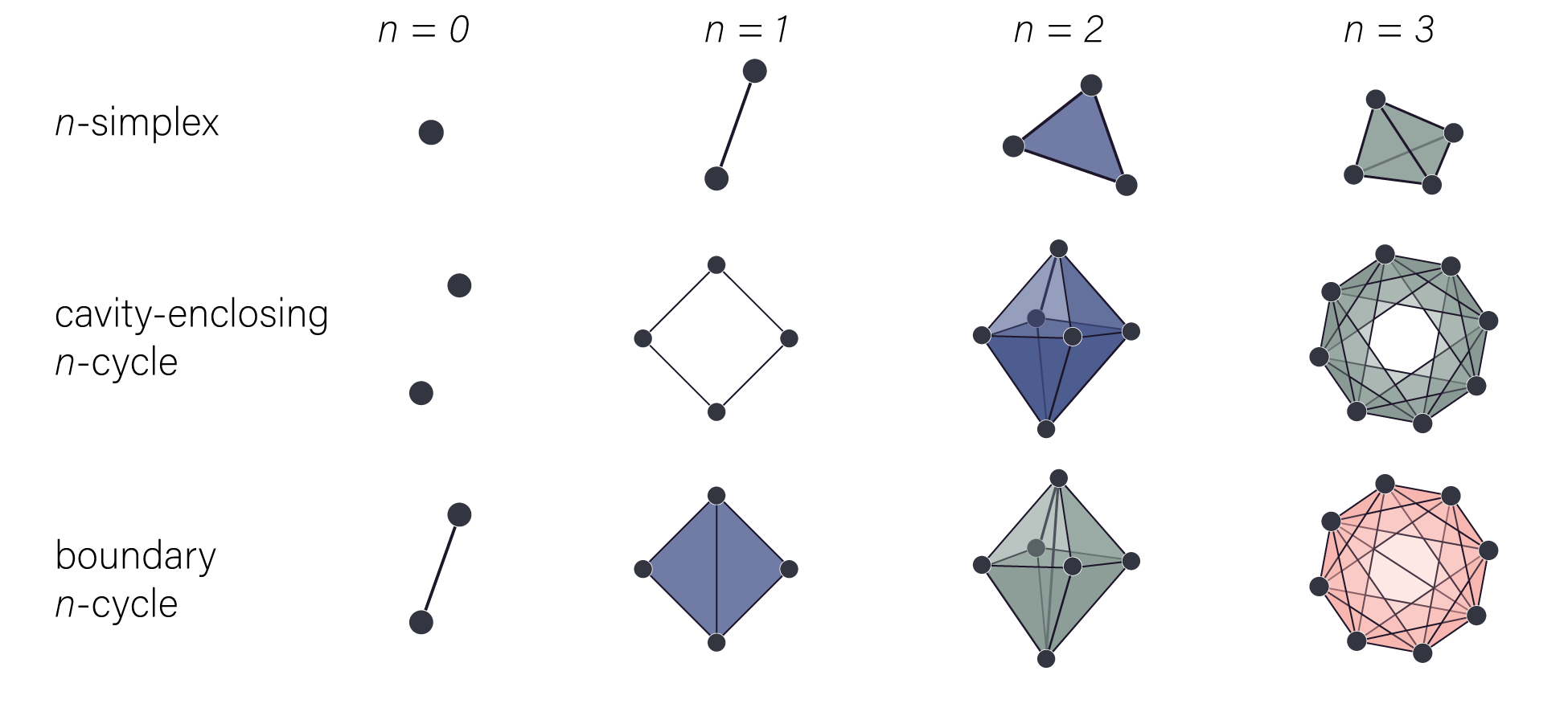}
	\caption{\textbf{Examples of $n$-cycles.} Simplices (top row), example of a cavity-enclosing $n$-cycle (middle row), and an example of a boundary $n$-cycle (bottom row) for small values of $n$.}
	\label{fig:cavities}
	
\end{figure}

While cycles composed of edges are often familiar, cycles, boundaries, and cavities of other dimensions are often less common. To expand upon our description of homology in the main text, here we present an example of one cavity-enclosing and one boundary $n$-cycle for $n = 0,1,2,3$. Shown in Fig.~\ref{fig:cavities}, we see first an example of a 0-cycle that is not a boundary (two disconnected nodes) and a 0-cycle that is also a 0-boundary (two connected nodes). Next we see the familiar cavity-enclosing and boundary 1-cycles in which the 1-simplices either surround a 2-dimensional void or form a boundary of two 2-simplices. The octahedron is an example cavity-enclosing 2-cycle, but filling the interior with 3-simplices then offers an example of a boundary 2-cycle. Finally we illustrate a projection of a cavity-enclosing 3-cycle which surrounds a 4-dimensional void with a shell of 3-simplices, and a 3-cycle that is also a boundary in which the shell of 3-simplices surrounds a collection of 4-simplices (peach).  

\subsubsection*{Example 2: Totally reorderable graphs} \label{totally_reorderable}

\begin{figure}
	\centering
	\includegraphics[width=6.5in]{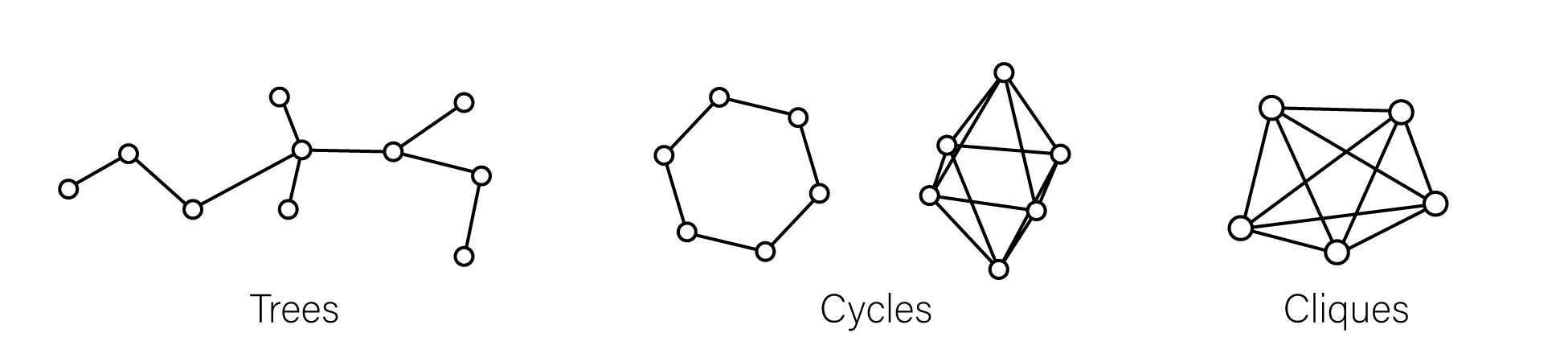}
	\caption{\textbf{Examples of completely reorderable graphs.} If we assume that we will form a filtered simplicial complex via the n-order complex construction, then the tree graph (\emph{left}) is completely reorderable in dimensions $\geq 1$, a minimal $n$-cycle (\emph{middle}) is completely reorderable in dimensions $\geq n$, and cliques (right) are totally reorderable in all dimensions.}
	\label{fig:totally_reorderable}
\end{figure}

In Fig.~\ref{fig:0a} we see the transposition of nodes $d$ and $f$ change the persistent homology of the growing graph, and in the main text we see that the reorderability of a growing graph depends on the topology of the binary graph $B$. In fact, some special binary graphs exist that will yield the same persistent homology in a particular dimension for \emph{any} ordering of nodes. First, if $B$ is a tree (Fig.~\ref{fig:totally_reorderable}, left) equipped with some node ordering $s:V\rightarrow \mathbb{N}$ then there will be no homology in dimension $n$ for $n\geq 1$ along the entire filtration. Indeed, any node addition order will yield the same persistent homology (here this means no persistent cavities emerge) in dimension $n$ for $n\geq 1$. Second, any minimal $n$-cycle (specifically an $(n+1)$-cross polytope, see \cite{giusti2015clique}) such as the minimal 1-cycle or minimal 2-cycle shown in Fig.~\ref{fig:totally_reorderable} (we assume that we always create the clique complex in this case) in which all nodes and $n$-simplices are required for the cycle will produce the same persistent homology regardless of node order for dimensions $\geq n$. Finally, an $n$-clique in $B$ as shown on the right of Fig.~\ref{fig:totally_reorderable} (again assuming that we take the clique complex) will produce the same persistent homology in all dimensions regardless of node order. These examples help us to understand the sort of topological homogeneity necessary for a binary network $B$ to exhibit complete reorderability.

\subsubsection*{Example 3: Rules of permuting nodes within a $k$-clique in $T_n$} \label{rules}

In the main text we stated that while finding a 3-clique in $T_n$ implies that any permutation of the nodes involved in the clique would produce the same Betti curves in dimension $n$ as the original ordering, and additionally that this fact does not hold for all permutations of nodes within a 4-clique of $T_n$. Here we illustrate a rule for permutations on $m$ nodes of an $m$-clique in $T_n$ that will guarantee conservation of $\beta_n$. 

Let $\{v_{i_1},\dots , v_{i_m}\}$ be the nodes of an $m$-clique found in $T_n$ with $i_j < i_{j+1}$. Note that in the original ordering $s_0$, $v_{i_{j+1}}$ may not follow directly after $v_{i_j}$, but for the following proof we can ignore any nodes added between each $v_{i_j}$, $v_{i_{j+1}}$ as they do not affect the outcome. We only require that $i_{j+1}>i_j$ so that in the original ordering $s_0$, $v_{i_{j+1}}$ comes after $v_{i_j}$. We can also ignore any nodes added before $v_{i_1}$ or after $v_{i_m}$ in $s_0$ as they will not cause a change in the Betti curves (see Fig.~\ref{fig:Supp_stability}). For the sake of clarity, we can rewrite our nodes of the $m$-clique as $v_1, v_2, \dots, v_m$ and only focus on the following portion of the graph filtration $\dots \rightarrow G_{v_1} \rightarrow \dots \rightarrow G_{v_m} \rightarrow \dots$ since each step is an induced subset of the ending binary graph $B$ on all nodes added up to and including that step. We call a permutation $\sigma$ of nodes $v_1, \dots, v_m$ \emph{admissible} if the permutation $s_{\sigma}$ of these nodes with respect to the original ordering $s_0$ results in a growing graph $(B,s_{\sigma})$ with the equal $\beta_n$ to that of $(B,s_0)$. Based on our cut-and-paste strategy illustrated in the main text, we propose the following rule for permutations $\sigma$ that if met, will show that the permutation $\sigma$ of nodes in an $m$-clique in $T_n$ is admissible.\\

\underline{\emph{Rule 1:}} Let $v_1, \dots, v_m$ form an $m$-clique in $T_n$ and  $\nu_k$ denote the first $k$ nodes in a reordering of $v_1, \dots, v_m$. Then a reordering of $v_1, \dots, v_m$ is admissible if for at least $(k-1)$ nodes $v_i \in \nu_k$, $i \leq k$, for all $k = 2, \dots, m$.\\ 

Informally, this rule states that if we take the first two nodes in the reordering, check that at least $v_1$ or $v_2$ is present, then move to the first three nodes and check that at least two of $v_1$, $v_2$, and $v_3$ is present, and so on, then the permutation will be admissible. 

In order to prove that an ordering following Rule 1 is indeed admissible, we fix the number of nodes $m$ and induct on $k$, so that for each $k$ we prove the modified rule:

\underline{\emph{Rule$_k$:}} Let $v_1, \dots, v_m$ form an $m$-clique in $T_n$ and $\nu_r$ denote the first $r$ nodes in the reordering. The reordering is admissible up to the $k^{th}$ node if for at least $(r-1)$ nodes we have $v_i \in \nu_r$, $i\leq r$, for all $r = 2,\dots, k$.

First, we show that the rule holds for $k=2$. Let $2 < q \leq m$. Then if $\nu_2 = (v_1, v_2)$, our filtration through the addition of $v_2$ is equal to that from $s_0$. If $\nu_2 = (v_1,v_q)$ or $\nu_2 = (v_q,v_2)$ then since $v_1, v_2, v_q$ are connected in $T_n$, $\beta_n$ must remain unchanged. Finally if $\nu_2 = (v_q,v_1)$ or $\nu_2 = (v_2,v_q)$, then using the cut-and-paste method we can show up through the second node added that the reordering will not change $\beta_n$. We show the $\nu_2 = (v_q,v_1)$ case below in Fig.~\ref{fig:rules}a to illustrate that by using the topological similarity of the $v_1, v_q$ pair and the $v_2,v_q$ pair, we can piecewise construct the relevant parts of the new graph filtration. In particular the filtration up through adding $v_q$ is the same as from $(B,s_{1,q})$ (peach) and adding $v_1$ in the new ordering will create the same binary graph as in $(B,s_{2,q})$ (blue) and thus must have conserved the Betti curve for at least the first two nodes added. We leave the $\nu_2 = (v_2,v_q)$ for the reader. 

Generally we see that for an admissible ordering, we will need to use no swaps, one swap, or two swaps with the cut-and-paste process. 

\begin{figure}
	\centering
	\includegraphics[width=6.5in]{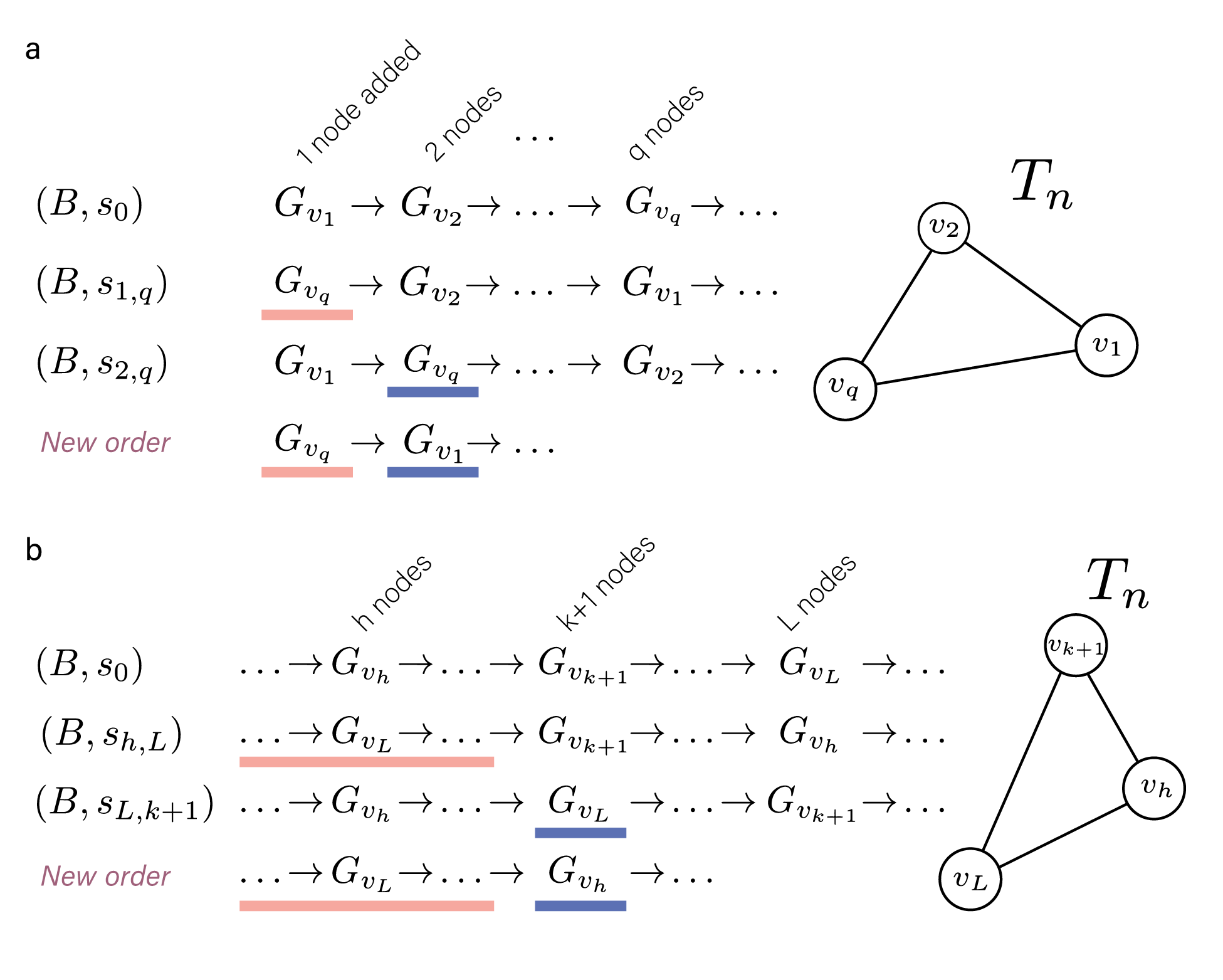}
	\caption{\textbf{Examining node swaps that conserve Betti curves.} \emph{(a)} Illustration of using the cut-and-paste process to show up through the first two nodes of the reordering that $\beta_n$ will be unchanged. \emph{(b)} Illustration detailing checking that $\beta_n$ is conserved through the first $k+1$ nodes as described in Case 2a.}
	\label{fig:rules}
\end{figure}

Now we need to show that given a reordering in which the first $k$ nodes of the reordering satisfy Rule$_k$, if Rule$_{k+1}$ is satisfied then the reordering up through the first $(k+1)$ nodes will be admissible. Assume the first $k$ nodes in the reordering follow Rule$_k$ so that for  $v_i \in \nu_r$, we have $i \leq r$ for at least $(r-1)$ of the nodes in $\nu_r$, for all $r = 2,\dots, k$. We need to consider the possible nodes that might come next, at the $k+1$ spot. We break this portion into two cases. As for the $k=2$ case, we will need to use the topological similarity of at most two node pairs.

\begin{itemize}

\item \emph{Case 1:} For all $k$ nodes $v_i \in \nu_k$, $i \leq k$. Then at the addition of the $k^{th}$ node in the reordering, we have the graph induced by the subset of nodes in $\nu_k$, which is equal to the graph after the $k^{th}$ node is added in the original ordering $s_0$. Let the next node added be $v_L$, and then $L \geq k+1$. Since $v_L$ is topologically similar to $v_{k+1}$ by the assumption, the ordering $s_{L,k+1}$ yields the same $\beta_n$, and we are done.

\item \emph{Case 2:} Exactly one of the $v_i \in \nu_k$ has $i\geq k+1$; call this node $v_L$. Then exactly one node $v_i$ with $i\leq k$ will be added later (after the $k^{th}$ node); call this node $v_h$. 

\begin{itemize}
\item \emph{Case 2a:} Let $L\geq k+2$. Then Rule$_{k+1}$ requires the $k+1$ node to be $v_h$. As illustrated in Fig.~\ref{fig:rules}b, we use the topological similarity of the $v_L$, $v_h$ pair and the $v_{k+1}$, $v_L$ pair to construct the binary graphs in the reordered filtration. Therefore our reordered growing graph has the same $\beta_n$ as the original up through the first $(k+1)$ nodes.

\item \emph{Case 2b:} Let $L = k+1$. Then the $k+1$ node could be any node, as we will still have that for $k$ of the $v_i\in \nu_{k+1}$, $i\leq k+1$. Denote the $k+1$ node by $v_M$. If $M = h$ then we are done as we only need to map into the original filtration. If $M\neq h$, then the reordered filtration will have equal $\beta_n$ up through the $(k+1)$ node to that emerging from the original ordering due to the topological similarity of the $v_h$, $v_L$ pair and the $v_h$, $v_M$ pair. 
\end{itemize}
\end{itemize}

This completes the induction step and therefore we can conclude that if the reordering of the $m$ nodes of an $m$-clique in $T_n$ adheres to Rule 1, then $\beta_n$ will remain unchanged.

\subsubsection*{Counterexample 1: Non-transitivity of the topological similarity relation}

\begin{figure}
	\centering
	\includegraphics[width=6.5in]{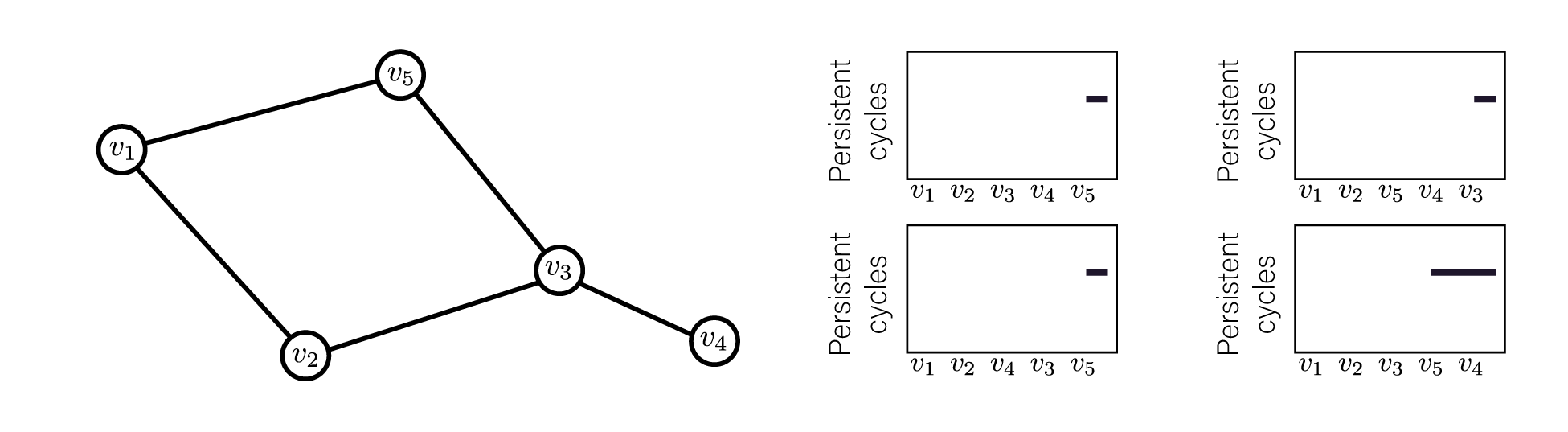}
	\caption{\textbf{Topological similarity is not a transitive relation.} The $v_3$ node is topologically similar to nodes $v_4$ and $v_5$, but $v_4$ and $v_5$ are not topologically similar as illustrated by the barcodes on the right.}
	\label{fig:transitive}
\end{figure}

Here we provide details proving that the binary topological similarity relation between node pairs is a dependency relation as opposed to an equivalence relation. Recall that we defined two nodes  $v_i$ and $v_j$ in a single growing graph $(B,s_0)$ as topologically similar if $d^n_{BN}(P_0,P_{i,j}) = 0$, and for this section we write $v_i \equiv_r v_j$ if this is so. First, this definition of topological similarity is trivially reflexive and symmetric. To show that topological similarity is not transitive, consider the growing graph in Fig.~\ref{fig:transitive} in which $v_3 \equiv_r v_4$ and $v_3 \equiv_r v_5$. Indeed, we see that the $v_4 \leftrightarrows v_5$ swap does not yield the same persistent homology as the original ordering, and so $v_4\not\equiv_r v_5$. Thus, topological similarity is not transitive, and therefore is also not an equivalence relation.

\subsubsection*{Counterexample 2: Not all swaps between four topologically similar nodes yield the same Betti curves}

\begin{figure}
	\centering
	\includegraphics[width=6.5in]{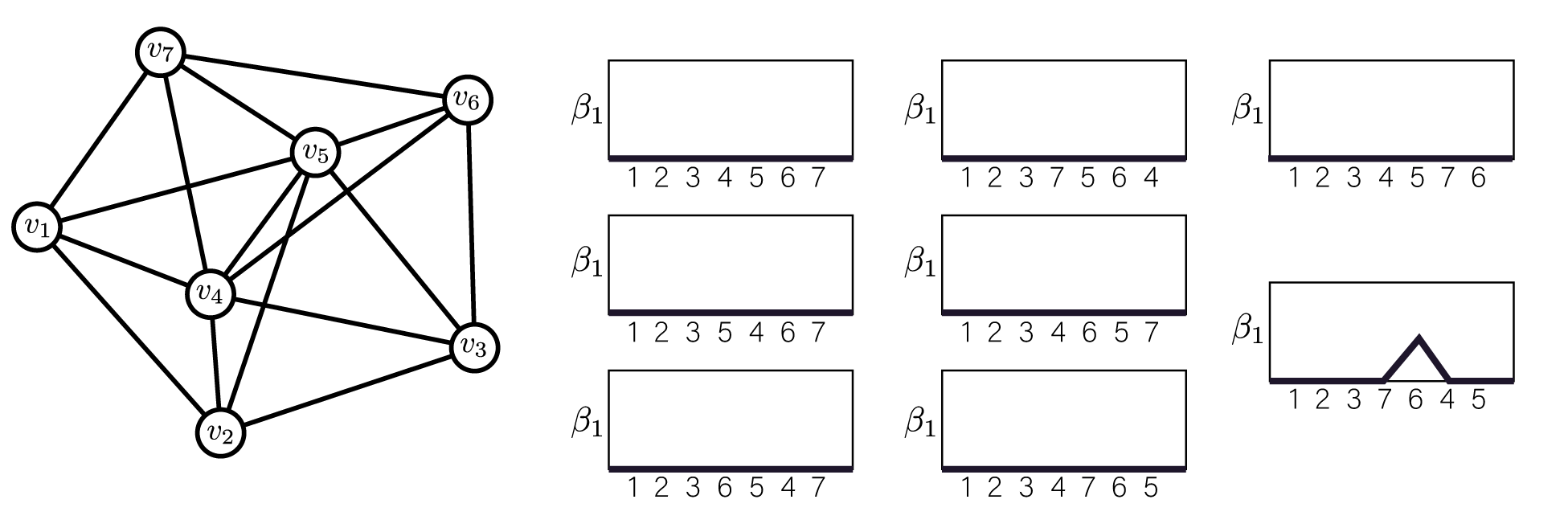}
	\caption{\textbf{Swapping nodes from a 4-clique in $T_n$ does not necessarily conserve Betti curves.} Nodes $v_4$, $v_5$, $v_6$, and $v_7$ are all pairwise topologically similar and all swaps of no more than three of these four nodes will not alter the Betti curves (right). However, the final ordering in which nodes $v_7$ and $v_6$ precede nodes $v_4$ and $v_5$ produces a different $\beta_1$ than the original ordering.}
	\label{fig:cex2}
\end{figure}

In the main text we consider how permutations of the $k$ nodes found in a $k$-clique in $T_n$ would affect the persistent homology. Here we provide a growing graph in which four nodes are all pairwise topologically similar, but a permutation of all four nodes does not result in the same persistent homology. Specifically, in the graph above (Fig.~\ref{fig:cex2}) one can show that nodes $v_4$, $v_5$, $v_6$, and $v_7$ form a 4-clique in $T_1$. As a consequence, all pairwise swaps of these nodes yield the same barcode as the original binary graph (Fig.~\ref{fig:cex2}, right), which shows no persistent cavities. However, we see that if we perform a reordering such that we add these four nodes in order $v_7$, $v_6$, $v_4$, and then $v_5$, $\beta_1$ has changed as now we see a peak at the addition of the fifth node ($v_6$).

\subsubsection*{Counterexample 3: Not all swaps between three topologically similar nodes yield the same persistent homology}

\begin{figure}
	\centering
	\includegraphics[width=6.5in]{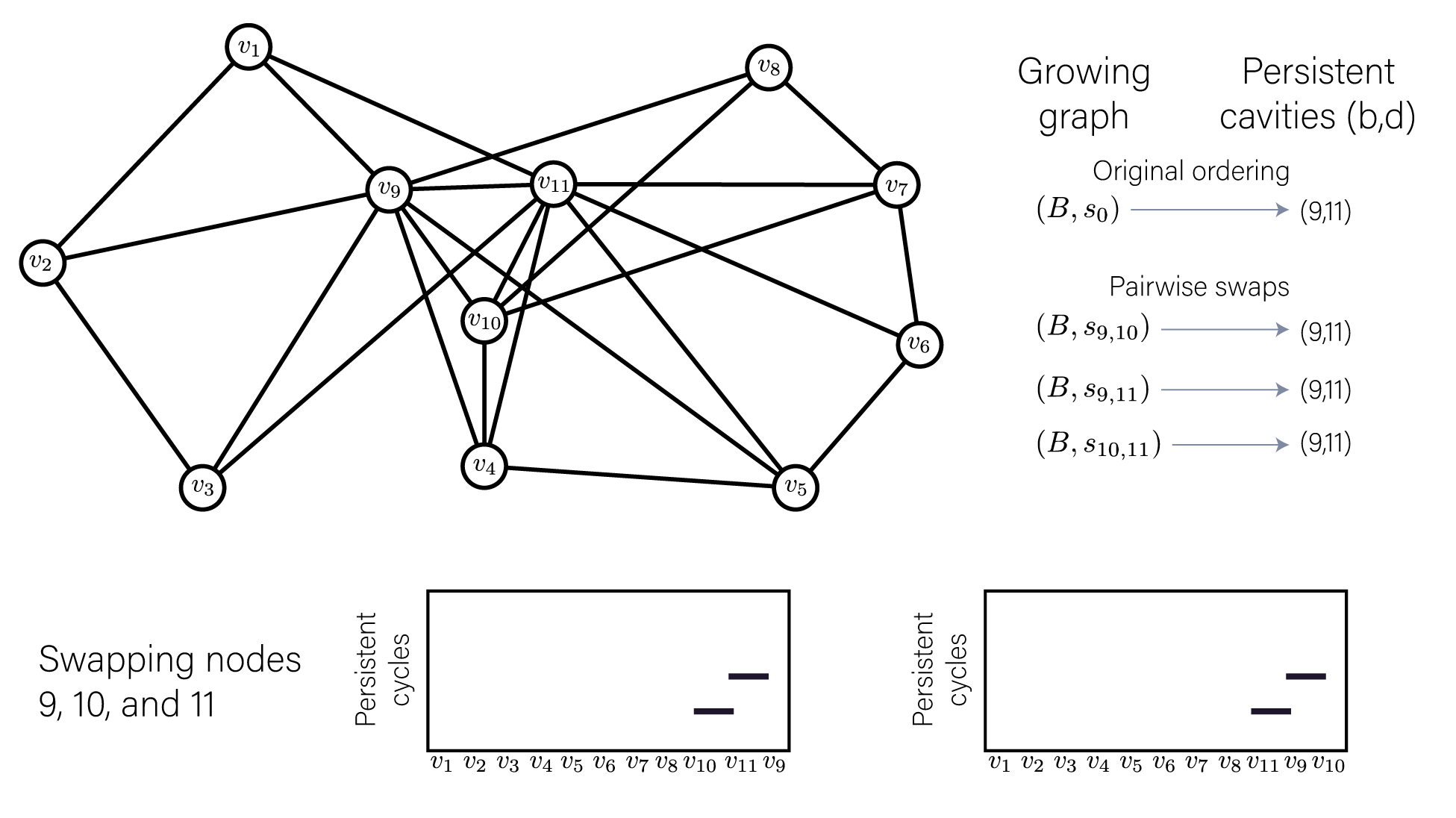}
	\caption{\textbf{Not all permutations of nodes within a 3-clique of $T_n$ conserve the persistent homology.} Nodes $v_9$, $v_{10}$, and $v_{11}$ are all pairwise topologically similar as shown by the persistent homology description (right). However, the permutations of these three nodes (shown in the bottom barcodes) produces different barcodes than those produced from the original ordering.}
	\label{fig:cex3}
\end{figure}

Continuing our investigation into how a clique in $T_n$ may determine the persistent homology after permutation of nodes within the clique, we now turn to observing the effect of such a permutation on the barcode generated by a growing graph. We see in the above 11-node graph (Fig.~\ref{fig:cex3}, left) that $v_9$, $v_{10}$, and $v_{11}$ are pairwise topologically similar, as swapping any pair results in the same barcode (Fig.~\ref{fig:cex3}, right). However, if these three nodes emerge in the network either in order $v_{10}$, $v_{11}$, $v_9$, or in order $v_{11}$, $v_9$, $v_{10}$, then we observe two bars of dimension 1 in the corresponding barcode (Fig.~\ref{fig:cex3}, bottom). Thus, we find that permutations of nodes' participation in the same clique of three nodes or more in $T_n$ does not necessarily preserve the persistent homology.

\subsection{Meet the team, extended}

\begin{figure}
	\centering
	\includegraphics[width=6.5in]{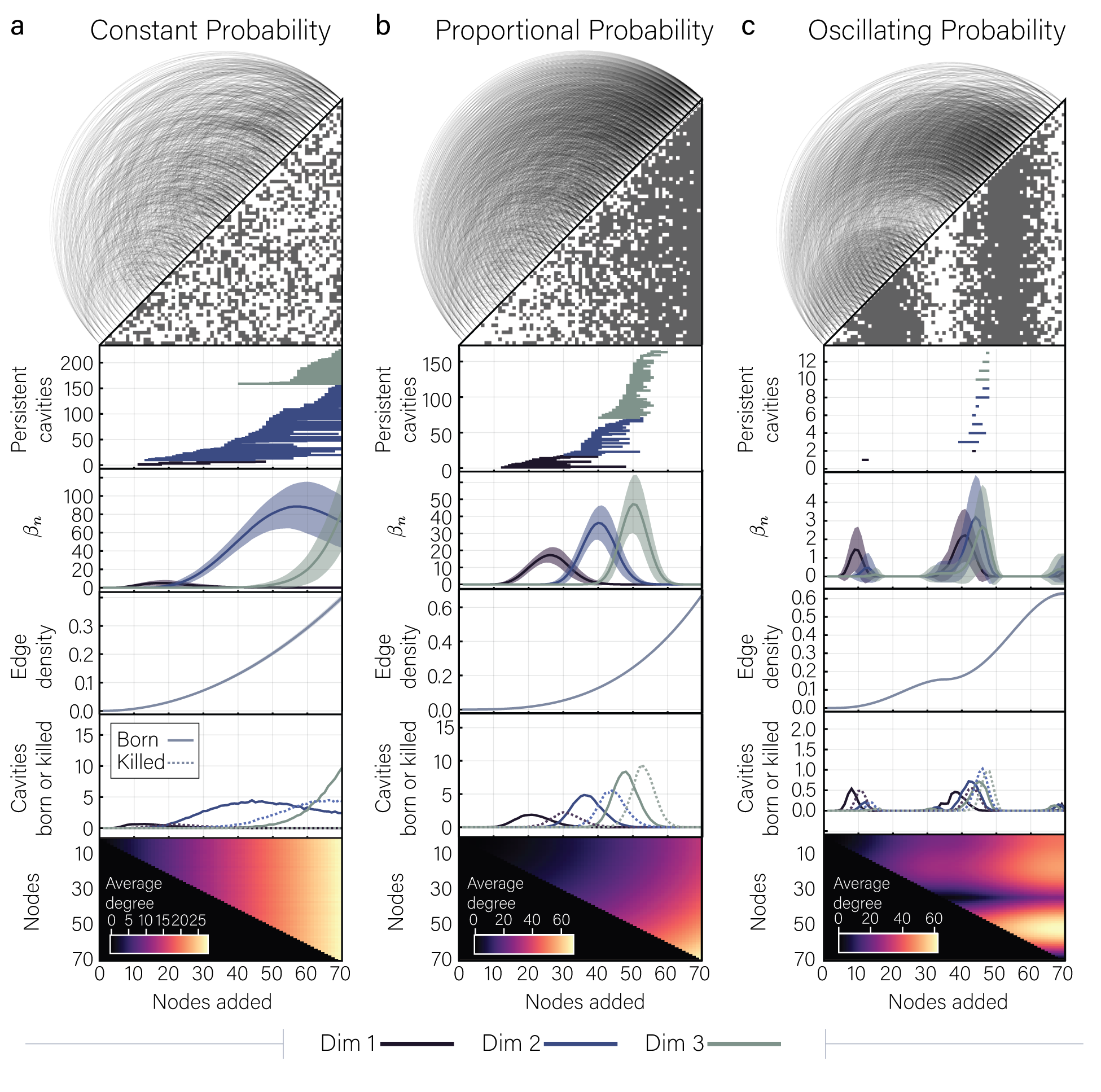}
	\caption{\textbf{Extended descriptions of the growing graph models: Part I.} Here we provide extended descriptions for the \emph{(a)} constant probability, \emph{(b)} proportional probability, and \emph{(c)} oscillating probability growing graph models. From top to bottom: fingerprint graph and binary adjacency matrix, barcode plot of one example growing graph, average Betti curves, edge density of the growing graph with respect to the final node count, plot showing the number of cavities born or killed at each node addition, and heatmap of evolving average degree for each node across the filtration. Results for dimension 1 shown in purple, dimension 2 shown in blue, and dimension 3 shown in green. }
	\label{fig:meet_the_team_extd1}
\end{figure}

\begin{figure}
	\centering
	\includegraphics[width=6.5in]{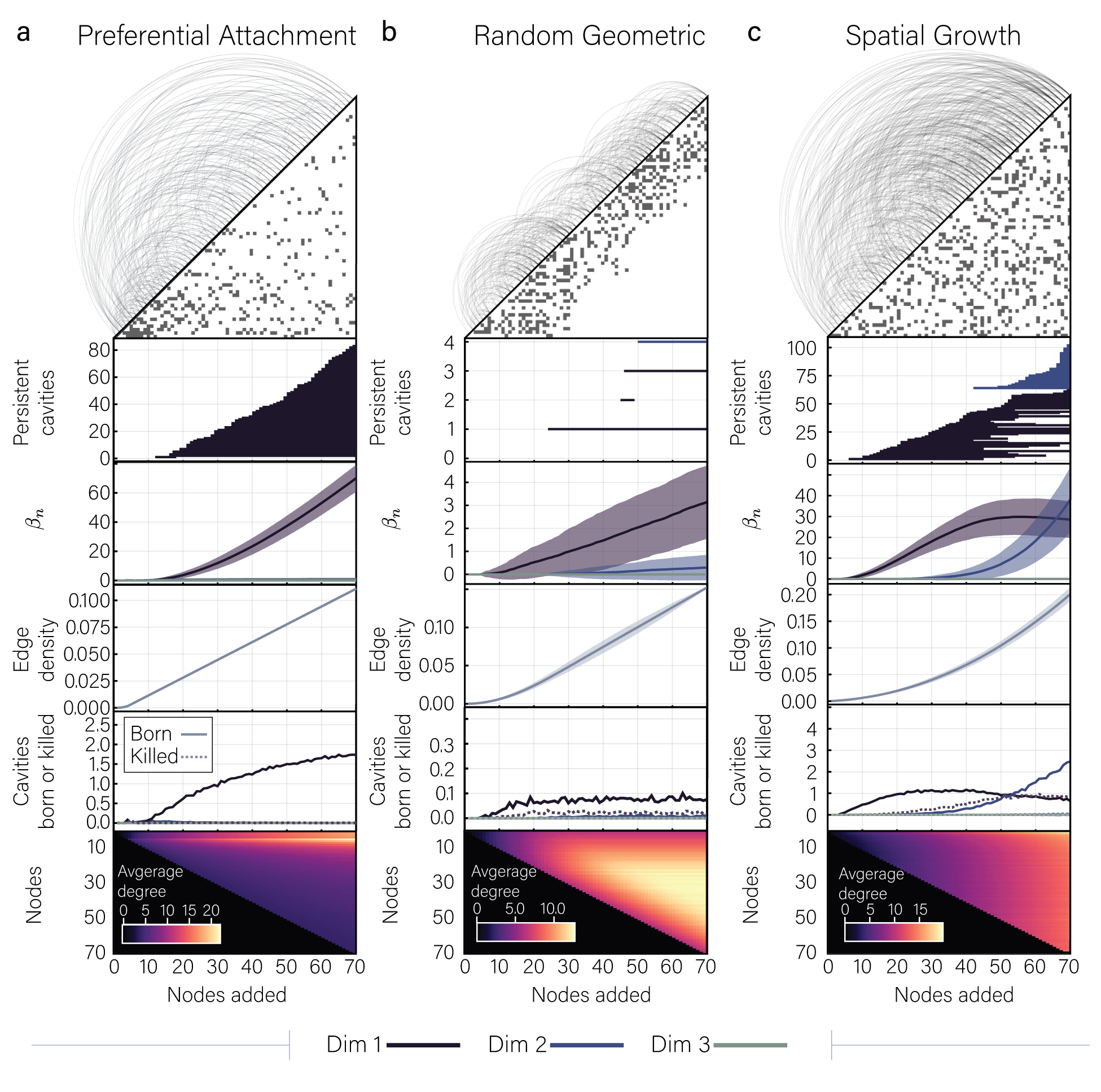}
	\caption{\textbf{Extended descriptions of the growing graph models: Part II.} Here we provide extended descriptions for the \emph{(a)} preferential attachment, \emph{(b)} random geometric, and \emph{(c)} spatial growth growing graph models. From top to bottom: fingerprint graph and binary adjacency matrix, barcode plot of one example growing graph, average Betti curves, edge density of the growing graph with respect to the final node count, plot showing the number of cavities born or killed at each node addition, and heatmap of evolving average degree for each node across the filtration. Results for dimension 1 shown in purple, dimension 2 shown in blue, and dimension 3 shown in green.}
	\label{fig:meet_the_team_extd2}
\end{figure}

In this section we include additional information about each of the six growing graph models defined and studied in the main text. We show results for the constant probability, proportional probability, and oscillating probability models in Fig.~\ref{fig:meet_the_team_extd1}, and for the preferential attachment, random geometric, and spatial growth models in Fig.~\ref{fig:meet_the_team_extd2}. We include the fingerprint graphs, adjacency matrix images, barcodes, and Betti curves from Fig.~\ref{fig:team1}, and we also add three more panels for each model. The first added panel records the average edge density of the growing graph after each node addition. Importantly, we compute the edge density at each step relative to the size of the final graph. That is, the edge density after the addition of $v_i$ is $\rho(i) = \text{edges added up through } v_i/\dbinom{N}{2}$ with $N  = 70$. The second additional panel shows the number of persistent cavities born (solid lines) or killed (dashed lines) after each node addition averaged across replicates. The final panel (last row in Fig.~\ref{fig:meet_the_team_extd1} and Fig.~\ref{fig:meet_the_team_extd2}), shows a heatmap of the evolving average degree of each node as the graph grows. More specifically, for each node $v_i$ we calculate the mean of the degree of $v_i$ after the addition of node $v_j$ across replicates. Repeating for all $i,j = 1,\dots, 70$, we recover the presented heatmap. For all plots, the horizontal axis follows graph growth.

\subsection{Additional experiments for investigating global and local reorderability}

\begin{figure}
	\includegraphics[width=6in]{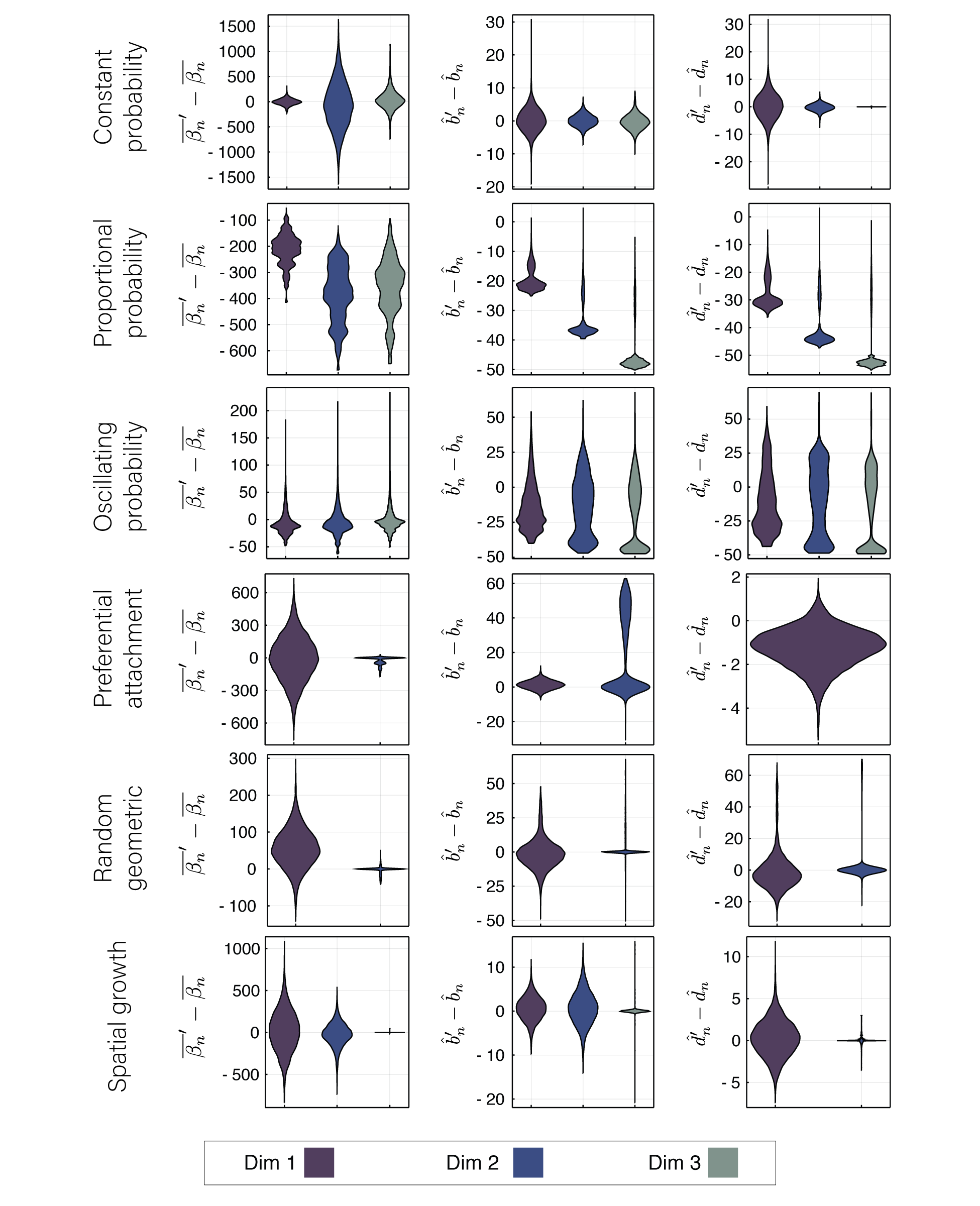}
	\caption{\textbf{Differences between Betti bar values, average birth, and average death times between the reordered and original growing graph persistent homology.} \emph{(Left)} Distributions of $\overline{\beta_n}' - \overline{\beta_n}$ values, \emph{(Middle)} distributions of $\hat{b}_n' - \hat{b}_n$ values, and \emph{(Right)} distributions of $\hat{d}_n' - \hat{d}_n$ values for each of the six growing graph models.}
	\label{fig:Supp3}
\end{figure}

After determining the reorderability of growing graph models, we naturally ask what types of changes to the barcodes or Betti curves lead to the distribution of distances shown in Fig.~\ref{fig:global_results1} and \ref{fig:global_results2}. Recall that $\overline{\beta}_n$ is the sum of all persistent cavity lifetimes. We first ask how $\overline{\beta_n}$ differs between the originally generated growing graphs and their associated reordered growing graphs. We show these distributions in the left column of Fig.~\ref{fig:Supp3}, specifically if $\overline{\beta_n}'$ is the Betti bar value for a reordered growing graph and $\overline{\beta_n}$ the original, we plot the distribution of $\overline{\beta_n}' - \overline{\beta_n}$ values. These plots help us to understand that the change in Betti curves of, for example, the globally reordered proportional probability graphs occurs in a manner that also decreases the value of $\overline{\beta_n}$ on average.

\begin{figure}
	\includegraphics[width=6in]{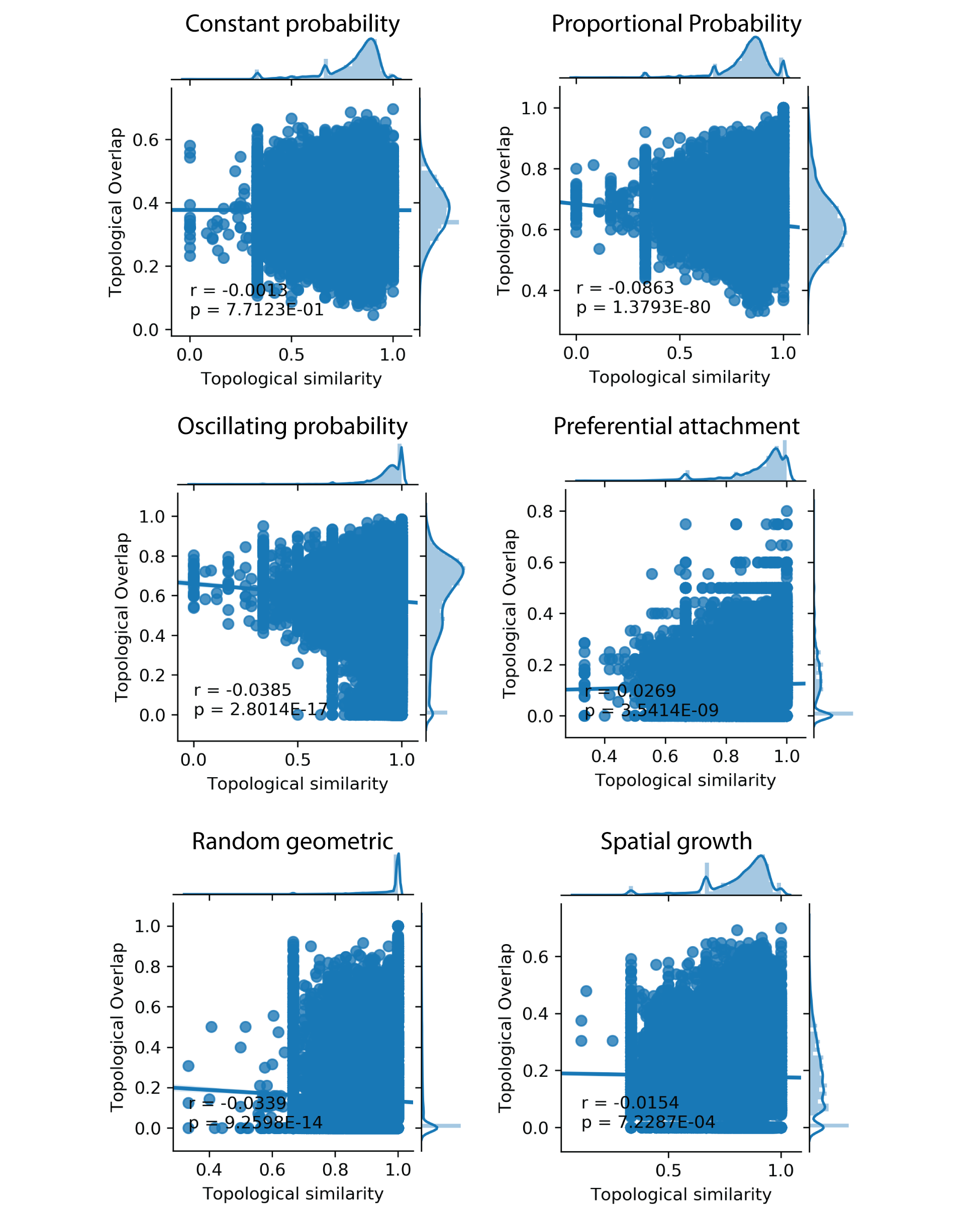}
	\caption{\textbf{Scatterplots and distributions of the averaged topological similarity and averaged topological overlap between node pairs.} Plots of topological similarity averaged over dimension between node pairs versus their computed topological overlap, calculated for each node pair in each growing graph model replicate. Line of best fit overlaid and Pearson correlation coefficient ($r$) and associated $p$-value $p$ displayed.}
	\label{fig:Supp1}
\end{figure}

Additionally we can perform the same calculation but for the average birth and death times of persistent cavities. Specifically, we calculate the average birth (death) time of all persistent cavities of dimension $n$ for the originally ordered growing graph $\hat{b}_n$ ($\hat{d}_n$), and we subtract this value from the average birth (death) time calculated from a random reordering of that growing graph $\hat{b}_n'$ ($\hat{d}_n'$). We repeat for all replicates and all reorderings and show the distributions as violin plots in the middle and right columns of Fig.~\ref{fig:Supp1}. As an example, for the preferential attachment model in dimension 1 these plots reveal that after reordering, on average we decrease the average death time of persistent cavities and often increase average birth times. 

When investigating how the persistent homology of growing graph models changes in response to node swaps, we intuitively expect node pairs with similar connection patterns to be the most topologically similar. In order to test this intuition, we compute the average topological overlap $O(v_i,v_j)$ of each node pair across binary networks generated from the growing graph model, and we plot this value against topological similarity averaged across dimensions, in Fig.~\ref{fig:Supp1}. For each growing graph model we also show the calculated Pearson correlation $r$ and associated $p$-value along with the line of best fit; on the top and right axes we plot the marginal distributions of each variable. We see in the topological overlap versus topological similarity scatterplots (Fig.~\ref{fig:Supp1}) that for most growing graph models, topological overlap is not a strong predictor of topological similarity.

Finally we note that for both the preferential attachment and proportional probability models, in Fig.~\ref{fig:LtoG_results1} we observe by eye that the degree of a node in $B$ looks to affect the overall topological similarity of that node. In order to investigate this potential relation further, we calculate the degree of each $v_i$ in each $B_{\alpha}$ and plot this value against the summed topological similarity (averaged over dimensions) of node $v_i$ for that replicate. Note that this statistic is equivalent to the strength or weighted degree of node $v_i$ in the averaged topological similarity graph constructed for each of the 20 replicates. For each growing graph model we also include the calculated Pearson correlation $r$ and $p$-value along with the line of best fit; on the top and right axes we plot the marginal distributions of each variable. We find that for most graph models, the average degree of a node poorly predicts the ability of that node to swap with all other nodes while preserving the persistent homology. Interestingly, the proportional probability model shows a clear U-shaped distribution of points suggesting that the lowest degree nodes and highest degree nodes can often swap with other nodes without changing the persistent homology much (relative to all nodes), while it is the mid-range degree nodes that are the least swappable.

\end{document}